\documentclass[longauth]{aa}

\usepackage{graphicx}
\usepackage{txfonts}
\usepackage{booktabs}
\usepackage[flushleft]{threeparttable}
\usepackage[table]{xcolor}
\usepackage{colortbl}
\usepackage{array}
\usepackage{subfig}
\usepackage{float}
\definecolor{cyan}{rgb}{0.88,1.0,1.0}
\definecolor{yellow}{rgb}{1.0,1.0,0.0}
\usepackage{multirow}
\usepackage{multicol}
\usepackage{blindtext}
\usepackage{longtable}

\newcommand{\teff}{$T\rm{_{eff}}$\xspace}
\newcommand{\logg}{$\log g$\xspace}
\newcommand{\kms}{km\,s$^{-1}$\xspace}
\newcommand{\vsini}{$v\,{\rm sin} i$\xspace}

\newcommand{\logtau}{$\log\,\tau_{\rm{5000}}$\xspace}
\newcommand{\ges}{\textit{Gaia}-ESO\xspace}
\newcommand{\logrhk}{$\log\,\rm{R^{\prime}_{HK}}$\xspace}

\begin{document}

    \title{The \textit{Gaia}-ESO Survey: A new approach to chemically characterising young open clusters\thanks{Based on observations collected with the FLAMES instrument at VLT/UT2 telescope
(Paranal Observatory, ESO, Chile), for the Gaia- ESO Large Public Spectroscopic Survey (188.B-3002, 193.B-0936).}}

   \subtitle{II. Abundances of the neutron-capture elements Cu, Sr, Y, Zr, Ba, La, and Ce}

   \author{M. Baratella\inst{\ref{unipd},\ref{oapd}}
   \and V. D'Orazi\inst{\ref{oapd},\ref{monash}}
   \and V. Sheminova\inst{\ref{sheminova}}
   \and L. Spina\inst{\ref{oapd},\ref{monash}}
   \and G. Carraro\inst{\ref{unipd}}
   \and R. Gratton\inst{\ref{oapd}}
   \and L. Magrini\inst{\ref{arcetri}}
   \and S. Randich\inst{\ref{arcetri}}
   \and M. Lugaro\inst{\ref{konkoly},\ref{ELTE},\ref{monash}}
   \and M. Pignatari\inst{\ref{hull},\ref{konkoly},\ref{nugrid},\ref{jina}}
   \and D. Romano\inst{\ref{bologna}}
   \and K. Biazzo\inst{\ref{oaro}}
   \and A. Bragaglia\inst{\ref{bologna}}
   \and G. Casali\inst{\ref{arcetri}}
   \and S. Desidera\inst{\ref{oapd}}
   \and A. Frasca\inst{\ref{oact}}
   \and G. de Silva\inst{\ref{macquire},\ref{macquire_2}}
   \and C. Melo\inst{\ref{eso_chile}}
   \and M. Van der Swaelmen\inst{\ref{arcetri}}
   \and G. Tautvai{\v s}ien{\. e}\inst{\ref{vilnius}}
   \and F. M. Jim\'{e}nez-Esteban\inst{\ref{fran_institution}}
   \and G. Gilmore\inst{\ref{uni_cambridge}}
   \and T. Bensby\inst{\ref{lund}}
   \and R. Smiljanic\inst{\ref{nicolaus}}
   \and A. Bayo\inst{\ref{uni_valparaiso},\ref{valparaiso}}
   \and E. Franciosini\inst{\ref{arcetri}}
   \and A. Gonneau\inst{\ref{uni_cambridge}}
   \and A. Hourihane\inst{\ref{uni_cambridge}}
   \and P. Jofr\'e\inst{\ref{diego_portales}}
   \and L. Monaco\inst{\ref{andres_bello}}
   \and L. Morbidelli\inst{\ref{arcetri}}
   \and G. Sacco\inst{\ref{arcetri}}
   \and L. Sbordone\inst{\ref{eso_chile}}
   \and C. Worley\inst{\ref{uni_cambridge}}
   \and S. Zaggia\inst{\ref{oapd}}
   }

 \institute{Dipartimento di Fisica e Astronomia {\it Galileo Galilei}, Vicolo Osservatorio 3, I-35122, Padova, Italy \label{unipd}\\
\email{martina.baratella.1@phd.unipd.it}
    \and 
    INAF -- Osservatorio Astronomico di Padova, vicolo dell'Osservatorio 5, 35122, Padova, Italy \label{oapd}
  \and
  Monash Centre for Astrophysics (MoCA), Monash University, School of Physics and Astronomy, Clayton, VIC 3800, Melbourne, Australia \label{monash}
  \and 
  Main Astronomical Observatory, National Academy of Sciences of Ukraine, Akademika Zabolotnoho 27, Kyiv, 03143 Ukraine \label{sheminova}
  \and 
  INAF -- Osservatorio Astrofisico di Arcetri, Largo E. Fermi 5, 50125, Firenze, Italy \label{arcetri}
  \and 
  Konkoly Observatory, Research Centre for Astronomy and Earth Sciences, E\"otv\"os Lor\'and Research Network (ELKH), Konkoly Thege Mikl\'{o}s \'{u}t 15-17, H-1121 Budapest, Hungary \label{konkoly}
  \and
  ELTE E\"{o}tv\"{o}s Lor\'and University, Institute of Physics, Budapest 1117, P\'azm\'any P\'eter s\'et\'any 1/A, Hungary \label{ELTE}
  \and 
  E.~A.~Milne Centre for Astrophysics, Department of Physics and Mathematics, University of Hull, HU6 7RX, United Kingdom \label{hull}
    \and 
  NuGrid Collaboration, \url{http://nugridstars.org} \label{nugrid}
    \and 
  Joint Institute for Nuclear Astrophysics - Center for the Evolution of the Elements \label{jina}
  \and
  INAF -- Osservatorio di Astrofisica e Scienza dello Spazio, via P. Gobetti 93/3, 40129 Bologna (Italy) \label{bologna}
  \and
  INAF -- Osservatorio Astronomico di Roma, via Frascati 33, I-00040, Monte Porzio Catone (RM), Italy \label{oaro}
  \and
  INAF–Osservatorio Astrofisico di Catania, via S. Sofia 78, 95123 Catania, Italy \label{oact}
  \and
  Australian Astronomical Optics, Faculty of Science and Engineering, Macquarie University, Macquarie Park, NSW 2113, Australia \label{macquire}
  \and
  Macquarie University Research Centre for Astronomy, Astrophysics $\&$ Astrophotonics, Sydney, NSW 2109, Australia \label{macquire_2}
  \and
  European Southern Observatory, Alonso de Cordova 3107, Vitacura, Santiago, Chile \label{eso_chile}
  \and 
  Institute of Theoretical Physics and Astronomy, Vilnius University, Sauletekio av. 3, 10257 Vilnius, Lithuania \label{vilnius}
  \and
  Departmento de Astrof\'{\i}sica, Centro de Astrobiolog\'{\i}a (INTA-CSIC), ESAC Campus, Camino Bajo del Castillo s/n, E-28692 Villanueva de la Cañada, Madrid, Spain \label{fran_institution}
  \and
  Institute of Astronomy, University of Cambridge, Madingley Road, Cambridge CB3 0HA, United Kingdom \label{uni_cambridge}
  \and
  Lund Observatory, Department of Astronomy and Theoretical Physics, Box 43, SE-221\,00 Lund, Sweden \label{lund}
  \and
  Nicolaus Copernicus Astronomical Center, Polish Academy of Sciences, ul. Bartycka 18, 00-716, Warsaw, Poland \label{nicolaus}
  \and
  Instituto de F\'isica y Astronom\'ia, Universidad de Valparaiso, Gran
Breta\~na 1111, Valpara\'iso\label{uni_valparaiso}
  \and
  N\'ucleo Milenio de Formaci\'on Planetaria, NPF, Universidad de
Valpara\'iso\label{valparaiso}
  \and
  N\'ucleo de Astronom\'{i}a, Facultad de Ingenier\'{i}a, Universidad Diego Portales, Av. Ej\'ercito 441, Santiago, Chile \label{diego_portales}
  \and
  Departamento de Ciencias Fisicas, Universidad Andres Bello, Fernandez Concha 700, Las Condes, Santiago, Chile \label{andres_bello}
}

   \date{Received ??; accepted ??}

 
  \abstract
   {Young open clusters (ages of less than 200\,Myr) have been observed to exhibit several peculiarities in their chemical compositions. These anomalies include a slightly sub-solar iron content, super-solar abundances of some atomic species (e.g. ionised chromium), and atypical enhancements of [Ba/Fe], with values up to $\sim+0.7$\,dex.  
   Regarding the behaviour of the other $s$-process elements like yttrium, zirconium, lanthanum, and cerium, there is general disagreement in the literature: some authors claim that they follow the same trend as barium, while others find solar abundances at all ages.}
   {In this work we expand upon our previous analysis of a sample of five young open clusters (\object{IC\,2391}, \object{IC\,2602}, \object{IC\,4665}, \object{NGC\,2516,} and \object{NGC\,2547}) and one star-forming region (\object{NGC\,2264}), with the aim of determining abundances of different neutron-capture elements, mainly Cu\,{\sc i}, Sr\,{\sc i}, Sr\,{\sc ii}, Y\,{\sc ii}, Zr\,{\sc ii}, Ba\,{\sc ii}, La\,{\sc ii}, and Ce\,{\sc ii}.
   For \object{NGC\,2264} and \object{NGC\,2547} we present  the measurements of these elements for the first time.}
   {We analysed high-resolution, high signal-to-noise spectra of 23 solar-type stars observed within the \ges survey. After a careful selection, we derived abundances of isolated and clean lines via spectral synthesis computations and in a strictly differential way with respect to the Sun.   }
   {We find that our clusters have solar [Cu/Fe] within the uncertainties, while we confirm that [Ba/Fe] is super-solar, with values ranging from +0.22 to +0.64\,dex. Our analysis also points to a mild enhancement of Y, with [Y/Fe] ratios covering values between 0 and +0.3\,dex. For the other $s$-process elements we find that [X/Fe] ratios are solar at all ages.}
   {It is not possible to reconcile the anomalous behaviour of Ba and Y at young ages with standard stellar yields and Galactic chemical evolution model predictions. 
   We explore different possible scenarios related to the behaviour of spectral lines, from the dependence on the different ionisation stages and the sensitivity to the presence of magnetic fields (through the Landé factor) to the first ionisation potential (FIP) effect. We also investigate   the possibility that they may arise from alterations of the structure of the stellar photosphere due to the increased levels of stellar activity that affect the spectral line formation, and consequently the derived abundances. These effects seem to be stronger in stars at ages of less than $\sim 100$\,Myr. However, we are still unable to explain these enhancements, and the Ba puzzle remains unsolved. With the present study we suggest that other elements, for example Sr, Zr, La, and Ce, might be more reliable tracer of the $s$-process at young ages, and we strongly encourage further critical observations.} 


   \maketitle
%

\section{Introduction}\label{sec:introduction}

Open clusters (OCs) are excellent tracers of the chemical properties of the Galactic disc and their time evolution. 
Thanks to dedicated spectroscopic surveys (e.g. the APO Galactic Evolution Experiment, APOGEE, \citealt{2016cunha}, \citealt{2018donor}, \citealt{2019carrera}; the Open Clusters Chemical Abundances from Spanish Observatories, OCCASO, \citealt{2019casamiquela}; GALactic Archaelogy with HERMES, GALAH, \citealt{2021spina}) we can analyse these systems with a large amount of data. In particular, within the \ges Survey \citep{ges,2013randich} almost 80 OCs with up to 100 members, spanning ages between a few million to several billion years, have been homogeneously analysed.
However, different studies over the last 15 years seem to indicate that young stars within 500 pc share a slightly sub-solar metal content (with [Fe/H] between $-0.05$ and $-0.10$\,dex), both in OCs, moving groups and associations (e.g. \citealt{2006james,2008santos,2011biazzoA,2014spinaB,2014spinaA,2017spina}). This is in contrast with what is expected from the standard Galactic chemical evolution (GCE) models that predict an enrichment of the interstellar medium of 0.10-0.15 dex over the last 4-5\,Gyr (e.g. \citealt{2013minchev}). 

Another intriguing aspect of young OCs (YOCs, i.e. OCs with ages $\lesssim$200\,Myr) is the behaviour of the elements mainly produced via the slow neutron-capture process \citep[hereafter $s$-process elements,][and references therein]{2011kappeler}.
Early analytical models found that the solar system abundances of the whole $s$-process elements could be explained by the contribution of the weak, the main and the strong components. The weak component accounts for the formation of elements up to the atomic mass A$\sim$90 (from Fe to Sr) and it takes place mostly in massive stars during convective He core and C shell burning phases (e.g. \citealt{the:07}, \citealt{2010pignatari}; \citealt{sukhbold:16}; \citealt{limongi2018}). 
Most of the  copper, gallium, and germanium in the solar system is made by the weak $s$-process in massive stars \citep[][]{2010pignatari}. In particular, copper was thought to be mostly made by thermonuclear supernovae since the $s$-process contribution was limited \citep[][]{matteucci:93}. However, thanks to a new generation of neutron-capture reaction rates the $s$-process production of copper in massive stars was revised \citep[][]{heil:08}. Therefore, present $s$-process calculations in massive stars means that the missing copper and the solar abundances can be explained \citep[][]{bisterzo:05,2007romano,2010pignatari}.  

Elements with A$\sim90-208$ traditionally belong to the main and strong components \citep[e.g.][]{1998gallino,bisterzo:14,2020kobayashi}, which are associated with low-mass asymptotic giant branch (AGB) stars ($\approx$ 1.5 - 4 M$_{\odot}$), during the thermally pulsating phase (e.g. \citealt{2012lugaro}; \citealt{karakas2014}). Rubidium, strontium, yttrium, and zirconium (with atomic number 37$\leq$ Z $\leq$40) belong to the first peak of the $s$-process in the solar abundance distribution; barium, lanthanum, cerium, praseodymium, and neodymium (56 $\leq$ Z $\leq$ 60) populate the second-peak; finally lead and bismuth are at the third peak. The three peaks correspond to the neutron magic numbers N=50, 82, and 126.

Starting from the pioneering work of \cite{2009dorazi}, it has been confirmed that the observed [Ba/Fe] ratios dramatically increase at decreasing ages, reaching values up to +0.60\,dex in very young clusters like \object{IC\,2391} and \object{IC\,2602} (ages of $\sim 30-50$ Myr). Conversely, older clusters with ages $\gtrsim$ 1\,Gyr exhibit solar-scaled abundances. These extraordinary enhancements cannot be explained, neither with non-local thermodynamic equilibrium (NLTE) effects nor with stellar nucleosynthesis and GCE models \citep{1999travaglio,2001busso}. As discussed in \cite{2009dorazi}, increasing the stellar yields by a factor $\sim$6 for AGB stars with masses of $1-1.5\,\rm{M}_{\odot}$, a GCE model is able to reproduce the observed abundances up to $500-600$\,Myr, but not the measured massive  overproduction of Ba in the last $50-100$\,Myr. The Ba overabundance has subsequently been confirmed by other studies (e.g. \citealt{2012yong, 2013jacobson}, among the others).

Interest has also moved toward the behaviour of other $s$-process elements (Y, Zr, La, and Ce). At present the abundance evolution of these elements with respect to age is matter of debate. \cite{2011maiorca} measured abundance ratios for these elements in a sample of 19 OCs, with ages from 0.7 to 8.4\,Gyr, and they found a steep growth at younger ages. Similar conclusions have been reached by \cite{2018magrini}, who analysed a sample of 22 OCs with ages spanning from 0.1 to 7\,Gyr. Recently, \cite{2019frasca} studied the young cluster ASCC\,123 (age $\sim 150$\,Myr) and found an overabundance of Sr, Y, and Zr, with values between +0.3 and +0.5\,dex. On the other hand, other studies have confirmed that young clusters and local moving groups display first and second peak elements with   different behaviour to Ba, in all cases showing solar values of Y, Zr, La, and Ce (e.g. \citealt{2012dorazi,2012yong,2013jacobson, 2017dorazi}). \cite{2015reddy} analysed stars belonging to five local associations ($5-200$\,Myr) and they found a large spread in [Ba/Fe] ratios, from +0.07 to +0.32\,dex. \cite{2015mishenina} again confirmed the trend of increasing Ba at decreasing ages from the analysis of giant stars in five OCs, together with solar-like abundances of La. From the stellar nucleosynthesis point of view the most puzzling signature to explain is not the intrinsic enrichment of Ba, but the production of Ba disentangled from La. For pure nuclear physics reasons, this cannot be done in $s$-process conditions \citep[e.g.][]{2011kappeler}. At the same time, observations of old metal-poor $r$-process-rich stars have confirmed that  the $r$-process co-produces Ba and La in similar amounts \citep[e.g.][and references therein]{sneden:08}. In the light of these considerations, \cite{2015mishenina} proposed the intermediate ($i$-) process as a possible explanation of the Ba enrichment in OCs.

The $i$-process was first introduced by \cite{1977cowan}, and it is characterised by neutron densities that are intermediate between the $s$-process and the $r$-process, of the order of 10$^{14-16}$ neutrons cm$^{-3}$. Under these conditions Ba production is disentangled from La, and it is indeed possible to reproduce the high [Ba/La] ratios seen in stars hosted by YOCs \citep[][]{2013bertolli,denissenkov:21}. Different types of stars have been proposed as possible stellar hosts of the $i$-process: post-AGB stars \citep[][]{herwig:11} and low-mass AGB stars \citep[e.g.][]{lugaro:15,cristallo:16,choplin:21}, super-AGB stars \citep[][]{jones:16}, rapidly-accreting white dwarfs \citep[e.g.][]{denissenkov:17,cote:18,denissenkov:19} and massive stars \citep[][]{roederer:16,clarkson:18,banerjee:18}.
However, in the context of YOCs, which stellar site where the $i$-process has become so relevant only in the last $\sim$Gyr is still a mystery. From the analysis of solar-twin stars, \cite{2017reddy} found a mild increase in La, Ce, Nd, and Sm with decreasing ages, while the trend for [Ba/Fe] is more evident and confirms all the previous findings.  They also provided an important piece of evidence in trying to solve the so-called barium puzzle. These authors detected a positive correlation between the activity index of the stars and their [Ba/Fe] ratios. A similar trend between [Ba/H] and chromospheric and accretion diagnostics were also found in the Lupus star-forming region (SFR) by \cite{2017biazzo}. 

In \cite{2020baratellaA} (hereafter Paper\,I) we demonstrated that the higher levels of stellar activity could affect the formation of spectral lines forming in the upper layers of the photosphere. The difference of the equivalent width (EW) of the strong Fe lines between a 30\,Myr solar analogue and the Sun increases at decreasing optical depth (i.e. moving up in the photosphere). The direct consequence is that the microturbulence velocity ($\xi$) parameter\footnote{This is a free fictitious parameter introduced in the 1D spectroscopic analysis to account for the difference between the observed and predicted EWs of atomic lines, mostly due to small-scale (compared to the mean free path of the photons) motions of matter in the stellar photospheric layers.} should be increased when derived by imposing that strong and weak Fe lines provide the same abundance. Values of the order of 2.0-2.5 \kms have been found for $\xi$ in young solar-type, main-sequence stars. The net effect is an underestimation of [Fe/H], with the various abundance ratios [X/Fe] rescaling accordingly. The same result was confirmed by \cite{2019galarza} and \cite{2020spina}, who proposed that the magnetic intensification could be responsible for the observed patterns. In both studies intermediate-age stars ($\sim 400$\,Myr) were analysed; in our case we are dealing with much younger stars and the effects of activity could be so intense that the solution of magnetic intensification is not sufficient.

\begin{figure*}[!htb]
     \centering
      \subfloat{\includegraphics[scale=0.32]{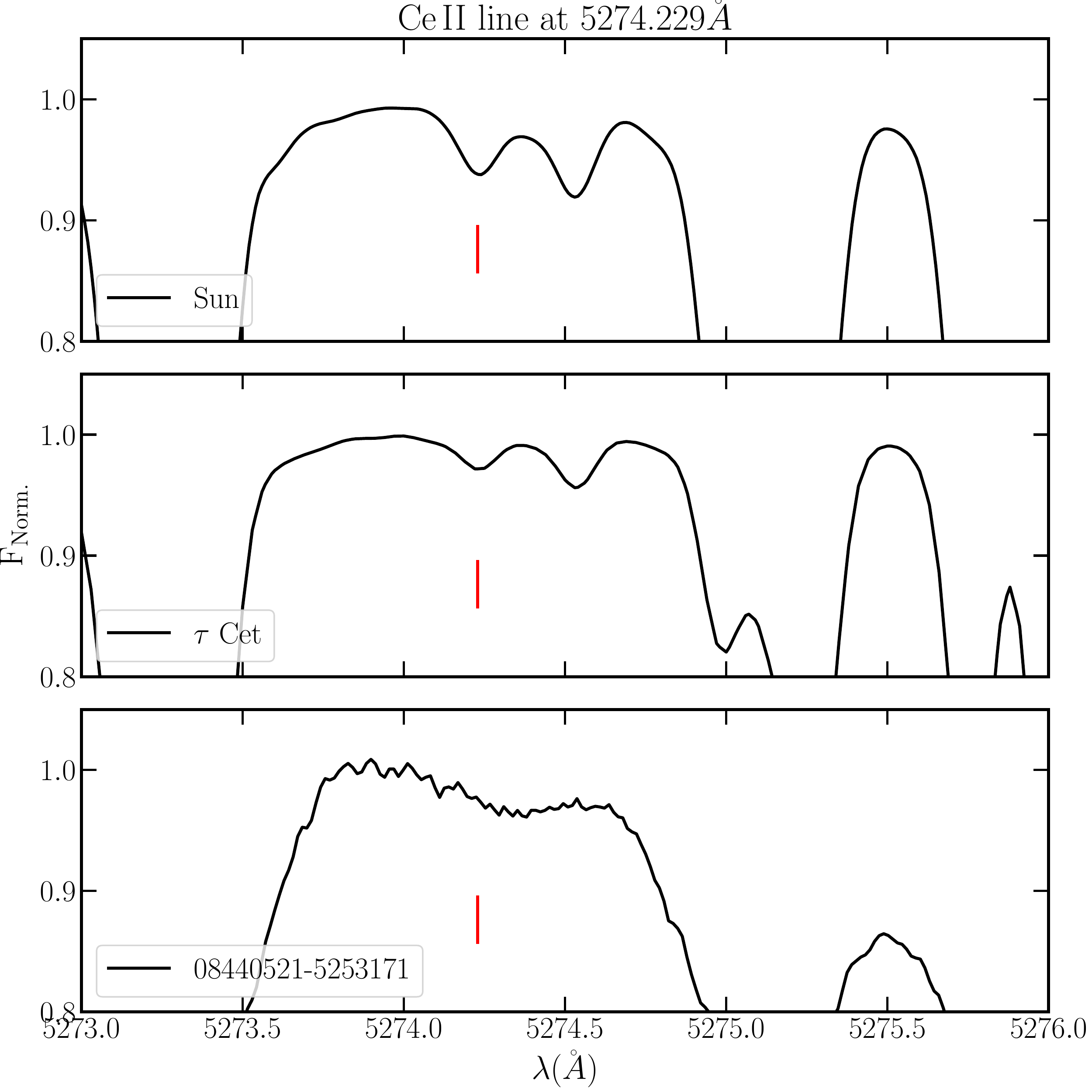}}
      \qquad
      \subfloat{\includegraphics[scale=0.32]{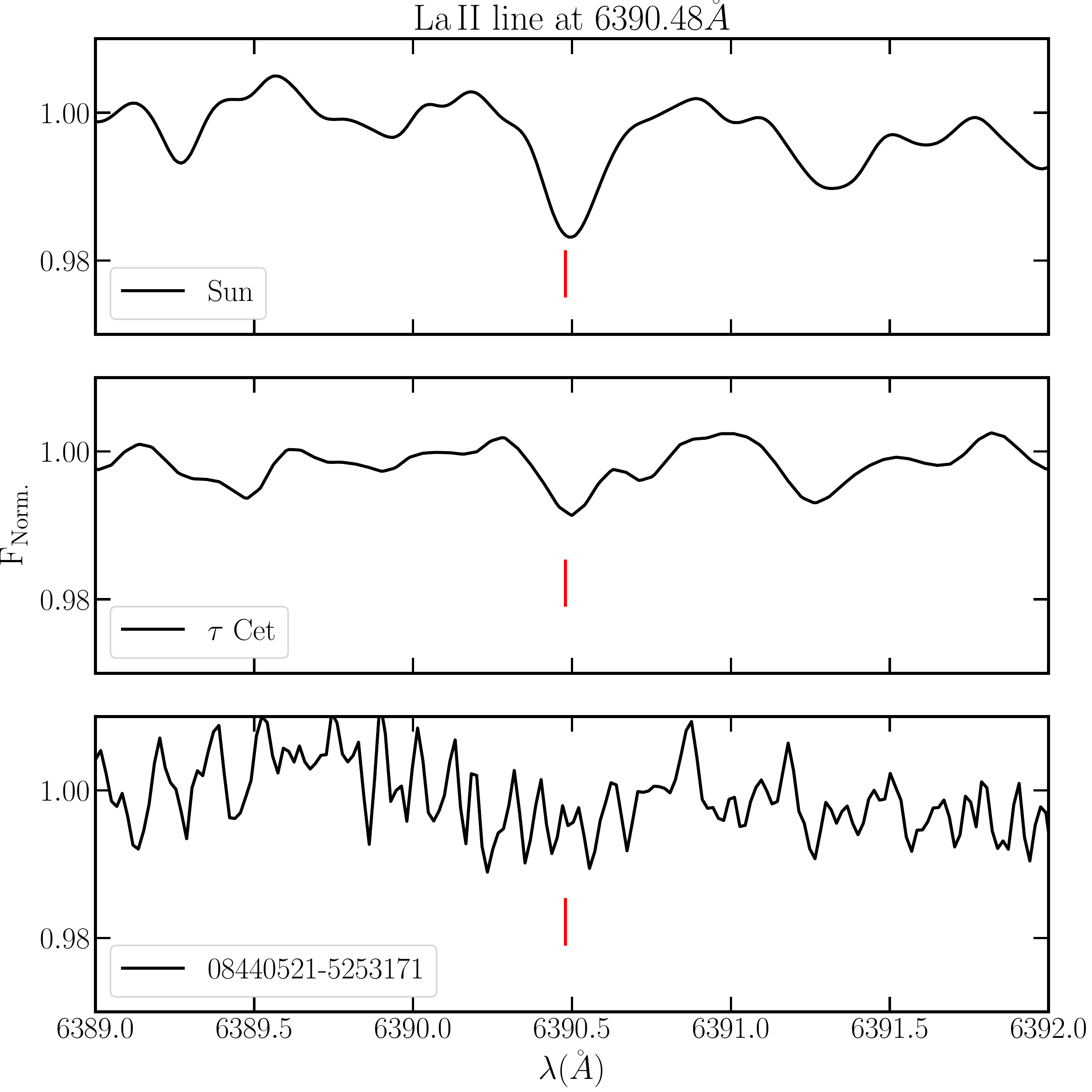}}
      \caption{ Comparison of spectra of the Sun (top), $\tau Cet$ (central), and the star 08440521-5253171 (bottom) of \object{IC\,2391} (S/N=260) in two spectral regions. Shown on the left a 3\,\AA\, window around the Ce\,{\sc ii} line 5274.23\,\AA, while on the right a region near the La\,{\sc ii} line at 6390.48\,\AA. }
        \label{lines}
\end{figure*}

\begin{table*}[]
\caption{List of selected spectral lines, with complete  atomic data. }   
\label{line_list}      
\centering  
\setlength\tabcolsep{11pt}
\small
\begin{tabular}{lccrcccr }    
\toprule
Element & $\lambda$\,(\AA) & E.P.\,(eV) & $\log gf$ & Ref. $\log gf$ &  $\log (\rm{X})_{\odot}$ &  $g_{\rm{L}}$ & FIP\,(eV)$^a$\\
\midrule
&&&&\textbf{3800 - 4800\,\AA (archive)}\\
Sr\,{\sc i} & 4607.33 & 0.00 & 0.28 & \cite{2012bergemann}& $2.72\pm0.07\pm0.07$(LTE) & 1.00 & 5.69  \\
 &  &  &  &  & $2.82\pm0.07\pm0.07$(NLTE)  \\
Sr\,{\sc ii} & 4215.52 & 0.00 & $-$0.16 & \cite{2012bergemann} & $2.87\pm0.18\pm0.1$(LTE) & 1.33 & 5.69 \\
Y\,{\sc ii } & 4398.01 & 0.13 & $-$1.00 &  \cite{1982hannaford}  & $2.26\pm0.10\pm0.08$ & 1.00 & 6.38\\
Zr\,{\sc ii} & 4050.32 & 0.71 & $-$1.06 & \cite{LNAJ} &  $2.55\pm0.08\pm0.06$ & 0.90 & 6.84\\
Zr\,{\sc ii} & 4208.98 & 0.71 & $-$0.51 & \cite{LNAJ} &  $2.58\pm0.11\pm0.05$ &0.86 & 6.84\\
La\,{\sc ii} & 3988.51 & 0.40 & 0.21 &  \cite{LBS}  & $1.05\pm0.07\pm0.07$ & 1.33 & 5.58\\
La\,{\sc ii} & 4086.71 & 0.00 & $-0.07$ & \cite{LBS} &  $1.10\pm0.10\pm0.06$ & 0.83 & 5.58\\
Ce\,{\sc ii} & 4073.47 &  0.48 & 0.21 & \cite{LSCI} &  $1.55\pm0.08\pm0.07$ & -$^{b}$ & 5.47\\
\hline
\hline
&&&&\textbf{4800 - 6800\,\AA (\ges)}\\
Cu\,{\sc i} & 5105.54 & 1.39 & $-$1.52 & \cite{2011kurucz}  & $4.21\pm0.12\pm0.07$ & 1.10 & 7.72\\ 
Y\,{\sc ii } & 4883.69 & 1.08 & 0.07 & \cite{1982hannaford}  & $2.19\pm0.08\pm0.06$ & 1.13 & 6.38 \\
Y\,{\sc ii } & 5087.42 & 1.08 & $-$0.17 & \cite{1982hannaford}  & $2.16\pm0.07\pm0.05$ & 1.25 & 6.38\\
Zr\,{\sc ii} & 5112.27 & 1.67 & $-$0.85 & \cite{LNAJ}  & $2.59\pm0.06\pm0.02$ & 0.80 & 6.84\\
Ba\,{\sc ii} & 5853.69 & 0.60 & $-$1.01 & \cite{1998mcwilliam_Ba}  & $2.31\pm0.08\pm0.10$ & 1.07 & 5.21\\
Ce\,{\sc ii} & 5274.23 & 1.04 & 0.13 & \cite{LSCI}  & $1.64\pm0.06\pm0.05$ & -$^b$ & 5.47\\
\bottomrule
\end{tabular}
\tablefoot{Atomic data of the lines used in the analysis, both for the bluer ($3800-4800$\,\AA) and the redder ($4800-6800$\,\AA) ranges. The solar abundances obtained from our analysis are shown in Column 6. The Landé factor and the FIP are reported in Columns 7 and 8, respectively.}
\begin{tablenotes}
\small
\item $^a$ : FIP values are taken from Table D.1 of \cite{1992gray}
\item $^b$ : Values of the Landé factor for Ce\,{\sc ii} lines have not been computed due to the complex term of its levels
\end{tablenotes}
\end{table*}

In Paper\,I we used the \ges available reduced spectra of solar-type stars belonging to five YOCs and one SFR, and developed a new spectroscopic approach to overcome the above-mentioned  issues. Here, we expand the analysis of these stars to derive the abundances of the heavy elements Cu, Sr, Y, Zr, Ba, La, and Ce. The main goal of this new investigation is to shed light on the behaviour of the $s$- process dominated elements. To our knowledge, for the cluster \object{NGC\,2547} and the SFR \object{NGC\,2264} no previous studies focusing on the heavy element abundances have been published to  date. The exceptions are \object{IC\,2391} and \object{IC\,2602}, which will be used as calibrators; \object{NGC\,2516}, for which only few elements have been investigated in the past; and \object{IC\,4665}, recently analysed by \cite{2021spina} within the GALAH DR3 \citep{2021buder}.

The paper is organised as follows. In Sect. \ref{sec:data} we present the stellar sample analysed, adopting the procedure described in Sect. \ref{sec:analysis}:  line-list selection, computation of the optical depth of line formation, and measurement of the element abundance. Our results, along with a comparison with literature estimates, are reported in Sect. \ref{sec:results}. We discuss our findings and their scientific implications in Sect. \ref{sec:discussion}, while Sect. \ref{sec:conclusions} contains our conclusions.


\section{Data}\label{sec:data}

In this work we analysed high-resolution, high signal-to-noise ratio (S/N)  spectra of 23 solar-type dwarf stars, with spectral types from F9 to K1. The selected targets are five YOCs (\object{IC\,2391}, \object{IC\,2602}, \object{IC\,4665}, \object{NGC\,2516,} and \object{NGC\,2547}) and the SFR \object{NGC\,2264}. The stellar spectra were acquired with the 580-setup (spectral coverage $4800-6800$\,\AA) of the FLAMES-UVES spectrograph (nominal resolution $R=47\,000$;  \citealt{uves}). The data reduction was performed by the \ges consortium (see \citealt{2014sacco}). In Paper\,I we selected only targets with rotational velocities \vsini$<20$\,\kms to avoid significant line blending, and with S/N per pixel $>$ 50. We also analysed spectra of the Sun and the four (old and slow-rotating) Gaia benchmark stars (hereafter GBS), namely \object{\object{$\alpha$\,Cen\,A}}, \object{$\tau$\,Cet}, \object{$\beta$\,Hyi,} and \object{18\,Sco}, exploiting UVES spectra taken from \cite{2014blancocuaresma}. Out of 34 GBS, our selection was restricted only to those targets with atmospheric parameters similar to our stars \citep{2015jofre,2018jofre}.

To analyse the largest set of spectral lines possible, we included in our analysis  lines found in the bluer region of the spectra. It is well known that the majority of strong, clean, and isolated atomic lines for heavy-element abundance determination are located in the wavelength range  $3800-4800$\,\AA, which is not accessible to our spectral setup. For this reason we searched through the ESO archive for further observational datasets of the cluster stars. We find that only star 08440521$-$5253171 (\object{IC\,2391}) and star 10440681$-$6359351 (\object{IC\,2602}) have been observed with the UVES, FEROS ($R\sim48\,000$, $\lambda = 3500-9200$\,\AA\, -- \citealt{1999kaufer}), or HARPS ($R\sim115\,000$, $\lambda = 3830-6900$\,\AA\, -- \citealt{2003mayor}) spectrographs.  In addition, we also re-analysed three stars of \object{IC\,2391} that had been previously published in \cite{2017dorazi}, namely PMM\,1142, PMM\,665, and PMM\,4362. These stars were   observed with UVES (blue setup $\lambda\lambda$=3900\,\AA) in the framework of the programme ID 082.C-0218 (PI Melo).

\section{Analysis}\label{sec:analysis}

In the following, we describe extensively the selected lines used in our analysis (Sect.\,\ref{sec:lines}), the procedure to compute the optical depths (Sect.\,\ref{sec:optical depth}), and the details of the abundance measurements (Sect.\,\ref{sec:measurements}). 

\subsection{Selection of the spectral lines}\label{sec:lines}

We carried out a careful selection of spectral lines in the redder part of the spectra, searching in the official \textit{Gaia}-ESO survey master line list \citep{geslinelist} for lines of Cu\,{\sc i}, Sr\,{\sc i} and Sr\,{\sc ii}, Y\,{\sc ii}, Zr\,{\sc i} and Zr\,{\sc ii}, Ba\,{\sc ii}, La\,{\sc ii,} and Ce\,{\sc ii}. From this list we selected only lines with highly accurate measurements of the atomic data ($gf\_flag$=Y) and those that were mostly unblended ($synflag$=Y or U).

The \ges line list covers the wavelength range $4750 -6850$\,\AA\, for the region of the UVES-580 setting. For the bluer part of the spectrum, we included lines that have been extensively used and that have been proven to be reliable (see e.g. \citealt{2017dorazi}). Despite the large number of lines available for each atomic species, we selected only those that are moderately strong and not blended in the solar spectrum. Since our stellar sample includes stars with \vsini up to 20\,\kms and some of the spectra are noisy, most of the pre-selected lines are too broad and not measurable. We show two examples for lines 5274.23\,\AA\, of Ce\,{\sc ii} and 6390.48\,\AA\, of La\,{\sc ii} in the left and right panels of Fig.\,\ref{lines}, respectively. In this figure the observed spectra of the Sun, \object{$\tau$\,Cet,} and star 08440521$-$5253171 of \object{IC\,2391} (\vsini=16.7\kms and S/N=260) are displayed. As  can be seen, both lines  are already relatively weak, though still usable, in the Sun, but they disappear at higher values of \vsini.

In the following we report the details of the lines used in the analysis (all the line lists are available upon request).\\
\\
\texttt{Copper:} In the solar spectrum only lines of neutral Cu were identified. For this element we only relied  on the line at 5105.54\,\AA\, since the one at 5700.24\,\AA\, is heavily blended in the Sun, whereas the line at 5782.13\,\AA\, falls in the UVES wavelength gap. For this element we considered the isotopic solar mixture of 69\% of $^{63}$Cu and 31\% of $^{65}$Cu \citep{2015grevesse}. Copper is affected by isotopic broadening and hyperfine structure (HFS), for which we adopted values from \cite{2011kurucz}. According to \cite{2014shi}, the NLTE corrections for line 5105.54\,\AA\, are small in the Sun, being of the order of +0.02\,dex.\\
\\
\texttt{Strontium:} We measured lines 4607.33\,\AA\, of Sr\,{\sc i} and 4215.52\,\AA\, of Sr\,{\sc ii}. According to \cite{2012bergemann} the Sr\,{\sc i} line has a NLTE correction of +0.10 dex in dwarf stars with solar metallicity, while line 4215.52\,\AA\, has negligible NLTE corrections. We note that the line of Sr\,{\sc ii} is very strong and is almost saturated. It is also blended with a nearby Fe\,{\sc i} line at 4215.42\,\AA\, and with the CN molecular lines. Both features have been accounted for in the spectral synthesis. \\
\\
\texttt{Yttrium:} For Y\,{\sc ii} we selected the lines at 4398.01\,\AA, 4883.69\,\AA, 4900.12\,\AA, 5087.42\,\AA, 5289.82\,\AA, and 5728.89\,\AA. The line at 4900.12\,\AA\, is blended with a nearby Ti\,{\sc i} line at 4899.91\AA, which becomes significant for \vsini$>4-5$\,\kms. The lines at 5289.82\,\AA\, and 5728.89\,\AA\ are  instead very weak in the Sun, both having EW$\sim4$\,m\AA, and thus we are not able to measure them in our targets. The HFS for Y can be neglected, as previously discussed in several papers (e.g. \citealt{2017dorazi}).\\
\\
\texttt{Zirconium:} Zirconium is present in the form of neutral and ionised species in the solar photosphere. However, the available reliable lines of Zr\,{\sc i} (at 6127.44\,\AA, 6134.55\,\AA, 6140.46\,\AA, 6143.2\,\AA, and 6445.74\,\AA\,) are too weak to be measured in our sample stars. In the bluer region we used lines 4050.32\,\AA\, and 4208.98\,\AA\, of Zr\,{\sc ii}. The line at 5112.27\,\AA\, of Zr\,{\sc ii} is also weak, and     we were able to measure it only in the Sun, in the GBS sample, and in one target. According to \cite{2010velichko}, Zr\,{\sc ii} lines form under LTE conditions in solar-type stars.\\
\\
\texttt{Barium:} For ionised barium, instead, we used only 5853.7\,\AA, which is not blended and does not experience severe HFS or isotopic shifts. To our knowledge this line is the best diagnostic to measure the Ba abundance. There are other lines of Ba\,{\sc ii} in our spectral range. However, the resonance Ba\,{\sc ii} line at 4554.03\,\AA\, is almost saturated; the line at 6141.7\,\AA\, is known to be blended with a strong Fe\,{\sc i} line; the line at 6496.9\,\AA\, is also blended with an iron line, and it is affected by strong NLTE effects. \cite{2015reddy} explored the possible detection of the line of neutral Ba at 5535\,\AA\, in a sample of F-G dwarfs. Even so, this line is blended with a strong Fe\,{\sc i} line that dominates the profile and its abundance shows a significant correlation with \teff (see their Fig.\,7), most likely caused by large NLTE effects. Therefore, as already pointed out by the authors, in the absence of NLTE corrections it is not suitable to derive accurate abundances and to solve the Ba puzzle. Nevertheless, to obtain more accurate abundances, we also considered   the HFS data from \cite{1998mcwilliam_Ba} and we adopted the isotopic solar mixture of 81\%  for
($^{134}$Ba +$^{136}$Ba +$^{138}$Ba) and 19\% for ($^{135}$Ba +$^{137}$Ba) (see \citealt{2015grevesse} for further details). According to \cite{2015korotin}, the NLTE corrections are small for stars in the parameter space covered by our sample. \cite{2020gallagher} derived NLTE corrections for the Sun and found that the $\Delta_{\rm{1D NLTE}}$=$-0.11$\,dex and $\Delta_{\rm{3D NLTE}}$=$0.03$\,dex. However, there are no available tables of NLTE corrections for stars with parameters similar to our sample. We note in this context that the NLTE corrections are not sufficient to solve the Ba puzzle. For these reasons, we report the LTE Ba abundances.  \\
\\
\texttt{Lanthanum:} For La\,{\sc ii} we selected lines at 4804.04\,\AA, 4920.98\,\AA, 5122.99\,\AA, and 6390.48\,\AA. Unfortunately, none of them is strong enough to be measured in our stars in the range $4800-6800$\,\AA. Instead, in the bluer part we relied on the measurements of lines 3988.51\,\AA\, and 4086.71\,\AA. Lanthanum has one single isotope $^{139}$La that accounts for 99.9$\%$ of the total La abundance in the solar material, and it is strongly affected by HFS. We followed the prescriptions by \cite{LBS}.\\
\\
\texttt{Cerium:} Finally, for Ce we measured only lines 4073.47\,\AA\, and 5274.23\,\AA. Cerium has four stable isotopes, all with zero nuclear spin: $^{136}$Ce (abundance of 0.185\%), $^{138}$Ce (abundance of 0.251\%), $^{140}$Ce (abundance of 88.450\%), $^{142}$Ce (abundance of 11.114\%). The isotopic splitting is negligible for both lines according to \cite{LSCI}. Thus, it is not affected by HFS.\\

In Table \ref{line_list} only the lines for which we obtained more than one measurement in our stellar sample are indicated; the element (Column 1), the corresponding wavelength (Column 2), the excitation potential energy (E.P., Column 3), the oscillator strength $\log gf$ (Column 4), references for the $\log gf$ values (Column 5), and the solar $\log (\rm{X})_{\odot}$ (Column 6) of each individual line are given (the average solar abundances are in Table\,\ref{solar_ab}). For each line we computed the Landé factor $g_{\rm{L}}$ following \cite{1982deglinnocenti} (Column 7). Instead, the first ionisation potential (FIP) values are taken from Table D.1 in \cite{1992gray} (Column 8).

\subsection{Computation of the optical depths of line formation } \label{sec:optical depth}

\begin{table*}[]
\caption{ Optical depths of line formation \logtau$_{\rm{core}}$ of the line core and \logtau$_{\rm{full}}$ of the full line profile.}   
\label{logtau}      
\centering  
\small
\begin{tabular}{lcccccccccr}    
\toprule  
$\lambda$\,(\AA) & El. & E.P.\,(eV) & $\log gf$   & EW$_{\mathrm{obs}}$\,(m$\AA$) &  R$_{\star}$   &  \logtau$_{\rm{core}}$ & \logtau$_{\rm{full}}$  & HFS  \\
\midrule
5105.54 & Cu\,{\sc i} & 1.39 & $-$1.52  & 88.0 & 0.51  & $-$3.4 & $-$2.4 & y  \\
4607.33 & Sr\,{\sc i} &  0.00 & 0.28  & 46.2 & 0.38  &   $-$2.1 & $-$1.6 & n \\
4215.52 & Sr\,{\sc ii} &  0.00 & $-$0.16  & 173.1 & 0.67  & $-$5.2 & $-$2.5 & n \\
4398.01 & Y\,{\sc ii} & 0.13 & $-$1.0 & 53.3 & 0.42  & $-$2.6 & $-$1.9 & n \\
4883.69 & Y\,{\sc ii} & 1.08 & 0.07  & 58.5 & 0.4  & $-$2.6 & $-$1.9 & n \\
5087.42 & Y\,{\sc ii} & 1.08 & $-$0.17  & 47.3 & 0.32  & $-$2.1 & $-$1.6 & n \\
4050.32 & Zr\,{\sc ii} & 0.71 & $-$1.06 & 22.0 & 0.23  & $-$1.4  & $-$1.2& n \\
4208.98 & Zr\,{\sc ii} & 0.71 & $-$0.51  & 45.3 & 0.39 & $-$2.1  & $-$1.6 & n \\
5112.27 & Zr\,{\sc ii} & 1.67 & $-$0.85  & 6.5 & 0.05  & $-$1.1 & $-$1.1 & n \\
5853.69 & Ba\,{\sc ii} & 0.60 & $-$1.01  & 66.4 & 0.36  & $-$3.2 & $-$2.3  & y \\
3988.51 & La\,{\sc ii} & 0.40 & 0.21  & 51.0 & 0.32  & $-$2.7 &  $-$1.9 & y \\
4086.71 & La\,{\sc ii}& 0.00 & $-$0.07  & 42.0 & 0.37  & $-$2.3 & $-$1.7 & y \\
4073.47 & Ce\,{\sc ii} & 0.48 & 0.21  & 20.3 & 0.16  & $-$1.5 & $-$1.3 & n \\
5274.23 & Ce\,{\sc ii} & 1.04 & 0.13  & 9.1 & 0.05  & $-$1.3 & $-$1.2 & n \\
\bottomrule
\end{tabular}
\end{table*}

Our working hypothesis is that lines forming in the upper layers of the photosphere are more influenced by the higher levels of the activity present in young stars. Therefore, these lines are stronger than those observed in the spectra of old and quiet stars, affecting the derivation of the stellar parameters and, finally, the abundances \citep{2019galarza,2020baratellaA,2020spina}. We computed the optical depth of line formation \logtau of all the selected lines in a consistent way following the prescriptions by \cite{2015gurtovenko}. 

Calculations of the average formation depth of the absorption line are based on the contribution function (CF),  which describes the contribution of the various layers of the stellar atmosphere to the absorption line (or line depression). \cite{1974gurtovenko} suggested to use the CF as the integrand of the emergent line depression in the solar disc centre, computed as

\begin{equation}
    R(0) = \int_0^{\infty} g(\tau_c)\eta(\tau_c) \exp(-\tau_l) d\tau_c = \int_0^\infty CF  d\tau_c,
\end{equation}

\noindent
where R=$1-I_l/I_c$ is the line $l$ depression and $\eta$ is the ratio of the coefficient of the selective absorption to the coefficient of continuum $c$ absorption. In the same formula, $g$ is the Unsold weighting function for LTE \citep{1932unsold}, multiplied by the emergent intensity in the continuum $I_c(\tau_c=0)$. This weighting function is expressed as 

\begin{equation}
    g(\tau_c) = \int_{\tau_c}^{\infty} B(\tau_c)*\exp(-\tau_c) d\tau_c - B(\tau_c)*\exp(-\tau_c),
\end{equation}

\noindent
where $B(\tau_c)$ is the Plank function.

When we are dealing with the interpretation of observed line profiles or line depth in  its centre or equivalent width, we use the average depth of the layers contributing to the absorption line. The average depth at a given wavelength position of the line profile $\Delta\lambda$ and at given position  on the stellar disc $ \mu=\cos\theta$ is calculated by the following formula:

\begin{equation}
     \langle {\tau}_{\Delta\lambda, \mu}\rangle = \int_{-\infty}^{\infty} \tau CF(\Delta\lambda, \mu , \tau) d \tau {/} \int_{-\infty}^{\infty} CF(\Delta\lambda, \mu, \tau) d \tau.
\end{equation}

If we consider the integrated line profile (i.e. the EW), its average optical depth is calculated as

\begin{equation}
    \langle \tau_{\Delta\lambda , \mu , EW}\rangle = \int_{\lambda_1}^{\lambda_2} \langle {\tau}_{\Delta\lambda, \mu} \rangle R(\Delta\lambda, \mu) d (\Delta\lambda){/} \int_{\lambda_1}^{\lambda_2}   R(\Delta\lambda, \mu) d (\Delta\lambda).
\end{equation}

Here $R(\Delta\lambda, \mu)$ is the  line depression at $\Delta\lambda$ and  $\mu$, while $\lambda_1$ and $\lambda_2$ are the initial and final wavelength positions of the line profile, respectively. To obtain the average depth of formation of the line depression observed in the stellar spectra at a given $\Delta\lambda$, we use the following formula:

\begin{equation}
    \langle \tau_{\Delta\lambda , \star}\rangle = \int_0^1
\langle {\tau}_{\Delta\lambda, \mu}\rangle \mu d\mu.
\end{equation}

Instead, to have the average formation depth of the whole line profile using EWs, we use

\begin{equation}
    \langle \tau_{EW, \star}\rangle =\int_{\lambda_1}^{\lambda_2} \langle {\tau}_{\Delta\lambda , \star}\rangle R_{\star}(\Delta\lambda) d(\Delta\lambda) {/} \int_{\lambda_1}^{\lambda_2}  R_{\star}(\Delta\lambda) d(\Delta\lambda)
,\end{equation}

\noindent
where $ R_{\star} $ is the line depression in the spectra of stellar flux.

In this work we computed the average depth of line formation of our lines both in the core (\logtau$_{\rm{core}}$) and in the whole profile (\logtau$_{\rm{full}}$), reported in Column 7 and 8 of Table\,\ref{logtau}, respectively. For this calculation we assumed the LTE approximation and we considered the damping constant associated with the van der Waals force between the absorbing and perturbing atoms to be equal to $\gamma_6$ according to the classical Unsold approximation. Our assumptions are acceptable to estimate the average depth of the formation of weak, moderate, and moderately strong lines, like those analysed here.  We measured the EW$_{\mathrm{obs}}$ and R$_{\star}$ (Column 5 and 6, respectively) of each line using the ARESv.2 software \citep{2015sousa} and performed the synthesis in the solar spectrum with the SPANSAT code of \cite{1988gadun}. The \logtau value was then derived from the abundance obtained when the EW of the synthetic line matches the EW$_{\rm{obs}}$. We adopted the MARCS solar atmosphere model with the chemical composition taken from \cite{2019lodders}, and with the stellar parameters reported in Table \ref{atmospheric_param}. We also considered the macroturbulent velocity equal to 2 km/s and \vsini = 1.84 \kms \citep{2019sheminova}. We considered the HFS indirectly in our computations for those lines labelled  \textit{y} in Column 9 of Table\,\ref{logtau}. 
Lines affected by strong HFS are split into multiple components, resulting in larger EWs. Thus, completely neglecting the HFS when deriving the abundances from EWs will result in overestimated values \citep{2015scott,2017jofre_paula}, and all lines form in higher layers of the photosphere since the overestimated abundances (corresponding to the larger EW$_{\rm{obs}}$) are used in the computation of the optical depth. At the same time, the abundance calculated from the fitting of central depths (R$_{\star}$) of the lines can be underestimated and the depths of the formation of the lines will be large (i.e. the line will form in deeper layers). Therefore, these values are indicative and should be considered with some caution. 
We believe that the possible errors due to the adopted approximations in the above-mentioned computations are small and that they can be neglected within the limits of specific calculations of the depths of line formation.

\subsection{Abundance measurements}\label{sec:measurements}

The abundances of the $s$-process elements were derived using  the technique of synthetic spectrum line profile fitting through the driver $synth$ in MOOG (version 2017, \citealt{1973sneden,2011sobeck}). We used 1D-LTE plane-parallel MARCS model atmospheres \citep{marcs}, fixing the atmospheric parameters to the values we found in Paper\,I, both for the cluster stars and the GBS. We report all the stellar parameters in Table \ref{atmospheric_param} for completeness. All abundances were computed in a strictly differential way (i.e. line-by-line) with respect to the Sun as [X/H]$_{\star}$=$\log(\rm{X})_{\star}-\log(\rm{X})_{\odot}$ (using the individual abundances $\log(\rm{X})_{\odot}$ in column 6 of Table \ref{line_list}). The final abundance ratios [X/Fe]=[X/H]$_{\star}-$[Fe/H]$_{\star}$ can be found in Tables \ref{gaia_benchmark}, \ref{abundances_480_680}, and \ref{abundances_blue} for the GBS, the stellar sample results in the $4800-6800$\,\AA\, range and in the $3800-4800$\,\AA\, range, respectively. 
The stars in our sample are in the main-sequence evolutionary phase; therefore, we set the carbon and magnesium isotopic ratio to the solar values, equal to $^{12}$C/$^{13}$C =89 \citep{2009asplund} and  $^{24}$Mg:$^{25}$Mg:$^{26}$Mg=80:10:10 \citep{2003fenner}, respectively. To compute the synthetic profiles we used the new tables of limb darkening coefficients (LDCs) by \cite{2019claret}, corresponding to the Gaia G$_{\rm{BP}}$ pass-band (their Table 6). The rotational broadening profiles were calculated using the \vsini measured by the \textit{Osservatorio Astrofisico di Catania} (OACT) Node of the \ges consortium, measured using the routine ROTFIT (see e.g. \citealt{2015frasca}, and references therein, for more details); the values are reported in Column 4 of Table \ref{abundances_480_680}. The values of \vsini of PMM\,665, PMM\,4362, and PMM\,1142 were taken from \cite{2013desilva} (hereafter DS13).  However, in some cases (e.g. for stars PMM\,1142, PMM\,665, PMM\,4362 of \object{IC\,2391}; the star in \object{NGC\,2264}; the stars 07544342$-$6024437 and 07574792$-$6056131 of \object{NGC\,2516}; and the star 08102854$-$4856518 of \object{NGC\,2547}), the profiles computed with the \vsini taken from literature did not reproduce well the line profiles. Then, we recomputed the \vsini by looking at 15-20 isolated and clean lines over the whole spectral range. Our measurements can be found in Column 3 of Table \ref{abundances_480_680}. Of the whole line list in the redder part, we measured only lines 5105.53\,\AA\, of Cu, 4883.69\,\AA\, and 5087.42\,\AA\, of Y and 5853.7\,\AA\, of Ba for all the stars. These are the strongest lines in our line list: the large values of \vsini (up to 18\,\kms) prevented us from measuring weaker lines. We measured Ce abundance from line 5274.23\,\AA\, only for two stars and Zr from line 5112.27\,\AA\, for one star, for which the uncertainty of the best fit model is 0.35\,dex (due to the relatively high rotational velocity).

\subsection{Error budget}\label{sec:errors}

There are two sources of internal uncertainties affecting the [X/Fe] ratios derived via spectral synthesis. The first kind of error, $\sigma_1$, is related to the best fit procedure, and spans values from $\pm0.06$ to $\pm 0.3$ dex depending on the quality of the spectra, mainly the S/N, which affects the continuum placement, and on the individual spectral features under consideration. 

The second kind of error, $\sigma_2$, is related to uncertainties in the stellar parameters (Table \ref{atmospheric_param}). We calculated these uncertainties in a conservative way by varying each quantity separately, leaving the others unchanged, and evaluating the abundance sensitivity to those changes as

\begin{equation}
\tiny{
\sigma_2=\sqrt{\left( \sigma_{T\rm{_{eff}}}  \frac{\partial \rm{[X/H]}}{\partial T\rm{_{eff}}}\right) ^2 + \left( \sigma_{\rm{\log\,g}} \frac{\partial \rm{[X/H]}}{\partial \rm{\log g}}\right) ^2 +\left( \sigma_{\xi} \frac{\partial \rm{[X/H]}}{\partial \rm{\xi}} \right) ^2+\left( \sigma_{\rm{[Fe/H]}} \frac{\partial \rm{[X/H]}}{\partial \rm{[Fe/H]}} \right) ^2}. }
\end{equation}

We report both errors in Tables \ref{line_list}, \ref{gaia_benchmark}, \ref{abundances_480_680}, and \ref{abundances_blue}.

\section{Results}\label{sec:results}

\subsection{The Sun and the Gaia benchmarks}

In Table\,\ref{solar_ab} we report the average solar abundance values and we compare them with the photospheric abundances from \cite{2015grevesse}, the meteoritic  abundances from \cite{2019lodders}, and the results reported in the \ges internal Data Release 4 (for La and Sr) and 5 (iDR4 and iDR5, respectively). For our values of Y, Zr, La, and Ce in Table\,\ref{solar_ab} we take the simple mean of abundances derived in the bluer and in the redder regions; the uncertainties are computed as the errors on the mean. For Sr the value in the Table\,\ref{solar_ab} is the average between the Sr\,{\sc i} (corrected for NLTE) and the Sr\,{\sc ii} results. For Ba and Cu, for which we analysed only one line, we report the individual abundance;  the uncertainty is the error on the fitting procedure ($\sigma_1$). As  can be seen from the comparison with the literature values, our mean solar abundances agree well with the photospheric values from \cite{2015grevesse}, and with the meteoritic abundances from \cite{2019lodders}. They are also in fair agreement with the \ges iDR5 and iDR4 results.

\begin{table}[h!]
\small{
\caption{Solar abundances.}
\centering
\begin{tabular}{lcccr}
\toprule
Species & This work & G15 & L19 & GES\\
\midrule
Cu& $4.21\pm0.12$ & $4.18\pm0.05$ & $4.21\pm0.03$ & $4.12\pm0.10$\\
Sr & $2.85\pm0.03$ & $2.83\pm0.06$ & $2.83\pm0.06$ & $2.87\pm0.1^{a}$ \\
Y & $2.20\pm0.03$ & $2.21\pm0.05$ & $2.20\pm0.05$ & $2.19\pm0.12$\\
Zr & $2.57\pm0.01$ & $2.59\pm0.04$ & $2.59\pm0.06$ & $2.53\pm0.13$\\
Ba & $2.31\pm0.09$ & $2.25\pm0.07$ & $2.19\pm0.07$ & $2.17\pm0.06$ \\
La & $1.08\pm0.03$ & $1.11\pm0.04$ & $1.14\pm0.03$ & $1.22\pm0.12^{b}$\\
Ce & $1.60\pm0.05$ & $1.58\pm0.04$ & $1.61\pm0.06$ & $1.70\pm0.11$ \\
\hline
\hline
\end{tabular}
\tablefoot{Mean solar abundances derived in this work, in \cite{2015grevesse} (G15), and meteoritic abundances from \cite{2019lodders} (L19). We also report the values derived by \textit{Gaia}-ESO (GES) in iDR5 and iDR4.}
\begin{tablenotes}
\small
\item $^{a}$ : value reported in  iDR4, derived from Sr\,{\sc i} lines only;
\item $^{b}$ : value reported in  iDR4.
\end{tablenotes}
\label{solar_ab}
}
\end{table}

\begin{table*}[!]
\caption{Abundances of the GBS of the neutron-capture process elements. }   
\label{gaia_benchmark}      
\centering  
\setlength\tabcolsep{8pt}
\small
\begin{tabular}{lc|cccr }    
\toprule  
El. & $\lambda$\,(\AA)   & [X/Fe]$_{\alpha\,\rm{Cen\,A}}$  & [X/Fe]$_{\tau\,\rm{Cet}}$  & [X/Fe]$_{\beta\,\rm{Hyi}}$  & [X/Fe]$_{18\,\rm{Sco}}$ \\
\midrule
&&& \textbf{3800 - 4800\,\AA}\\
Sr\,{\sc i}(NLTE) & 4607.33 & $-0.02\pm0.08\pm0.07$ & $-0.05\pm0.08\pm0.07$ & $-0.05\pm0.11\pm0.07$ &  $+0.05\pm0.1\pm0.06$ \\
Sr\,{\sc ii} & 4215.52 & $0.00\pm0.2\pm0.1$ & $-0.07\pm0.19\pm0.1$ &$-0.07\pm0.22\pm0.09$ & $+0.06\pm0.19\pm0.09$  \\
Y\,{\sc ii } & 4398.01 & $-0.03\pm0.15\pm0.09$ & $-0.14\pm0.11\pm0.07$ & $-0.04\pm0.1\pm0.07$ & $+0.01\pm0.13\pm0.07$ \\ 
Zr\,{\sc ii} & 4050.32 & $+0.02\pm0.12\pm0.06$ & $-0.07\pm0.14\pm0.08$ & $-0.02\pm0.13\pm0.08$  & $-0.01\pm0.11\pm0.06$ \\
Zr\,{\sc ii} & 4208.98 & $+0.01\pm0.15\pm0.07$ & $-0.05\pm0.1\pm0.09$ & $-0.04\pm0.12\pm0.05$ & $+0.01\pm0.10\pm0.07$  \\
La\,{\sc ii} & 3988.51 & $-0.04\pm0.1\pm0.06$ & $-0.01\pm0.15\pm0.06$ & $-0.02\pm0.15\pm0.06$  & $+0.10\pm0.1\pm0.07$ \\
La\,{\sc ii} & 4086.71 & $-0.08\pm0.15\pm0.08$ & $-0.05\pm0.17\pm0.07$ & $-0.06\pm0.16\pm0.07$ & $+0.06\pm0.12\pm0.08$ \\
Ce\,{\sc ii} & 4073.47 & $0.00\pm0.13\pm0.06$ & $0.03\pm0.15\pm0.05$  & $0.00\pm0.15\pm0.05$ & $+0.08\pm0.08\pm0.06$  \\
\hline
\hline
&&& \textbf{4800 - 6800\,\AA}\\
Cu\,{\sc i} & 5105.54  & $+0.15\pm0.20\pm0.09$ & $-0.08\pm0.20\pm0.11$  & $-0.01\pm0.15\pm0.11$ & $-0.02\pm0.10\pm0.10$ \\ 
Y\,{\sc ii } & 4883.69 & $+0.02\pm0.15\pm0.13$ & $-0.08\pm0.13\pm0.09$ & $-0.06\pm0.15\pm0.08$ & $-0.09\pm0.09\pm0.07$\\
Y\,{\sc ii } & 5087.42 & $+0.05\pm0.15\pm0.1$ & $-0.11\pm0.13\pm0.07$ & $-0.05\pm0.11\pm0.07$ & $-0.10\pm0.10\pm0.07$\\
Zr\,{\sc ii} & 5112.27 & $-0.06\pm0.20\pm0.06$ & $-0.01\pm0.1\pm0.07$ & $-0.10\pm0.07\pm0.06$ & $+0.01\pm0.08\pm0.07$\\
Ba\,{\sc ii} & 5853.69 & $+0.04\pm0.15\pm0.12$ & $-0.12\pm0.12\pm0.07$ & $-0.08\pm0.17\pm0.11$ & $-0.04\pm0.15\pm0.1$ \\
Ce\,{\sc ii} & 5274.23 & $-0.02\pm0.14\pm0.08$ & $-0.08\pm0.1\pm0.05$ & $-0.06\pm0.08\pm0.09$ & $+0.03\pm0.1\pm0.05$\\
\bottomrule
\end{tabular}
\tablefoot{The first source of uncertainty is due to the best fit procedure, while the second is related to uncertainties in the stellar parameters (details on the computations can be found in Sect.\,\ref{sec:errors}).}
\end{table*}

\begin{sidewaystable*}
\renewcommand\arraystretch{1.0}
\caption{Abundances of the neutron-capture process elements derived for the whole stellar sample.}  
\setlength\tabcolsep{3pt}
\label{abundances_480_680}
\centering
\scalebox{0.85}{
\begin{tabular}{lccccccccr}
\toprule
CNAME & S/N & \vsini$\pm\sigma$ & \vsini$_{\rm{lit.}}$ & [Cu/Fe]$_{\rm{I,5105\AA}}\pm\sigma_1\pm\sigma_2$ & [Y/Fe]$_{\rm{II,4883\AA}}\pm\sigma_1\pm\sigma_2$ & [Y/Fe]$_{\rm{II,5087\AA}}\pm\sigma_1\pm\sigma_2$ & [Zr/Fe]$_{\rm{II, 5112\AA}}\pm\sigma_1\pm\sigma_2$ & [Ce/Fe]$_{\rm{II,5274\AA}}\pm\sigma_1\pm\sigma_2$ &  [Ba/Fe]$_{\rm{II,5853\AA}}\pm\sigma_1\pm\sigma_2$\\
\midrule
\textit{\object{IC\,2391}}\\
08365498$-$5308342 & 159 & $8.3\pm0.3$ & $8.5\pm0.8$ & $-0.02\pm0.10\pm0.09$ & $+0.26\pm0.10\pm0.08$ & $+0.25\pm0.17\pm0.07$ & - & -& $+0.57\pm0.12\pm0.08$\\
08440521$-$5253171 & 259 & $18.4\pm0.4$ & $16.7\pm0.8$ & $-0.12\pm0.10\pm0.07$ & $+0.24\pm0.09\pm0.07$ & $+0.22\pm0.15\pm0.09$ & - & - & $+0.52\pm0.09\pm0.08$\\
PMM1142$^{a}$ & 100 & $7.5\pm0.4$ & 7.3$^{\ast}$ & $-0.02\pm0.11\pm0.10$ & $+0.26\pm0.10\pm0.07$ & $+0.21\pm0.10\pm0.07$ &  - & $-0.02\pm0.15\pm0.09$ & $+0.61\pm0.08\pm0.07$\\
PMM665$^{a}$ & 60 & $8.2\pm0.3$ & 7.47$^{\ast}$ & $-0.10\pm0.12\pm0.10$ & $+0.19\pm0.10\pm0.07$ & $+0.14\pm0.15\pm0.08$ &  -& - & $+0.46\pm0.10\pm0.11$\\
PMM4362$^{a}$ & 58 & $9\pm0.3$ & 8.61$^{\ast}$ & $-0.09\pm0.13\pm0.08$ & $+0.19\pm0.15\pm0.09$ & $+0.23\pm0.15\pm0.07$ & -  & - & $+0.48\pm0.10\pm0.09$\\
\\
\textit{\object{IC\,2602}}\\
10440681$-$6359351 & 92 & $12.1\pm0.4$ & $12.2\pm0.7$ & $-0.15\pm0.12\pm0.11$ & $+0.19\pm0.16\pm0.11$ & $+0.16\pm0.17\pm0.11$ & -& - & $+0.40\pm0.12\pm0.13$\\
10442256$-$6415301 & 110&  $11.1\pm0.3$ & $10.9\pm0.8$  & $-0.08\pm0.13\pm0.11$ & $+0.16\pm0.15\pm0.09$ & $+0.15\pm0.12\pm0.07$ & $0\pm0.35\pm0.10$ & - & $+0.36\pm0.11\pm0.08$\\
10481856$-$6409537 & 99 & $14.5\pm0.2$ & $14.4\pm0.8$ & $-0.08\pm0.13\pm0.10$ & $+0.22\pm0.15\pm0.08$ & $+0.15\pm0.15\pm0.08$ &  -& - & $+0.44\pm0.15\pm0.11$\\
\\
\textit{\object{IC\,4665}}\\
17442711+0547196 &53 & $13.8\pm0.3$ & $14.5\pm1.0$ & $-0.02\pm0.25\pm0.09$ & $+0.49\pm0.20\pm0.09$ &  -& -&- & $+0.64\pm0.15\pm0.07$\\
17445810+0551329 &64& $10.3\pm0.4$ & $10.0\pm0.9$  & $+0.04\pm0.10\pm0.07$ & $+0.26\pm0.12\pm0.08$ & $+0.24\pm0.16\pm0.09$ & - &- & $+0.56\pm0.10\pm0.10$\\
17452508+0551388 & 54& $13.5\pm0.2$ & $13.5\pm0.7$ &$+0.13\pm0.2\pm0.10$ & $+0.27\pm0.17\pm0.12$ &  -& -& -& $+0.61\pm0.15\pm0.12$\\
\\
\textit{\object{NGC\,2264}}\\
06405694+0948407 & 135 & $18\pm0.6$ & $15\pm3.8$ & $-0.17\pm0.15\pm0.09$ & $+0.31 \pm0.15\pm0.07$ & $+0.24\pm0.1\pm0.08$ & - & -& $+0.41\pm0.12\pm0.08$\\
\\
\textit{\object{NGC\,2516}}\\
07544342$-$6024437 & 73& $4.2\pm0.5$ & 3$^{\ast\ast}$ & $-0.03\pm0.11\pm0.09$ & $+0.12\pm0.10\pm0.07$ & $+0.03\pm0.10\pm0.08$ & - & $+0.03\pm0.25\pm0.10$ &$+0.23\pm0.16\pm0.08$\\
07550592$-$6104294 & 110& $12.2\pm0.3$ & $11.8\pm0.8$ & $-0.17\pm0.15\pm0.1$ & $+0.09\pm0.12\pm0.09$ & $+0.06\pm0.13\pm0.08$ &  -& -& $+0.34\pm0.10\pm0.08$\\
07551977$-$6104200 & 129& $14.5\pm0.4$ & $14.6\pm1.1$  & $-0.15\pm0.10\pm0.09$ & $+0.17\pm0.15\pm0.08$ & $+0.15\pm0.13\pm0.07$ & -&-& $+0.22\pm0.10\pm0.07$\\
07553236$-$6023094 &82& $9.6\pm0.4$ & $9.2\pm1.2$  & $+0.08\pm0.15\pm0.11$ & $+0.05\pm0.13\pm0.09$ & $+0.06\pm0.17\pm0.10$ & -&- & $+0.34\pm0.09\pm0.10$\\
07564410$-$6034523 &56& $8.7\pm0.4$ & $7.9\pm0.8$ & $-0.12\pm0.20\pm0.09$ & $+0.10\pm0.20\pm0.08$ & $+0.05\pm0.20\pm0.10$ &  -&-& $+0.36\pm0.15\pm0.09$\\
07573608$-$6048128 &50& $5.4\pm0.8$ & $4.6\pm1.0$ & $-0.02\pm0.16\pm0.10$ & $+0.08\pm0.20\pm0.10$ & $+0.06\pm0.17\pm0.09$ &  -&-& $+0.34\pm0.15\pm0.12$\\
07574792$-$6056131 &68& $5\pm0.4$ & $3.2\pm1.7$ & $-0.04\pm0.17\pm0.07$ &  $+0.15\pm0.13\pm0.08$ & $+0.13\pm0.15\pm0.09$ & - & - & $+0.25\pm0.10\pm0.10$ \\
07575215$-$6100318 &51& $6\pm0.8$ & $5.5\pm0.9$ & $+0.05\pm0.18\pm0.12$ & - & $+0.02\pm0.15\pm0.12$ &-&-&$+0.24\pm0.11\pm0.09$\\
07583485$-$6103121 & 100& $12.5\pm0.4$ & $11.7\pm1.3$  & - & $+0.15\pm0.20\pm0.07$ & $+0.03\pm0.20\pm0.08$ & -&-& $+0.30\pm0.10\pm0.10$\\
07584257$-$6040199 &83& $9.2\pm0.3$ & $8.8\pm0.7$ & $-0.12\pm0.12\pm0.07$ & $+0.13\pm0.20\pm0.08$ & $+0.11\pm0.20\pm0.08$  &-&-& $+0.38\pm0.10\pm0.09$\\
08000944$-$6033355 &81& $8.2\pm0.4$ & $7.2\pm1.1$  & $-0.13\pm0.12\pm0.09$ & $+0.11\pm0.10\pm0.07$ & $+0.07\pm0.13\pm0.08$ &  -& -& $+0.34\pm0.15\pm0.07$\\
08013658$-$6059021 &69& $9.4\pm0.3$ & $8.9\pm0.9$ & $-0.17\pm0.15\pm0.08$ &$+0.11\pm0.20\pm0.07$& $+0.04\pm0.20\pm0.09$ & -&-& $+0.32\pm0.12\pm0.10$\\
\\
\textit{\object{NGC\,2547}}\\
08102854$-$4856518 &90& $16\pm0.4$ & $15.3\pm0.8$  & $-0.12\pm0.13\pm0.09$ & $+0.19\pm0.17\pm0.09$ & $+0.12\pm0.18\pm0.09$ & - & -& $+0.54\pm0.15\pm0.13$\\
08110139$-$4900089 &77& $8.8\pm0.3$ & $8.8\pm0.8$ & $-0.02\pm0.12\pm0.09$ & $+0.17\pm0.15\pm0.11$ & $+0.15\pm0.13\pm0.09$ &  -&-& $+0.52\pm0.12\pm0.10$\\
\bottomrule
\end{tabular}}
\tablefoot{Abundances of the $s$-process elements derived with the lines in the 480-680nm range for the whole stellar sample (Cu line at 5105\,\AA, Y lines at 4883\,\AA\, and 5087\,\AA, Zr line at 5112\,\AA, Ce line at 5274\,\AA, Ba line at 5853\,\AA). The first source of uncertainty ($\sigma_1$) is due to the best fit procedure, while the second ($\sigma_2$) is related to uncertainties in the stellar parameters (details of the computations can be found in Sect.\,\ref{sec:errors}).}
\begin{tablenotes}
\small
\item $^{a}$: The identification name are   from \cite{2013desilva}
\item $\ast$: Velocities  are from \cite{2013desilva}
\item $\ast\ast$: Values reported in both  the iDR5 and iDR6, no error provided because it is an upper limit (VSINI FLAG=1). 
\end{tablenotes}
\label{s-process}
\end{sidewaystable*}

\begin{sidewaystable*}
\renewcommand\arraystretch{1.0}
\caption{Abundances of the neutron-capture elements in the bluer region of the spectrum (3800-4800\,\AA).}
\setlength\tabcolsep{3.0pt}
\label{abundances_blue}
\centering
\begin{tabular}{l|cccccc|r}
\toprule
\midrule
 & & &\textit{\object{IC\,2391}} & & & &\textit{\object{IC\,2602}} \\
CNAME  & 08440521$-$5253171 & 08440521$-$5253171 & 08440521$-$5253171 &  PMM1142 & PMM665 & PMM4362 & 10440681$-$6359351 \\
Instrument &  UVES & FEROS & HARPS & UVES & UVES & UVES & HARPS\\
S/N & 70 & 68 & 53& 66 & 25 & 29 & 46 \\
\vsini$\pm\sigma$ & $18.4\pm0.4$ & $18.4\pm0.4$ & $18.4\pm0.4$ & $7.5\pm0.4$ & $8.2\pm0.3$ & $9\pm0.3$  & $12.1\pm0.4$  \\

\vsini$_{\rm{lit.}}$ & $16.7\pm0.8$ & $16.7\pm0.8$ & $16.7\pm0.8$ & 7.3$^{\ast}$ & 7.47$^{\ast}$ & 8.61$^{\ast}$ & $12.2\pm0.7$   \\

[Sr/Fe]$_{\rm{I}}\pm\sigma_1\pm\sigma_2$ & - & $-0.05\pm0.15\pm0.08$ & $-0.1\pm0.20\pm0.07$ & -& -&-& $+0.03\pm0.15\pm0.09$\\

[Sr/Fe]$_{\rm{II}}\pm\sigma_1\pm\sigma_2$ & - & $-0.08\pm0.17\pm0.09$ & $-0.05\pm0.20\pm0.09$ & - & $+0.03\pm0.17\pm0.10$ & $-0.01\pm0.17\pm0.08$ & $-0.04\pm0.18\pm0.08$ \\

[Y/Fe]$_{4398\AA}\pm\sigma_1\pm\sigma_2$ &   $+0.3\pm0.2$ & $+0.25\pm0.20\pm0.08$ &  $+0.3\pm0.25\pm0.09$&  $+0.31\pm0.12\pm0.10$ & $+0.29\pm0.16\pm0.10$ & $+0.20\pm0.15\pm0.11$ & $+0.37\pm0.30\pm0.09$ \\

[Zr/Fe]$_{4050\AA}\pm\sigma_1\pm\sigma_2$& - & - & - & $-0.01\pm0.10\pm0.08$ & $+0.01\pm0.17\pm0.09$ & - & -\\

[Zr/Fe]$_{4208\AA}\pm\sigma_1\pm\sigma_2$ & - & - & - & $+0.03\pm0.12\pm0.07$  &  $+0.04\pm0.20\pm0.07$ & $+0.03\pm0.15\pm0.08$  &  - \\

[La/Fe]$_{3988\AA}\pm\sigma_1\pm\sigma_2$ & $-0.1\pm0.10\pm0.07$ & $-0.10\pm0.15\pm0.07$ & $-0.05\pm0.20\pm0.07$ & $-0.11\pm0.10\pm0.08$ & $-0.01\pm0.12\pm0.06$ & $-0.06\pm0.12\pm0.07$ & $-0.03\pm0.25\pm0.08$ \\ 

[La/Fe]$_{4087\AA}\pm\sigma_1\pm\sigma_2$ & - &  - & - & $-0.06\pm0.15\pm0.06$ & $-0.09\pm0.13\pm0.07$ & $-0.08\pm0.16\pm0.08$ & $+0.02\pm0.30\pm0.08$\\

[Ce/Fe]$_{4073\AA}\pm\sigma_1\pm\sigma_2$ & - & - & - & $-0.03\pm0.20\pm0.07$ & $+0.01\pm0.15\pm0.08$ &  - & -\\
\bottomrule
\end{tabular}
\tablefoot{Individual abundances of the $s$-process elements derived with the lines in the 380-480 nm range. The first source of uncertainty ($\sigma_1$) is due to the best fit procedure, while the second ($\sigma_2$) is related to uncertainties in the stellar parameters (details of the computations can be found in Sect.\,\ref{sec:errors})}
\begin{tablenotes}
\item $\ast$: Velocities are from \cite{2013desilva}.
\end{tablenotes}
\label{s-process_blue}
\end{sidewaystable*}

\begin{table*}[]
\tiny{
\caption{Mean values of the abundance ratios [X/Fe] of the GBS and the YOCs, and comparison with the literature values.}
\setlength\tabcolsep{7pt}
\centering
\begin{tabular}{lccccccc}
\toprule
Star/Cluster & [Cu/Fe] & [Sr/Fe] & [Y/Fe] & [Zr/Fe] & [Ba/Fe] & [La/Fe] & [Ce/Fe]\\ 
\midrule
\rowcolor[gray]{0.9}\textbf{\object{$\alpha$\,Cen\,A} }& $+0.15\pm0.20$ & $-0.01\pm0.01$ & $+0.01\pm0.04$ & $-0.01\pm0.04$ & $+0.04\pm0.15$ & $-0.06\pm0.03$ & $-0.01\pm0.01$ \\
\cite{2018luck} & +0.18$\pm0.07$ & $-0.08\pm0.17$ & $-0.01\pm0.04$ & $-0.07\pm0.04$ & $-0.12\pm0.03$ & $-0.07\pm0.02$ & $-0.04\pm0.03$\\
\cite{2020casamiquela} & $+0.07\pm0.02$ & - & $-0.07\pm0.02$ & - & $-0.11\pm0.02$ & $-0.01\pm0.03$ & $-0.11\pm0.03$\\
\cite{2020casali} & $+0.11\pm0.03$ & $+0.03\pm0.01$ & $-0.01\pm0.01$ & $-0.07\pm0.02$ & $-0.07\pm0.01$ & - & $-0.06\pm0.02$ \\
\\
\rowcolor[gray]{0.9}\textbf{\object{$\tau$\,Cet} }& $-0.08\pm0.20$ & $-0.06\pm0.01$ & $-0.11\pm0.03$ & $-0.04\pm0.03$ & $-0.12\pm0.12$ & $-0.03\pm0.03$ & $-0.03\pm0.08$ \\
\cite{2018luck} & $+0.08\pm0.08$ & $-0.12$ & $-0.09\pm0.02$ & $+0.07\pm0.05$ & $-0.08\pm0.01$ & $+0.17\pm0.06$ & $+0.07\pm0.04$\\
\cite{2020casamiquela} & $+0.05\pm0.02$ & - & $-0.02\pm0.02$ & - & $-0.09\pm0.02$ & $+0.11\pm0.04$ & $-0.14\pm0.08$\\
\\
\rowcolor[gray]{0.9}\textbf{\object{$\beta$\,Hyi} }& $-0.01\pm0.15$ & $-0.06\pm0.01$ & $-0.05\pm0.01$ & $-0.05\pm0.04$ & $-0.08\pm0.17$ & $-0.04\pm0.02$ & $-0.03\pm0.03$ \\
\cite{2015jofre_emiliano} & - & - & - &- & $+0.09\pm0.10$ & - & -\\
\cite{2014bensby} & - & -& $-0.12\pm0.11$ & - & $+0.03\pm0.05$ & -& -\\
\\
\rowcolor[gray]{0.9}\textbf{\object{18\,Sco}} & $-0.02\pm0.10$ &  $+0.055\pm0.005$ & $-0.06\pm0.05$ & $0.00\pm0.01$ & $-0.04\pm0.15$ & $+0.08\pm0.03$ & $+0.06\pm0.04$ \\
\cite{2018luck} & $-0.06\pm0.04$ & $-0.04\pm0.17$ & $+0.02\pm0.02$ & $+0.01\pm0.09$ & $+0.02\pm0.01$ & $+0.01\pm0.02$ & $+0.02\pm0.03$ \\
\cite{2020casamiquela} & $-0.04\pm0.02$ & - & $+0.04\pm0.01$ & - & $+0.02\pm0.01$ & $-0.06\pm0.03$ & $-0.01\pm0.04$\\
\cite{2020casali} & $-0.03\pm0.02$ & $+0.09\pm0.01$ & $+0.07\pm0.01$ & $+0.06\pm0.01$ & $+0.04\pm0.01$ & - & $+0.07\pm0.01$\\

\\
\hline
\rowcolor[gray]{0.9}\textbf{\object{IC\,2391}} & $-0.07 \pm 0.02$ & $-0.04 \pm 0.05$ & $+0.23\pm0.03$ & $+0.02\pm0.02$ & $+0.53\pm0.02$ & $-0.07\pm0.03$ & $-0.01\pm0.02$\\
\cite{2009dorazi} & -&-&-&-& $+0.68\pm0.07$ & -&-\\
\cite{2013desilva} & -&-&-&-& $+0.62\pm0.07$ & -&-\\
\cite{2017dorazi} & -&- & $+0.08\pm0.02$ & $+0.09\pm 0.03$ & - &  $+0.14\pm0.02$ & $+0.15\pm0.05$\\
\\
\rowcolor[gray]{0.9}\textbf{\object{IC\,2602}} & $-0.10\pm0.02$ & $-0.01\pm0.05$ & $+0.19\pm0.04$  & 0.0 $\pm$ 0.35** & $+0.40\pm0.02$ & $-0.01\pm0.04 $** & \\
\cite{2009dorazi} & -&-&-&-& +0.64 $\pm$ 0.07 & -&-\\
\cite{2017dorazi} & -&- & +0.08 $\pm$ 0.02 & +0.06 $\pm$ 0.01 & - &  +0.10 $\pm$ 0.02 & +0.02 $\pm$ 0.02\\
\cite{2021spina} & $-0.15 \pm 0.05$ & - & $+0.17 \pm 0.11$ & - & $+0.34 \pm 0.14$ & - & -\\
\\
\rowcolor[gray]{0.9}\textbf{\object{IC\,4665}} & $+0.05\pm0.04$ & - & +0.34 $\pm$ 0.1  & - &  $+0.60\pm0.02$ & - & -\\
\cite{2021spina} & $-0.27 \pm 0.05$ & - & $+0.29 \pm 0.16$ & - & $+0.37 \pm 0.16$ & $+0.48 \pm 0.05$***\\
\\
\rowcolor[gray]{0.9}\textbf{\object{NGC\,2264}}* & $-0.17\pm0.15$ & -  & $+0.28\pm0.05$ & - & $+0.41\pm0.12$ & - & -\\
\\
\rowcolor[gray]{0.9}\textbf{\object{NGC\,2516}} & $-0.06\pm0.03$ & - & +0.09 $\pm$ 0.04 &  - & $+0.31\pm0.03$ & - & $+0.03\pm0.25$** \\
\cite{2009dorazi}  & - & -& - & -& +0.41 $\pm$ 0.04& - & -\\
\cite{2011maiorca} & - & -& +0.16 $\pm$ 0.03 & -& -& - & +0.18 $\pm$ 0.02\\
\cite{2018magrini} & -&-& +0.11 $\pm$ 0.06 & +0.58 $\pm$ 0.07 & +0.20 $\pm$ 0.08 & - & +0.38 $\pm$ 0.10 \\
\\
\rowcolor[gray]{0.9}\textbf{\object{NGC\,2547}} & $-0.07\pm0.05$ & - & +0.16 $\pm$ 0.01 & - & $+0.53\pm0.01$  & -& -\\
\bottomrule
\end{tabular}
\tablefoot{ The uncertainties for the GBS stars is the standard deviation of different lines (for Cu and Ba, for which we only analysed one line, the error is the quadratic sum of $\sigma_1$ and $\sigma_2$). The uncertainties of the mean abundances computed for our clusters are the errors on the mean, representing the star-to-star variation.}
\begin{tablenotes}
\small
\item *: Since we analysed only one star in this SFR, we report the individual abundance values and the uncertainties on the fitting procedure.
\item **: Measurement based on only one target.
\item ***: Values affected by strong blends \citep{2021buder}.
\end{tablenotes}
\label{mean_values}
}
\end{table*}

For the GBS we list in Table \ref{gaia_benchmark} the abundance ratios [X/Fe] obtained for each line separately in the blue and red spectral ranges, while the average values and the comparison with the literature values are provided in Table \ref{mean_values} (the GBS atmospheric parameters can be found in the Appendix; see Table \ref{atmospheric_param}). For Cu and Ba, for which we have only one line, the uncertainties in the table represent the total uncertainties (computed as the square root of the sum of the squares of $\sigma_1$ and $\sigma_2$). Instead, for the other elements the error is the standard deviation of the abundances obtained from the different lines.  As  can be seen, for the GBS that are old and for quiet stars we obtain solar-scaled abundances for all the heavy elements. Our estimates for \object{$\alpha$\,Cen\,A}, \object{$\tau$\,Cet,} and \object{18\,Sco} are in fair agreement with \cite{2018luck}, who analysed high-resolution (R$=115\,000$), high S/N HARPS spectra of a sample of 907 F-G-K dwarf stars in the solar neighbourhood. Our estimates are also similar to the results by \cite{2020casali}, who performed a detailed spectral analysis of HARPS spectra of 560 solar-type stars. The marginal discrepancies of the individual abundances could be related to the different line lists (inclusion of HFS and isotopic splitting) and methods used (i.e. EW versus spectral synthesis).  We also find that our values are in good agreement with \cite{2020casamiquela}, who analysed a sample of GBS. They used the same spectra and \ges line list as we did, but they analysed different lines (e.g. the Cu\,{\sc i} lines at 5218.20\,\AA\, and at 5220.07\,\AA). Thus, the small discrepancies that can be seen in Table \ref{mean_values} (for example for [Cu/Fe] of \object{$\alpha$\,Cen\,A} or for [La/Fe] of \object{$\tau$\,Cet}) can be related to the different lines used. \object{$\beta$\,Hyi} is a slightly evolved GBS, which is in the sub-giant branch phase of its evolution. In the literature we find few measurements of the $s$-process elements: our [Ba/Fe] estimate is in fair agreement with \cite{2014bensby} and \cite{2015jofre_emiliano}. To our knowledge these are the only results of the heavy elements abundances for \object{$\beta$\,Hyi}.

\subsection{The young clusters}

As already mentioned in Sect.\ref{sec:data}, in addition to  the stars observed within the GES, we analysed three stars of \object{IC\,2391} taken from \cite{2017dorazi}. Firstly, we re-determined the atmospheric parameters by applying our new spectroscopic approach using titanium lines, as described in Paper\,I. Our final values can be found in Table \ref{ic2391_do17}. The largest differences are seen for the $\xi$ parameter, for which we obtained a difference $\Delta\xi$(our-DS13)=$-0.59$\,\kms for star PMM\,665, $-0.43$\,\kms for PMM\,4362, and $-0.55$\,\kms for star PMM\,1142. Instead, for star PMM\,1142 our \teff value is 300\,K lower than the value found  by DS13, mostly due to the different line lists used. However, our spectroscopic \teff is corroborated by the estimates derived using 2MASS photometry \citep{2003cutri} and using the relations by \cite{2010casagrande}, which are equal to $T(J-K)=5378\pm105$\,K, $T(V-J)=5386\pm135$\,K, $T(V-H)=5408\pm153$\,K, and $T(V-K)=5380\pm155$\,K. An increase of $\sim0.6$\,\kms in the $\xi$ values results in a decrease of $\sim0.2$\,dex in [Fe/H] and of $\sim0.3$\,dex in the Y abundance, which is derived from moderately strong lines. Instead, the same variation produces a negligible change in La abundance, for which only weak lines have been used. Thus, the sensitivity to variations in the $\xi$ parameter depends mainly on the strength of the line.  

The final abundances of the individual stars can be found in Table \ref{abundances_480_680} for the red spectral setup and Table \ref{abundances_blue} for the blue range. The uncertainties $\sigma_1$ and $\sigma_2$ indicated in the two tables are calculated as described in Sect. \ref{sec:errors}. The mean cluster abundances for each elements are given in Table \ref{mean_values}; in this case, the uncertainties were computed as the error on the mean. We note that these mean values come from the averaged abundances of redder and bluer spectral ranges. For \object{NGC\,2264} we analysed only one star; therefore, we assumed as a conservative error value the uncertainty in the fitting procedure. For completeness, we also report  the \vsini measured by us along with values derived by the OACT node of the \ges consortium (\vsini$_{\rm{lit.}}$), as well as the S/N. The mean difference (and error on the mean) between our \vsini and the OACT values is equal to 0.6 $\pm$ 0.2\,\kms, with a standard deviation $\sigma$=0.8\,\kms. Overall, our measurements are in good agreement with the OACT node results. 

As  can be seen in Table \ref{mean_values}, only \object{IC\,2391}, \object{IC\,2602,} and \object{NGC\,2516} have been  extensively studied.  Regarding the first two, our mean values of [Ba/Fe] of these clusters confirm the overabundance already pointed out by \cite{2009dorazi} (for \object{IC\,2391} and \object{IC\,2602}) and \cite{2013desilva} (for \object{IC\,2391} alone). We note that our estimate is lower by 0.15\,dex for \object{IC\,2391} and by 0.24\,dex for \object{IC\,2602} than the value reported by \cite{2009dorazi}. This difference could be related to the different techniques and lines used, since \cite{2009dorazi} derived the Ba abundance from the EWs of the lines at 5853.7\,\AA\, and at 6496.9\,\AA, while here we use the spectral synthesis only for the line at 5853.7\,\AA. Moreover, there is a difference in the adopted solar abundances between the two studies: \cite{2009dorazi} derived a $\log$(Ba)$_{\odot}$ equal to 2.22\,dex, which is 0.09\,dex lower than our adopted value. For \object{IC\,2391} we find a value of [Ba/Fe] that is  0.09\,dex lower than that found by \cite{2013desilva}, and that is in fair agreement within the observational uncertainties. 

When focusing on the other $s$-process elements, we detect a mild enhancement for [Y/Fe], at $\sim$ 0.3 dex level, higher with respect to the values reported by \cite{2017dorazi} by $\Delta$[Y/Fe]=+0.15\,dex and $\Delta$[Y/Fe]=+0.11\,dex for IC 2391 and IC 2602, respectively. This can be simply explained by the difference in the $\xi$ values, which impacts the abundances derived from strong lines. Instead, [Zr/Fe], [La/Fe], and [Ce/Fe] exhibit solar values, as in \cite{2017dorazi}. For [La/Fe] of \object{IC\,2602}, the mean value is calculated only on the values derived from the two lines in the blue range measured in the star 10440681$-$6359351. Regarding the abundances in the red for Zr in \object{IC\,2602}, we rely only on one measurement, for the star 10442256$-$6415301. For this star we obtain [Zr/Fe]=$0.0 \pm 0.35 \pm 0.1$, where the large uncertainty on the fit is due to the poor quality of the spectrum. On the other hand, the determination of Zr abundance measurement for the star 10440681$-$6359351 (in the blue wavelength domain) is hampered by the very low S/N value in the HARPS spectrum.
For both \object{IC\,2391} and \object{IC\,2602} we present here the first estimates of [Cu/Fe] and [Sr/Fe].\\   
For \object{NGC\,2516}, the third most studied cluster in our list, our mean [Ba/Fe] estimate of $+0.31\pm0.03$ lies  between the results of \cite{2009dorazi} (+0.41 $\pm$ 0.04,  determined through EWs) and \cite{2018magrini} (+0.20 $\pm$ 0.08, the average value of the results in the \ges iDR5 catalogue). Our [Y/Fe] is in very good agreement within the errors with \cite{2011maiorca} (+0.16 $\pm$ 0.03), while we see larger differences for [Ce/Fe] between the two studies. Our findings confirm a solar [Ce/Fe] for \object{NGC\,2516}, while \cite{2011maiorca} and \cite{2018magrini} reported +0.18 $\pm$ 0.02 and +0.38 $\pm$ 0.10, respectively. However, we measured Ce only in one star,  07544342$-$6024437. The discrepancy between our results and the literature values could be due to the different techniques used. 

The cluster \object{IC\,4665} was recently used by \cite{2021spina} in his study of Galactic OCs observed within the GALAH survey \citep{2021buder} (for which we considered only the results obtained with the dwarf stars). As  can be seen in Table\,\ref{mean_values}, there is a large discrepancy between our [Cu/Fe] value and that of \cite{2021spina}, possibly due to the different line lists, techniques, stars per clusters analysed, and the lower resolution of the spectra analysed, all of which can affect the measurements in crowded regions. The measured [Y/Fe] ratios in the two studies agree very well, and we also found fairly good agreement  for [Ba/Fe]. In the table we also report  the value of [La/Fe], for which  caution should be used, however, as stated in \cite{2021buder}. These measurements could be affected by heavy blends of La lines used in the analysis in the GALAH survey.

Finally, for \object{NGC\,2547} and the SFR \object{NGC\,2264}, this is the first time the abundances of the $s$-process elements have been determined. We obtained super-solar abundances of [Ba/Fe], mild enhancement of [Y/Fe], and solar values of [Cu/Fe].

\subsection{Trends with stellar parameters}

\begin{figure*}[htbp]
     \centering
      \subfloat{\includegraphics[scale=0.3]{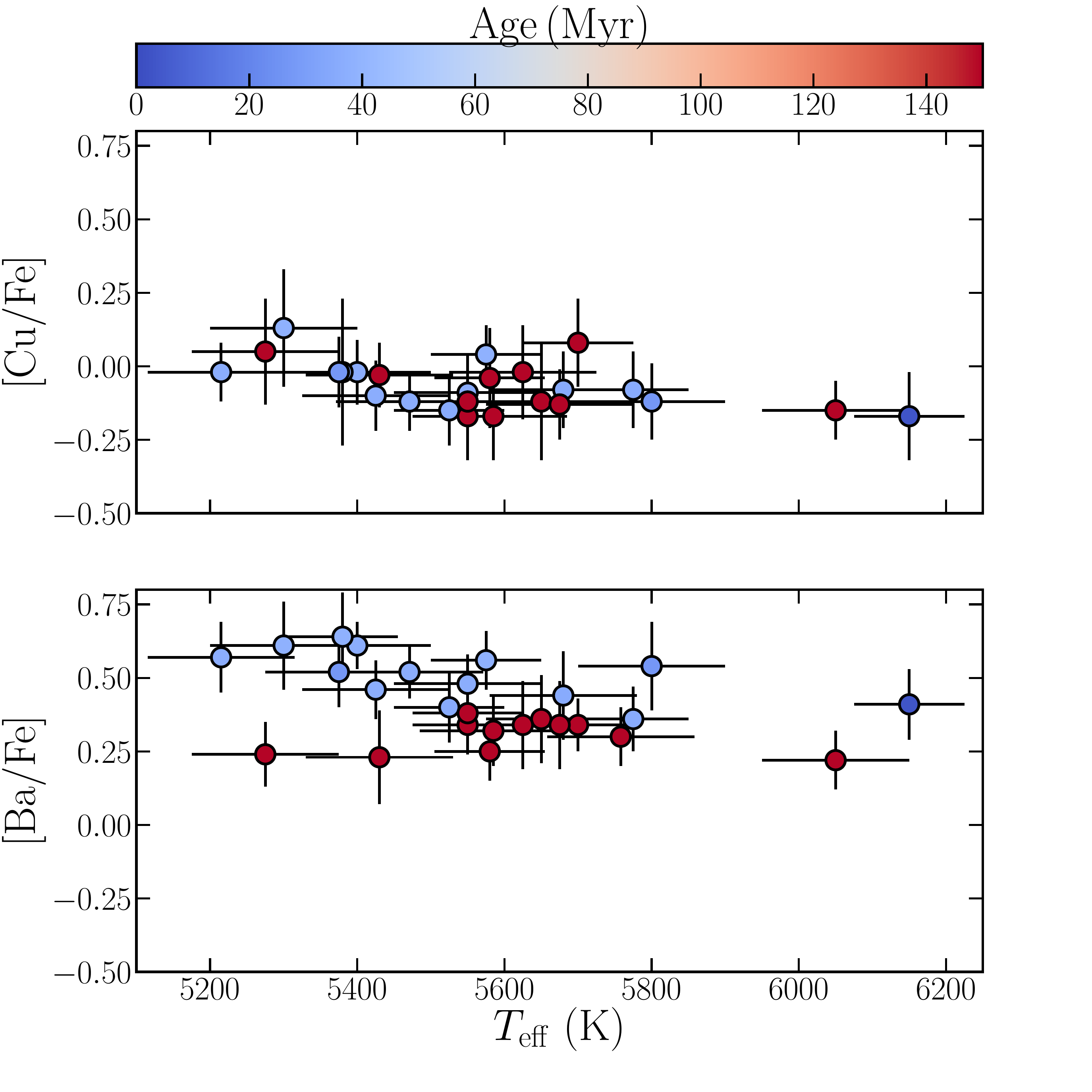}}
      \qquad
      \subfloat{\includegraphics[scale=0.3]{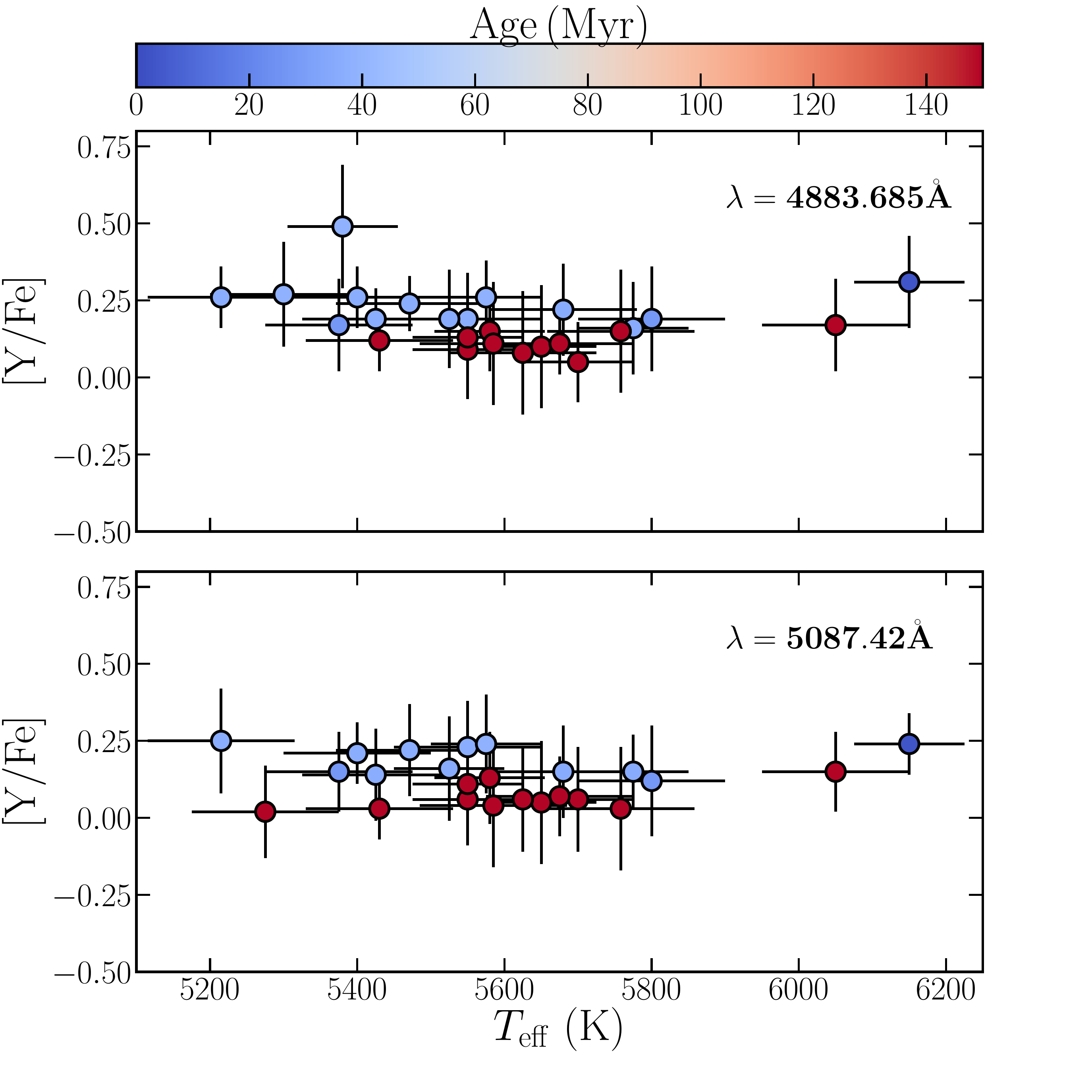}}
      \caption{ [X/Fe] as a function of \teff;  the points are colour-coded according to   age. }
        \label{trend_teff_abb_age}
\end{figure*}

\begin{figure*}[htbp]
     \centering
      \subfloat{\includegraphics[scale=0.3]{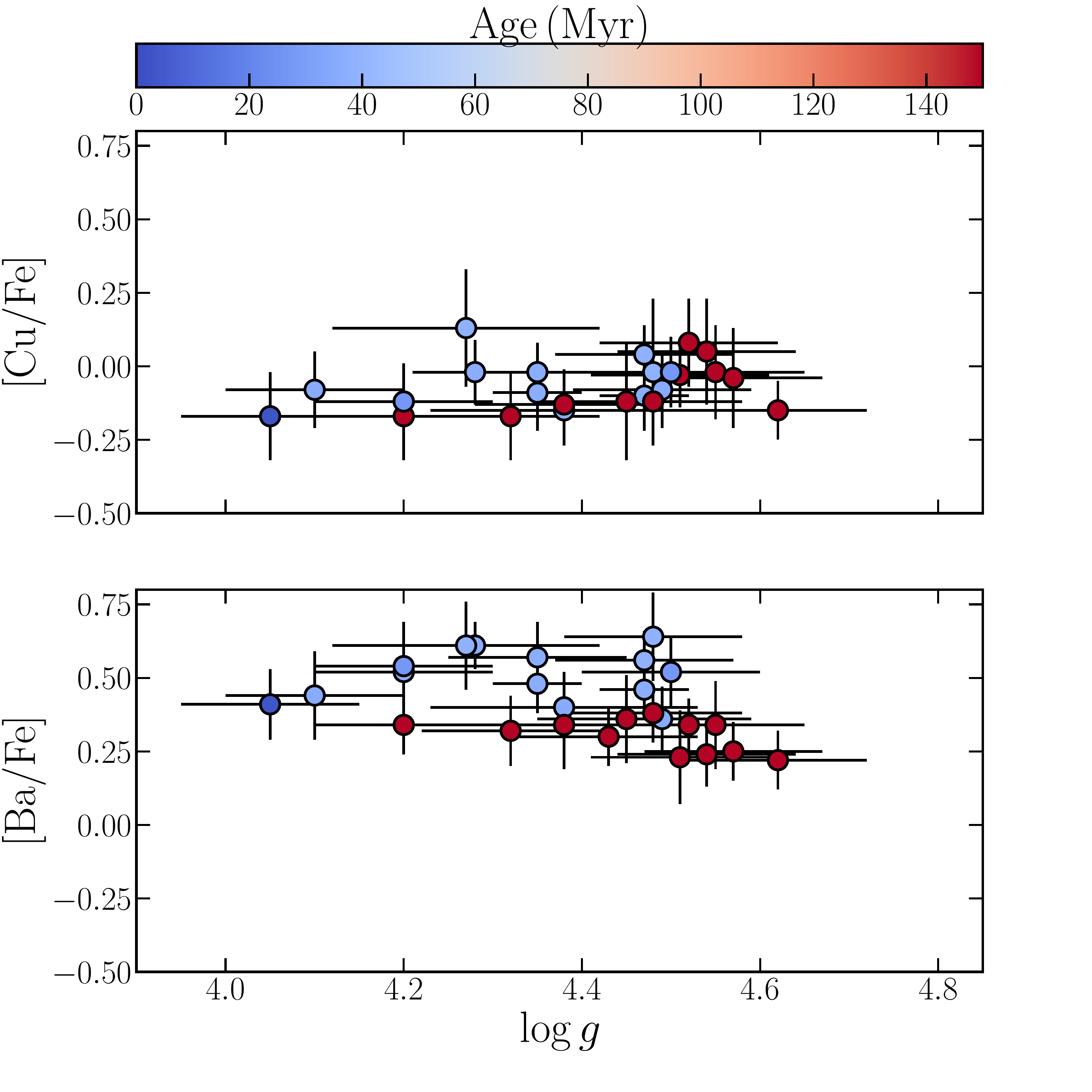}}
      \qquad
      \subfloat{\includegraphics[scale=0.3]{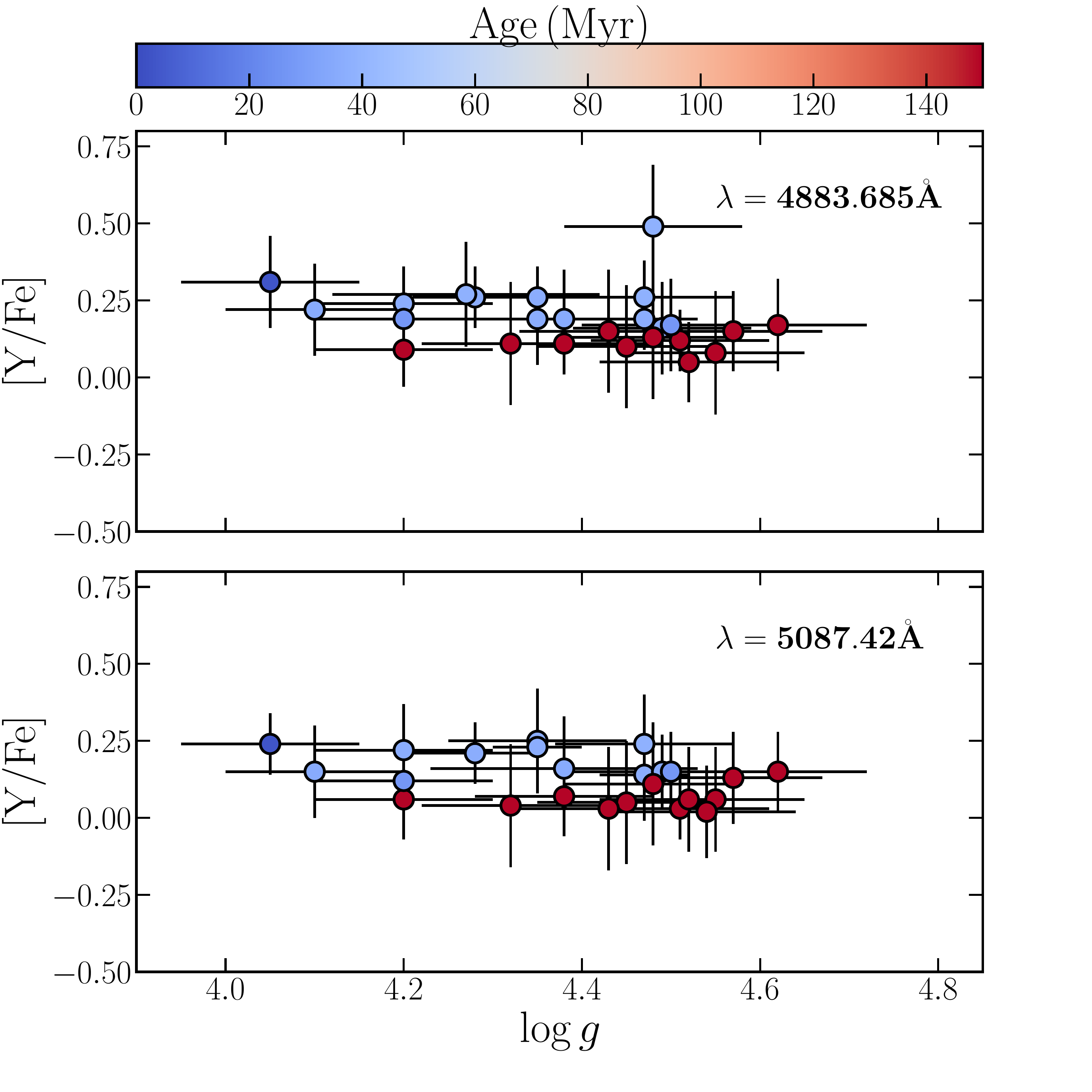}}
      \caption{ [X/Fe] as a function of \logg;   the points are colour-coded according to   age. }
        \label{trend_logg_abb_age}
\end{figure*}

\begin{figure*}[htbp]
     \centering
      \subfloat{\includegraphics[scale=0.3]{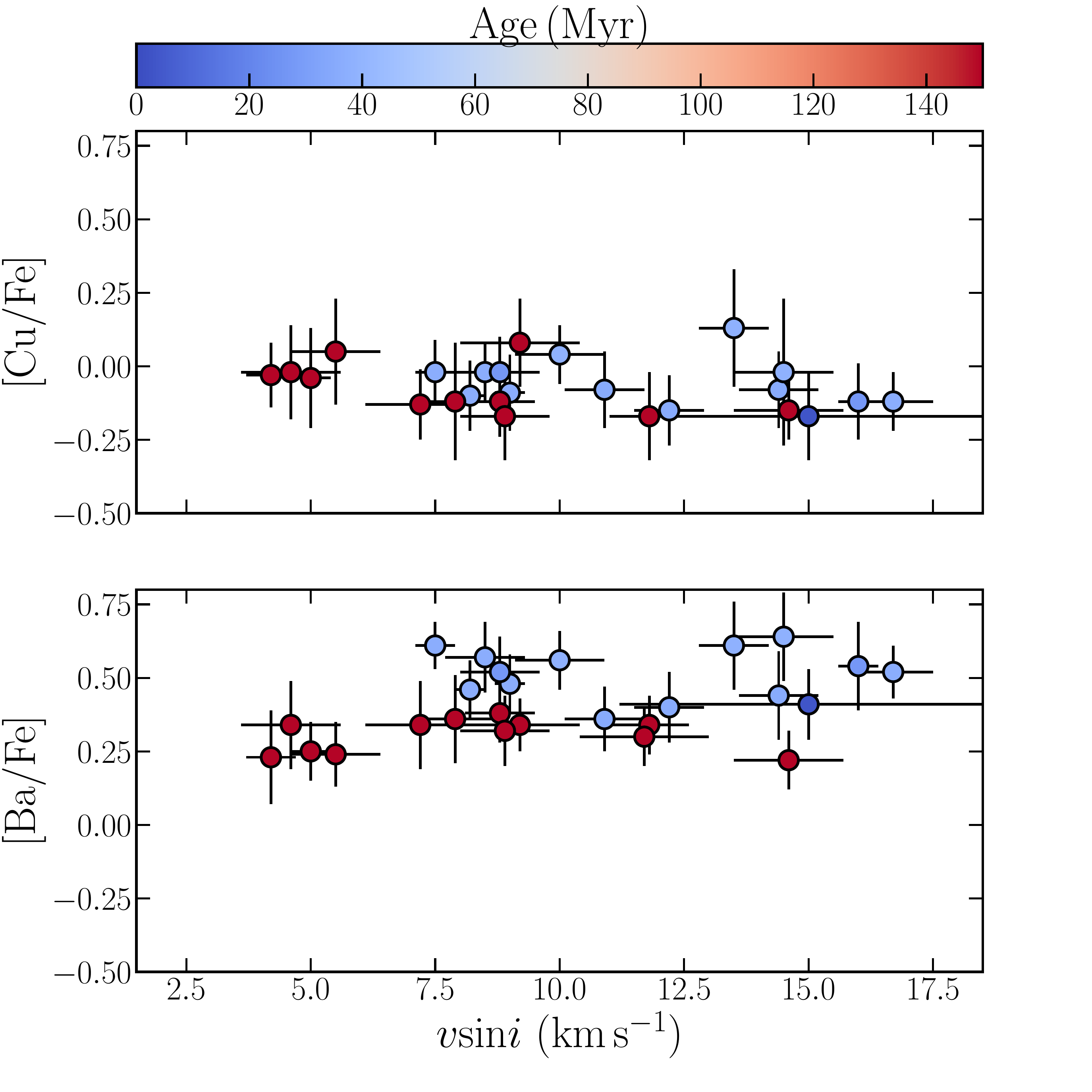}}
      \qquad
      \subfloat{\includegraphics[scale=0.3]{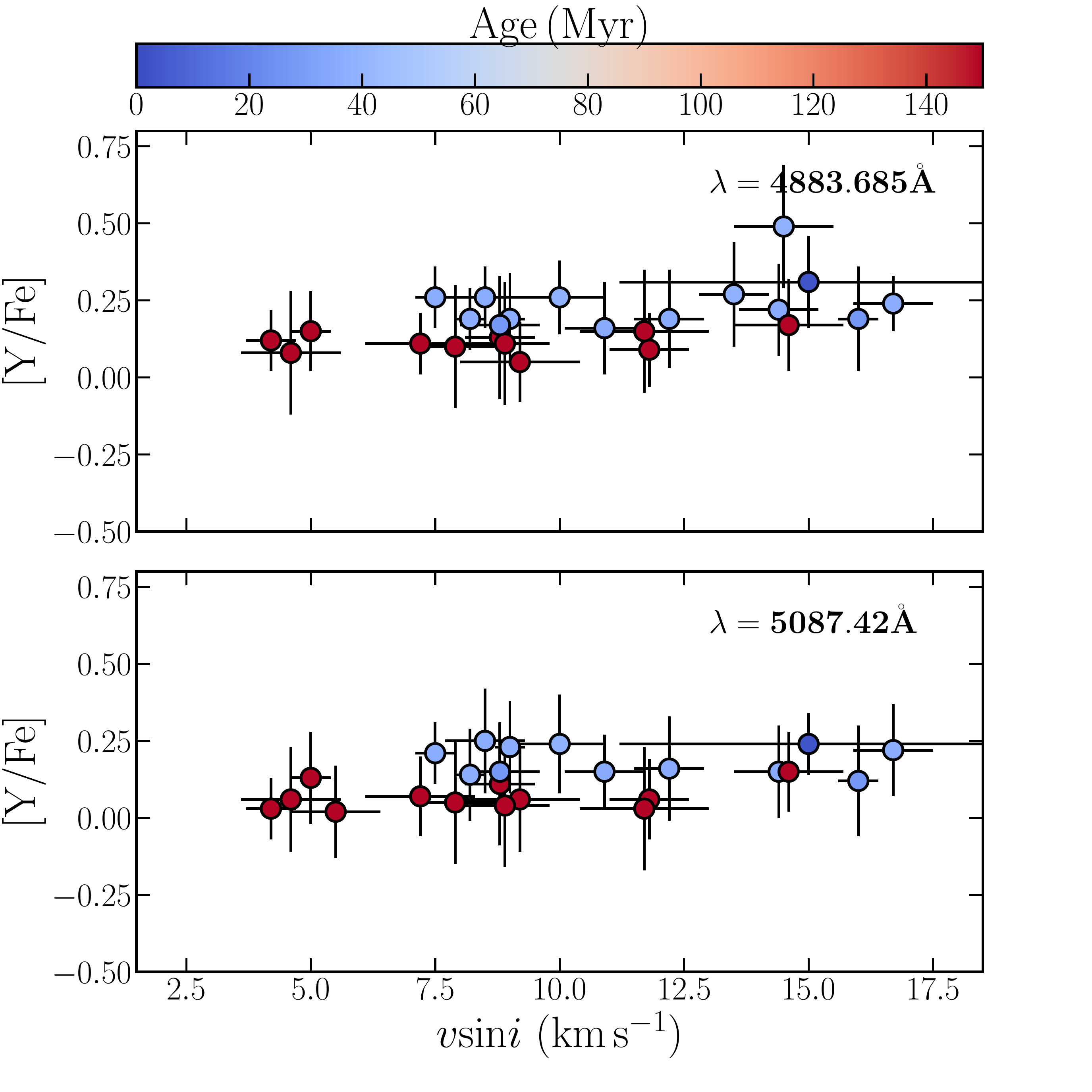}}
      \caption{ [X/Fe] as a function of \vsini;   the points are colour-coded according to   age. }
        \label{trend_vsini_abb_age}
\end{figure*}

In Figs.\,\ref{trend_teff_abb_age}, \ref{trend_logg_abb_age}, and \ref{trend_vsini_abb_age}, we plot the [X/Fe] ratios as a function of the stellar parameters \teff, \logg, and \vsini, respectively. In these plots the points are colour-coded according to  age. There are no significant trends with the stellar parameters, which validates our spectroscopic analysis. As  can be seen in the bottom left panel, we obtain the super-solar [Ba/Fe] ratios, more evident at younger ages, with values between roughly +0.25 and 0.65-0.70\,dex. Curiously, our [Y/Fe] estimates (top right and bottom right panels) also indicate  a mild enhancement going toward younger ages, ranging between +0.1 up to +0.25\,dex, at variance with the solar values previously found by some authors (e.g. \citealt{2012dorazi,2017dorazi}). Regarding Y, none of the trends with the stellar parameters is significant. 

For both elements (Y and Ba) a sharp separation between the blue dots (the youngest stars, with ages of less than 50 Myr) and the red dots (the oldest stars in our sample, with   ages of $\sim$150 Myr) can be seen. For Cu, instead, the results are homogeneously distributed with the age. 

\begin{table*}[!]
\caption{Atmospheric parameters and abundances of Fe and Ti of stars in \object{IC\,2391}.}
\setlength\tabcolsep{6pt}
\small
\centering
\begin{tabular}{lccccccc}
\toprule
ID & \teff & \logg & $\xi$ & [Fe/H]$_{\rm{I}}$ & [Fe/H]$_{\rm{II}}$ & [Ti/H]$_{\rm{I}}$ & [Ti/H]$_{\rm{II}}$\\
\midrule
PMM\,1142 & 5400 $\pm$ 100 & 4.28 $\pm$ 0.07 & 0.95 $\pm$ 0.10 & 0.00 $\pm$ 0.02 $\pm$ 0.07&  +0.11 $\pm$ 0.03 $\pm$ 0.06& +0.04 $\pm$ 0.02 $\pm$ 0.10& +0.06 $\pm$ 0.03 $\pm$ 0.05\\
PMM\,665 & 5425 $\pm$ 100 & 4.47 $\pm$ 0.05 & 1.01 $\pm$ 0.15 & +0.08 $\pm$ 0.02 $\pm$ 0.07& +0.14 $\pm$ 0.05 $\pm$ 0.06 & +0.10 $\pm$ 0.02 $\pm$ 0.11& +0.11 $\pm$ 0.05 $\pm$ 0.05 \\
PMM\,4362 & 5550 $\pm$ 100& 4.35 $\pm$ 0.05& 0.97 $\pm$ 0.15& +0.13 $\pm$ 0.02 $\pm$ 0.08& +0.14 $\pm$ 0.04 $\pm$ 0.06 &+0.09 $\pm$ 0.03 $\pm$ 0.11 &+0.12 $\pm$ 0.02 $\pm$ 0.05\\
\bottomrule
\end{tabular}
\tablefoot{Atmospheric parameters and abundances of Fe and Ti derived with the same methodology as in Paper\,{\sc i} of the stars from  \cite{2017dorazi}. The uncertainties are due to the dispersion among different atomic lines (first value) and to uncertainties on the stellar parameters (second value).}
\label{ic2391_do17}
\end{table*}

\begin{figure}[!]
\centering
      \subfloat{\includegraphics[width=0.47\textwidth]{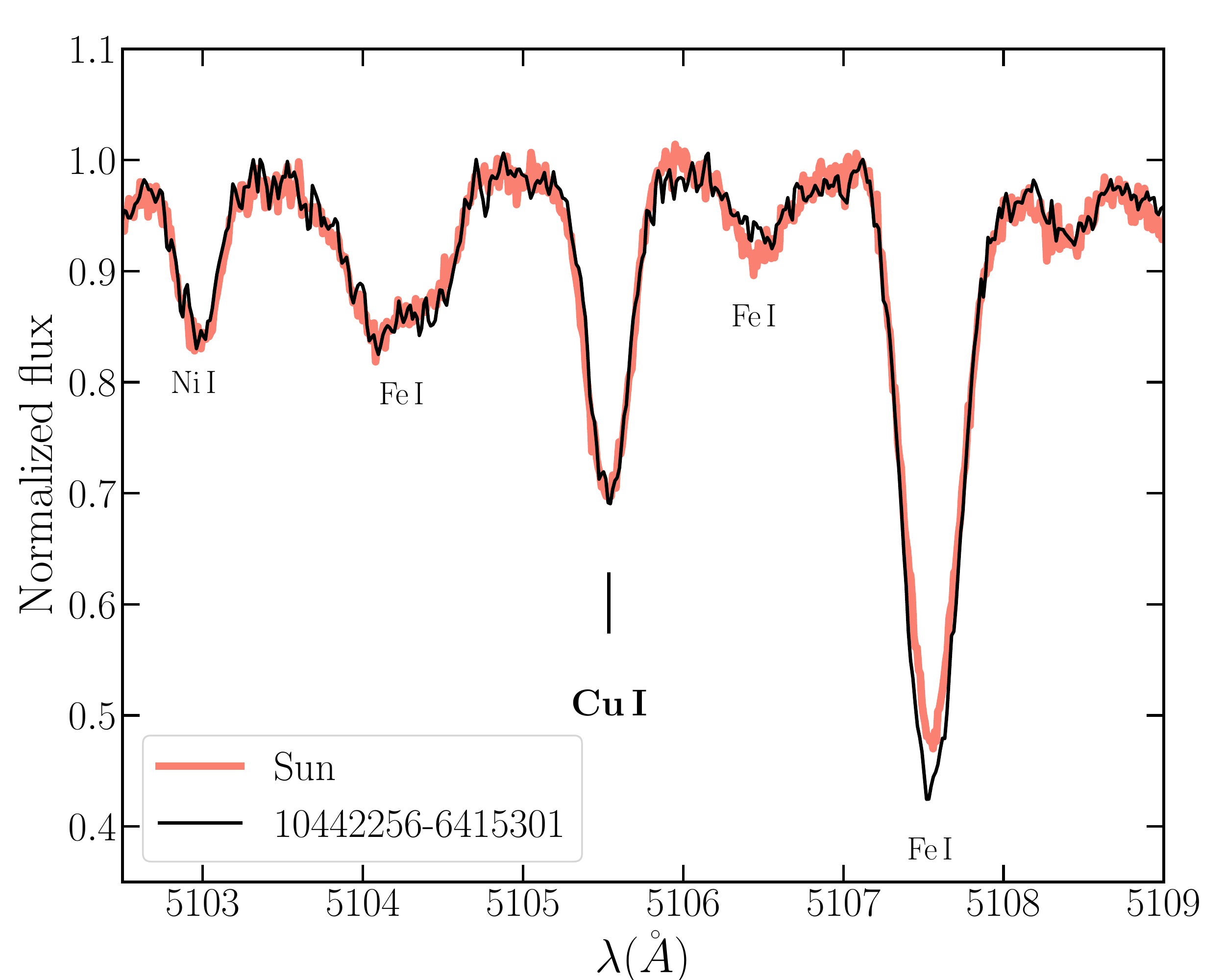}}
      \qquad
      \subfloat{\includegraphics[width=0.47\textwidth]{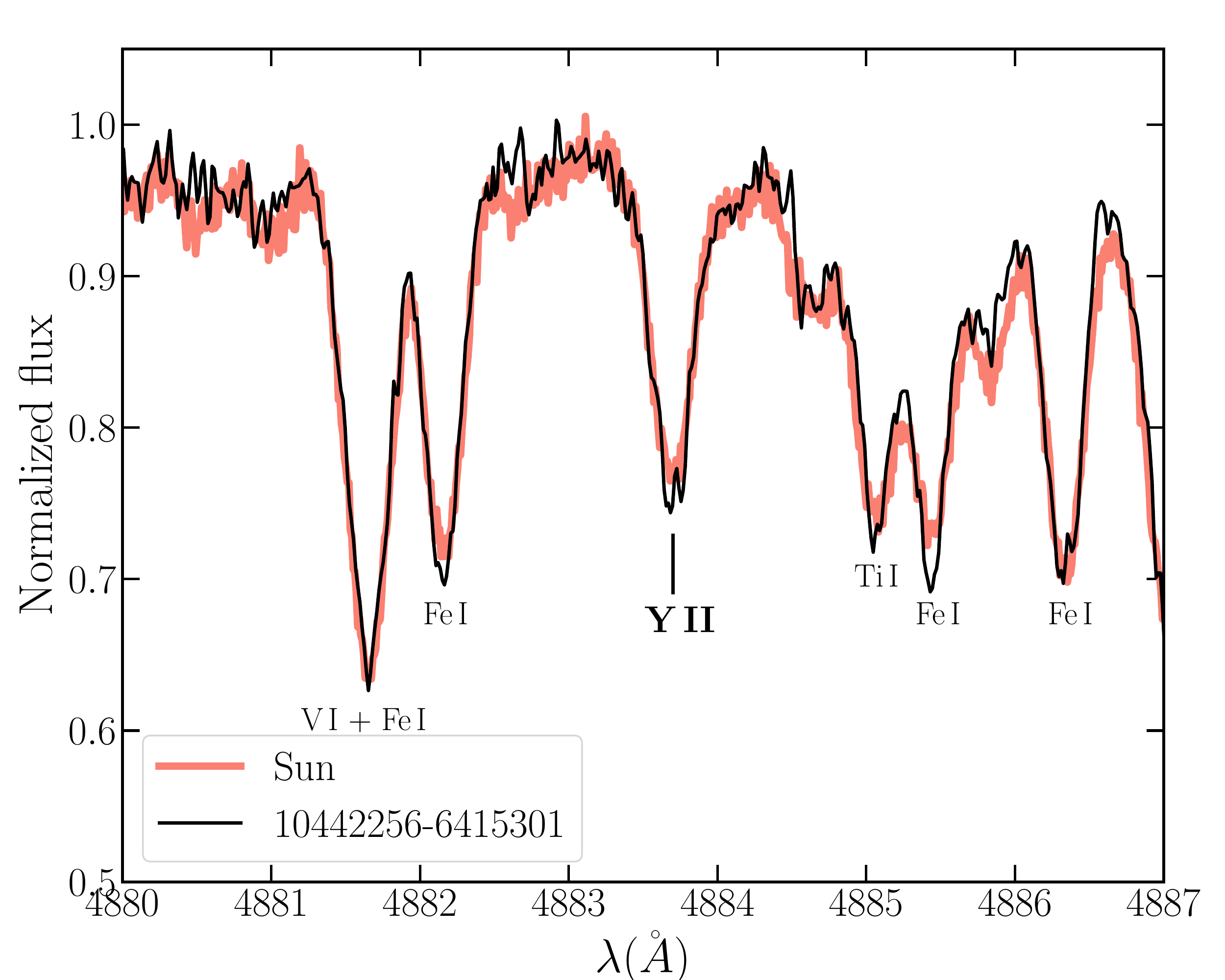}}
      \qquad    
      \subfloat{\includegraphics[width=0.47\textwidth]{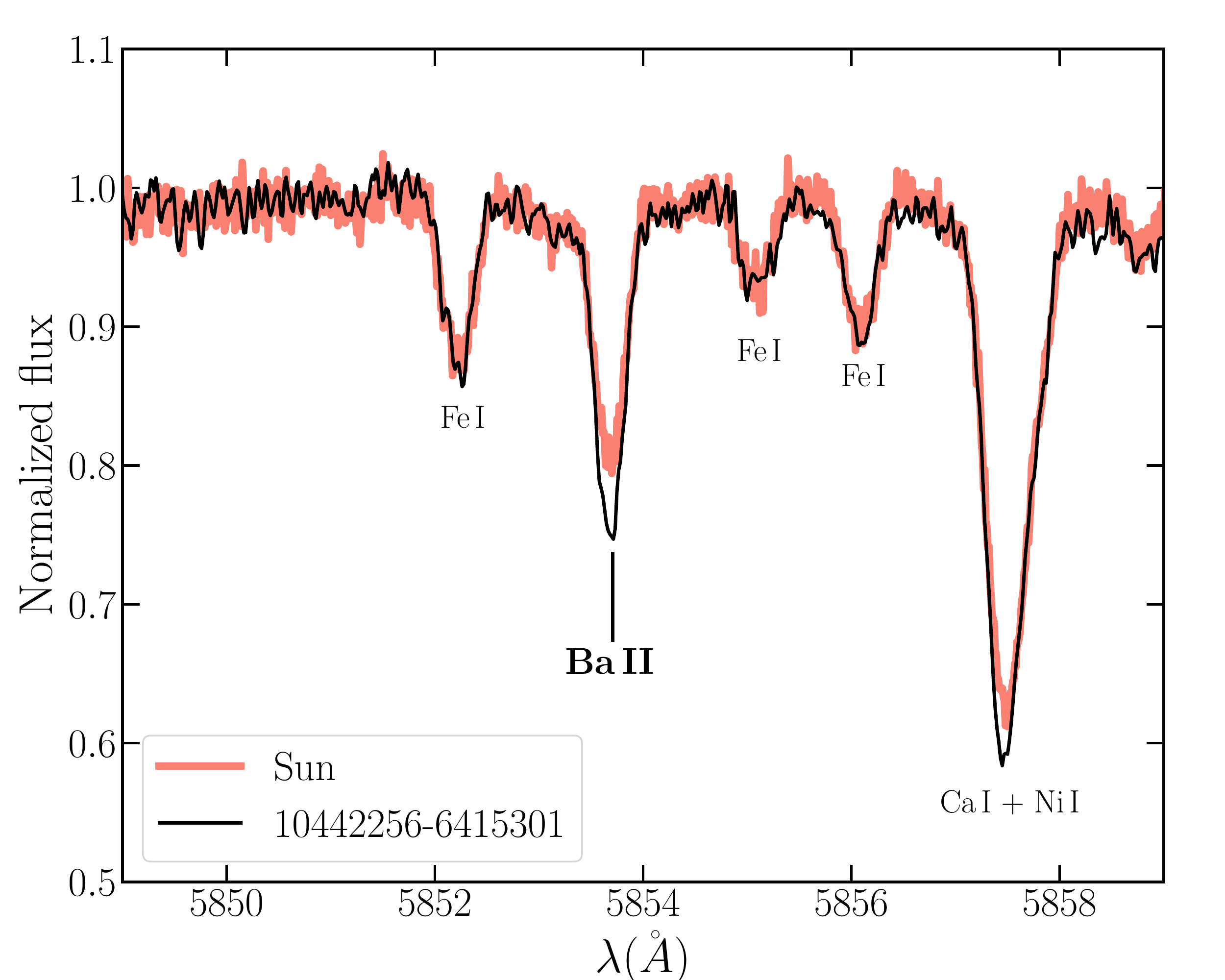}}
\caption{Comparison of Cu\,{\sc i} line at 5105.54\,\AA\, (top panel), Y\,{\sc ii} line 4883.69\AA\, (central panel), and Ba\,{\sc ii} line at 5853.7\,\AA\,(bottom panel) between the Sun (light pink line, age $\sim$4 Gyr) and the young star 10442256$-$6415301 (black line, age $\sim$35 Myr). }
\label{Ba_Y_line}
\end{figure}

\section{Discussion}\label{sec:discussion}

As already mentioned in Sect.\,\ref{sec:introduction}, it is not possible to reconcile the super-solar abundances of Ba together with solar La and Ce abundances with the predictions of the $s$-process and $r$-process nucleosynthesis models (e.g. \citealt{2020kobayashi} and references therein),
without invoking other processes. On the other hand, the enrichment of Y with respect to Sr and Zr is less clear and needs to be studied in more detail. As discussed in this work, the Y enhancement could be caused by some observational issue. Moreover, a large variety of processes can contribute to the production of elements in the Sr-Y-Zr region, which could have caused the variations observed in YOCs. Together with the nucleosynthesis processes mentioned in the previous sections (the $s$-process, the $i$-process, and the $r$-process), elements in this mass region may  also be made in different neutrino-wind components from CCSNe \citep[e.g.][]{farouqi:09,roberts:10,arcones:11} and in electron-capture supernovae \citep[e.g.][]{wanajo:11,jones:19}. 

From the observational point of view, the difficulty may be related to the spectral analysis and mechanisms that are at work on the photosphere of a young star to magnify the Ba abundance or to the nucleosynthesis processes that produce the elements heavier than Fe in the Galaxy. We discuss both possibilities below;  in Sect.\,\ref{sec:spectral_lines} the problems that may be related to the spectra lines and in Sec.\,\ref{sec:evolution} the potential issues with GCE models.

\subsection{Behaviour of spectral lines}\label{sec:spectral_lines}

Since we cannot reconcile the observed Ba overabundance and the slightly super-solar values of Y with nucleosynthesis models predictions, we believe that the key to understanding the Ba puzzle might rely on the age of the stars. The younger the star, the higher the levels of its activity,   at the level of the chromosphere or of photospheric magnetic fields. This in turn can result in an alteration of the photospheric structure, and consequently in an alteration of the profile (i.e. strengths) of the spectral lines. Thus, in young stars it is important to know where the line forms in the photosphere and how it is affected by magnetic activity, as already demonstrated by \cite{2020spina}. 

Looking at Table \ref{logtau}, where the optical depths of line formation are listed, we infer that the line at 5105.54\,\AA\, of Cu and line at 5853.7\,\AA\, of Ba form at similar depths, with \logtau$_{\rm{core}}$ of $-3.4$ and $-3.2$, respectively. Thus, we would expect to observe the same effects in the two elemental abundances derived through these lines. Nonetheless, we obtain solar values of [Cu/Fe], while [Ba/Fe] is enhanced (between +0.25 and +0.70\,dex). This is also confirmed by Fig.\,\ref{Ba_Y_line}, where we compare the spectrum of the Sun (light pink line) with the solar analogue 10442256$-$6415301 (black line) that belongs to \object{IC\,2602} (age $\sim 35$\,Myr).  In these plots the solar spectrum was convolved with a rotational profile with \vsini=11\,\kms, to match the \vsini of the young star, and we added  Gaussian noise to obtain S/N=110 (we use the iSpec tool by \citealt{2014blanco-ispec}). Both lines are strong, but the Ba line (bottom panel) is deeper in the younger stars, while the profiles of the Cu line (top panel) are identical. For   star 10442256$-$6415301 we obtain [Ba/Fe]=$+0.36\pm0.11$, while [Cu/Fe]=$-0.08\pm0.13$ is solar within the uncertainties. Nevertheless, the different ionisation stages of the two species (i.e. neutral Cu versus singly ionised Ba) could explain (at least partially)  the observed behaviour. In the following we investigate neutral and ionised lines of other species. 

It has been observed that strong ionised lines of Fe, Ti, and Cr yield large abundances in young and cool (\teff<5400\,K) stars (see e.g. \citealt{2009dorazi,2010schuler,2019tsantaki,2020baratellaB}). This is referred to as the ionisation balance problem. \cite{2017aleo} explain this effect as being due to the presence of undetected blends in the ionised lines (in particular of Fe) that become more severe in the cool regime. We selected a set of seven strong Fe\,{\sc ii} lines that were   initially excluded from our analysis in Paper\,{\sc i}. For each line in Fig.\,\ref{comp_FeII_lines_v11} we compare the spectra of the Sun and star 10442256$-$46415301, as in Fig.\,\ref{Ba_Y_line}. Most  of the Fe\,{\sc ii} lines are deeper in the young star than in the Sun, as observed for the Ba\,{\sc ii} line. This is also corroborated by the measured EWs and abundances obtained from the Fe\,{\sc ii} lines (computed by adopting the stellar parameters of Paper\,{\sc i}), as  can be seen in Table \ref{feII_abundances}.  

However, we note that even lines of neutral species show a behaviour similar to that of  the ionised lines. In Fig.\,\ref{comp_FeII_lines_v11} it is evident that Fe\,{\sc i} (in the first and second panels of the figure) is stronger in the young star than in the Sun. In the bottom panel of Fig.\,\ref{Ba_Y_line} the blend Ca\,{\sc i}+Ni\,{\sc i} at 5857.5\,\AA\, behaves similarly to Fe\,{\sc i}. We then compare the two spectra in small windows around two Mg\,{\sc i} lines in Fig.\,\ref{comp_Mg}, and eight Ca\,{\sc i} lines in  Fig.\,\ref{comp_Ca}. As  can be seen, the profiles of the  Mg lines are almost identical. Instead, for Ca we note that the weak lines (panels on the left) have similar depths, while the strong lines (panels on the right) are deeper in the young star (black line) than in the Sun (light pink line).    

Barium and strontium belong to the same group in the periodic table, so their outermost electron shells have similar configurations. Furthermore, they share similar nucleosynthetic origins. Hence, we should witness some effects also on Sr abundances, for which we exploit lines at 4607.33\,\AA\, of Sr\,{\sc i} ($\Delta_{\rm{NLTE}}$=+0.1\,dex) and 4215.52\,\AA\, of Sr\,{\sc ii}.  As  can be seen in Table\,\ref{abundances_blue}, we find good agreement between the Sr\,{\sc i} and Sr\,{\sc ii} abundances, at least in stars 0844052$-$5253171 (\object{IC\,2391}) and 10440681$-$6359351 (\object{IC\,2602}). In general, we find that [Sr/Fe] is solar in both cases. Looking at the depth of formation (Table \ref{logtau}), the Sr\,{\sc i} line forms deeper in the photopshere than the Ba line. This might in principle explain why we obtain solar-scaled abundances. Conversely, the Sr\,{\sc ii} line forms at \logtau=$-5.2$,  in the upper layers. However, when deriving the abundance of Sr\,{\sc ii}, we obtain again a solar composition, as  can be seen from Fig.\,\ref{best_fit_sr4215} where we plot the best fit models of Sr\,{\sc ii} lines in the Sun, star 08440521-5253171 (\object{IC\,2391}), and \object{$\tau$\,Cet}. 

Regarding the Y lines at 4883.69\,\AA\, and at 5087.42\,\AA, we found that they form at a similar depth, with \logtau$_{\rm{core}}$ equal to $-2.6$ and $-2.1$, respectively. From the comparison of the solar and the young solar analogue spectra in Fig.\,\ref{Ba_Y_line} (central panel), it can be seen that the Y\,{\sc ii} line at 4883.69\,\AA\, is stronger in the young star than in the Sun. This is in agreement with our derived abundance estimates.  
Interestingly, the La\,{\sc ii} lines analysed in this work should exhibit a behaviour similar to the Y\,{\sc ii} features (they share formation depths, ionisation stages, and nucleosynthesis channels). Nevertheless, the La abundances are solar, whereas [Y/Fe] values are enhanced.

Finally, in Fig.\,\ref{comp_Sc_lines_v11} we compare Sc\,{\sc i} and Sc\,{\sc ii} lines in the same way as in Fig.\,\ref{Ba_Y_line}. Scandium has the same electronic configuration as Y, and is similar to La, so we expect   these elements to show  similar behaviours. Instead, from Fig.\,\ref{comp_Sc_lines_v11} it is evident how the profile of both neutral and ionised Sc lines are the same in the Sun and in the young star. In this case there are no differences between the two spectra, for any line, with the exception of  5658.361\,\AA. The small difference we see in this line may be due to a blend with a nearby Fe\,{\sc i} line.

\begin{figure*}[h!]
\centering
\includegraphics[scale=0.38]{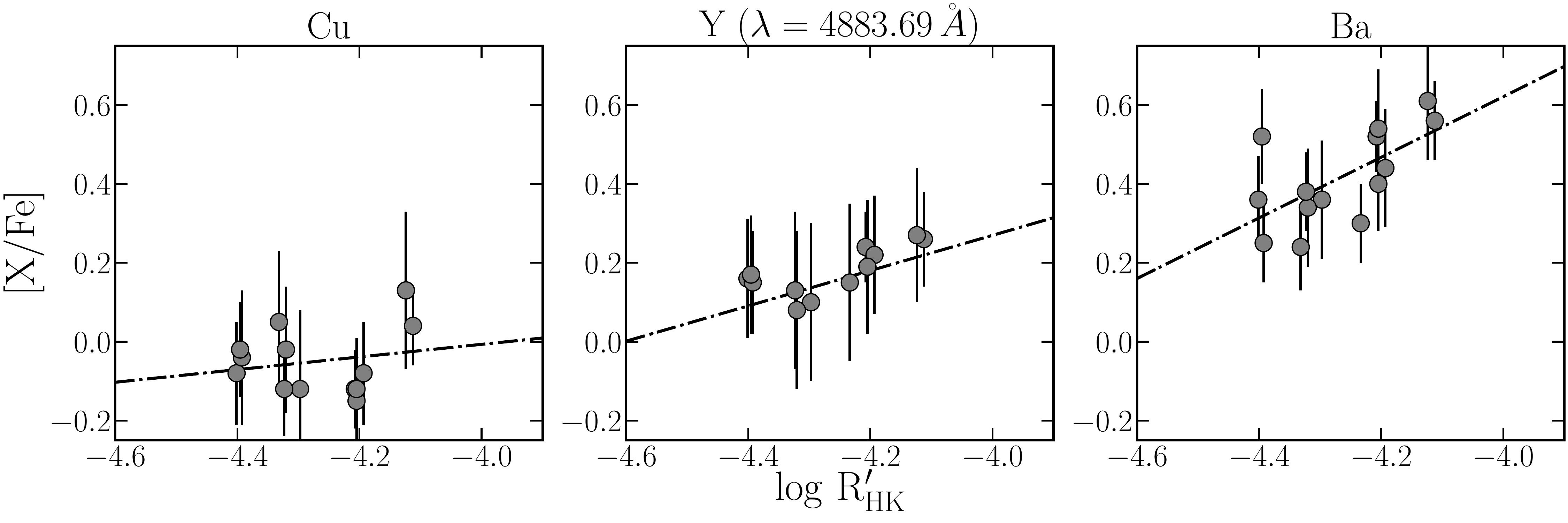}
\caption{[X/Fe] as a function of \logrhk, computed following \cite{2008mamajek} from the $\log\,(L_{\rm{X}}/L_{\rm{bol}})$ (details on the computation can be found in Paper I). The dot-dashed lines are the linear fits of the measurements.  } 
\label{activity}
\end{figure*}

We conclude that line formation depth and ionisation stages of the elements are not able to fully explain the very peculiar pattern of $s$-process elements in young open clusters. On the other hand, we cannot exclude that different (conspiring) mechanisms could be simultaneously at work.  Figure\,\ref{activity} shows that there seems to be a correlation between the larger abundances of Y and Ba with decreasing \logrhk (i.e. higher levels of stellar activity). We considered the \logrhk values already computed in Paper\,I, which were derived from the X-ray luminosities found in the literature and using the conversion relation by \cite{2008mamajek}. We note that the X-ray luminosities (and hence the \logrhk values) are not synchronous to our spectra, and consequently, to the derived abundances. Nevertheless, globally, we can conclude that indeed there is an {indication} of a correlation between the [Y/Fe] and [Ba/Fe] and \logrhk (not so evident for [Cu/Fe]). This plot surely deserves an in-depth investigation, and further observations are needed to study the behaviour at \logrhk$>-4.0$ and \logrhk$<-4.4$.

\cite{2020spina} proposed magnetic intensification as a possible explanation of the anomalous abundances of Ba  (and of other elements). In \cite{2017biazzo}, this possibility is also explored. Given the young age of our targets, this seems to be a promising solution. The presence of magnetic fields causes the atomic levels to be split into different components according to the Zeeman effect. This results in a broadening of the spectral line, with increased EWs and reduced line depths, which however is not seen in our lines (see Fig.\ref{Ba_Y_line}). The amount of splitting is directly proportional to $g_{\rm{L}}$, the square of the wavelength and the magnitude B of the magnetic field. To evaluate the sensitivity to magnetic fields of each line, we compute the Landé factor $g_{\rm{L}}$ (Column 7 of Table \ref{line_list}), as described in Sect.\,\ref{sec:lines}. Overall, our lines have  $g_{\rm{L}}<1.3$, which is relatively low. Looking at the Ba and Cu lines, we note that they have very similar $g_{\rm{L}}$ values, 1.07 and 1.10, respectively. Nevertheless, we obtained super-solar Ba abundance and solar Cu. The La line at 3988.5\,\AA\,has $g_{\rm{L}}=1.33$, which is the highest value, but we obtain solar abundance. Thus, we conclude that the magnetic intensification cannot explain the high values we obtain. 

Another possibility is the first ionisation potential (FIP) effect;  our lines, in particular Ba, seem to be  suitable candidates to show this. It has been shown that the coronal abundances derived from lines with FIP values below 10\,eV in the Sun are enhanced with respect to the photospheric values (see e.g.  the review by \citealt{2015laming}).  \cite{1999Sheminova} explored the idea that the gas exhibiting the FIP effect in the corona is connected to the photosphere through magnetic flux tubes, generated from magnetic elements or sunspots present on the surface. In principle, the same enhancement observed in the corona could also be found for  photospheric abundances. We retrieve the FIP values (Column 8 of Table \ref{line_list}) from \cite{1992gray}. As  can be seen, Ba and La lines have similar FIP values; therefore, this does not explain their discrepant abundances. However, the higher levels of activity due to the very young ages of our stars could be completely different than what is observed in other active, older stars. 

In summary, all the possible effects described above may play a role; however, there is no convincing evidence that any of them provides a definitive solution, yet.

\begin{table*}[]
\caption{Abundances of the individual strong Fe\,{\sc ii} lines.}
\setlength\tabcolsep{7pt}
 \small
\centering
\begin{tabular}{lccccccr}
\toprule
$\lambda$  & E.P.\,(eV) & $\log gf$ & EW$_{\odot}$\,(m\AA) & $\log$(Fe)$_{\odot}$ & EW$_{\star}$\,(m\AA) & $\log$(Fe)$_{\star}$\\
\midrule
4923.92 &  2.89 & -1.26 & 157.00 & 7.06 & 186.00 & 7.24\\
5197.57  & 3.23 & -2.22 & 89.60 & 7.37 & 104.00 & 7.55\\
5234.62  & 3.22 & -2.18 & 85.57 & 7.36 & 95.24 & 7.48\\
5325.55  & 3.22 & -3.16 & 40.38 & 7.34 & 43.25 & 7.39 \\
5534.84  & 3.25 & -2.87 & 62.66 & 7.45 & 73.40 & 7.61\\
6456.38  & 3.90 & -2.19 & 64.30 & 7.51 & 68.00 & 7.54\\
6516.08  & 2.89 & -3.31 & 63.83 & 7.64 & 76.70 & 7.85\\
\bottomrule
\end{tabular}
\tablefoot{Abundances of the individual strong Fe\,{\sc ii} lines from EW measurements in the Sun (Cols. 4 and 5) and the young solar analogue 10442256$-$6415301 (\object{IC\,2602}, age $\sim 35$\,Myr, Cols. 6 and 7).}
\label{feII_abundances}
\end{table*}

\begin{figure*}[!]
\centering
      \subfloat{\includegraphics[width=0.48\textwidth]{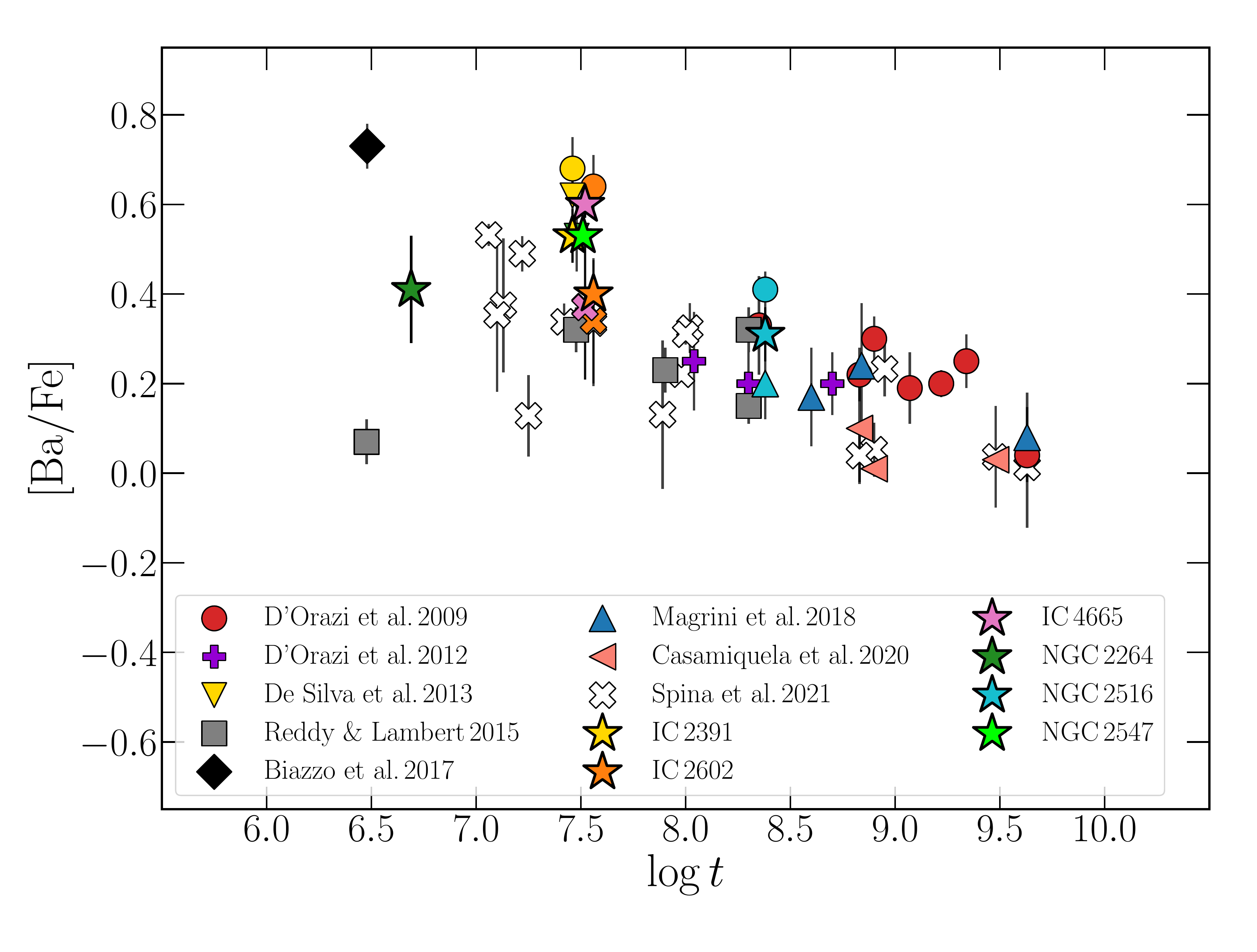}}
      \qquad
      \subfloat{\includegraphics[width=0.48\textwidth]{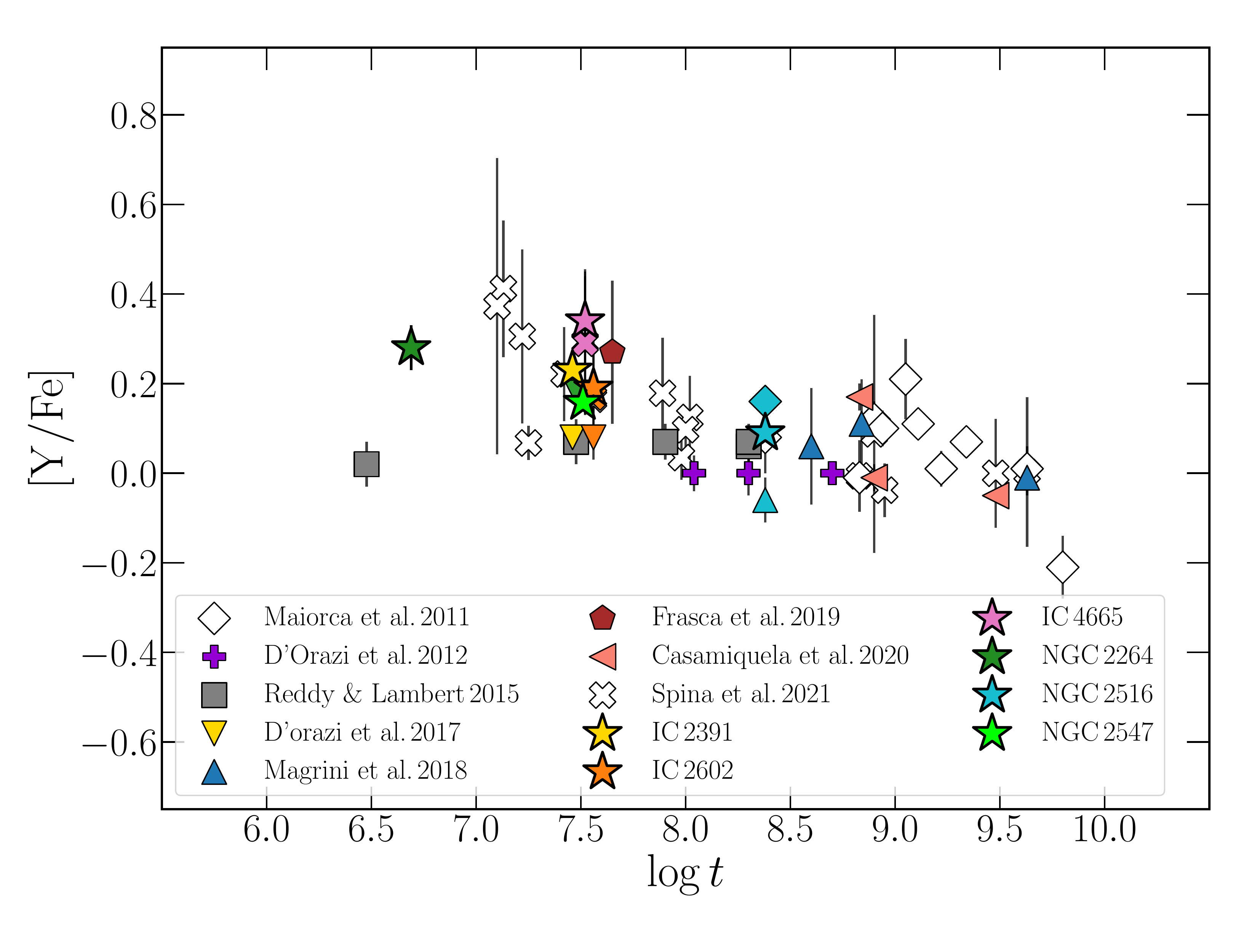}}
      \qquad
      \subfloat{\includegraphics[width=0.48\textwidth]{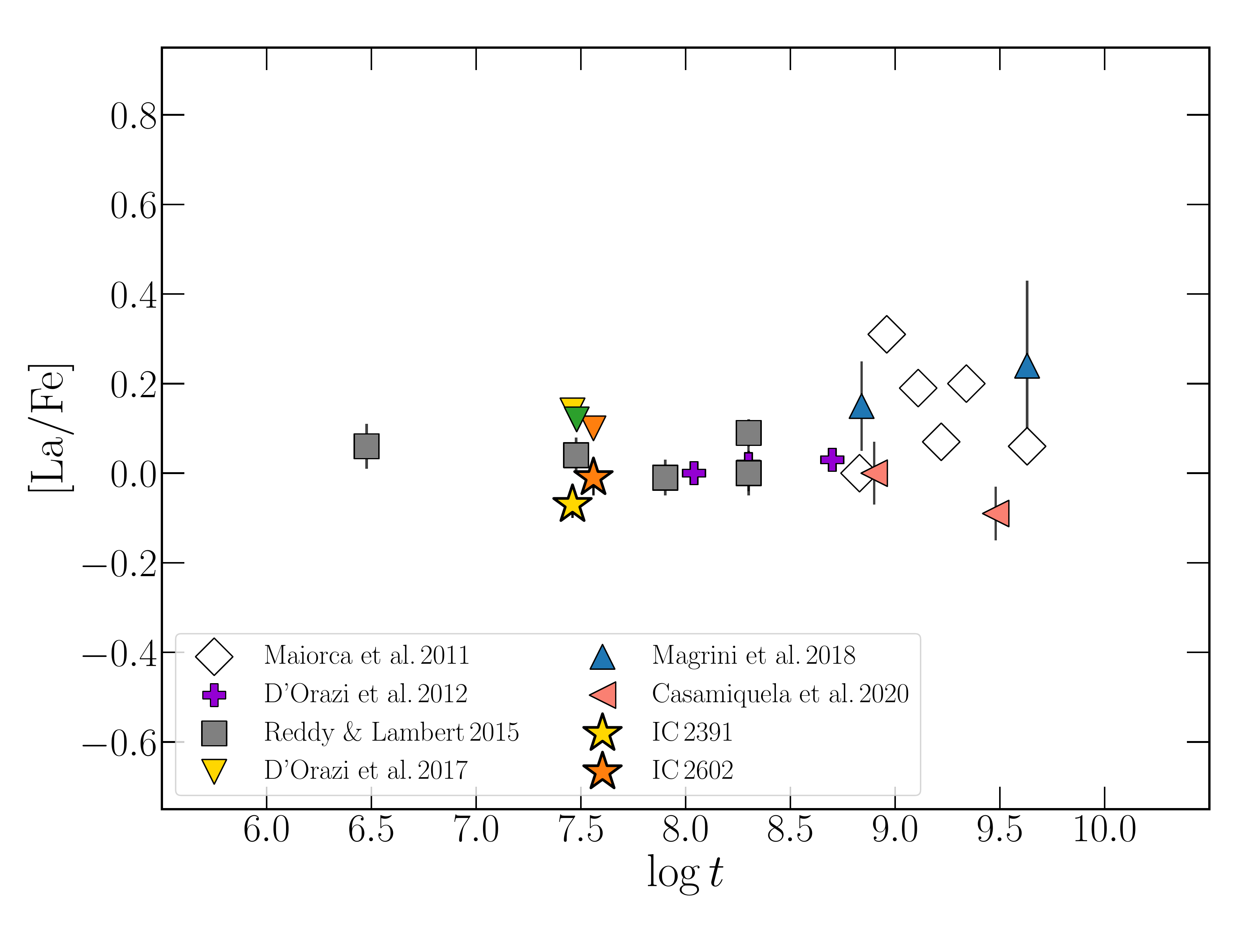}}
      \qquad
      \subfloat{\includegraphics[width=0.48\textwidth]{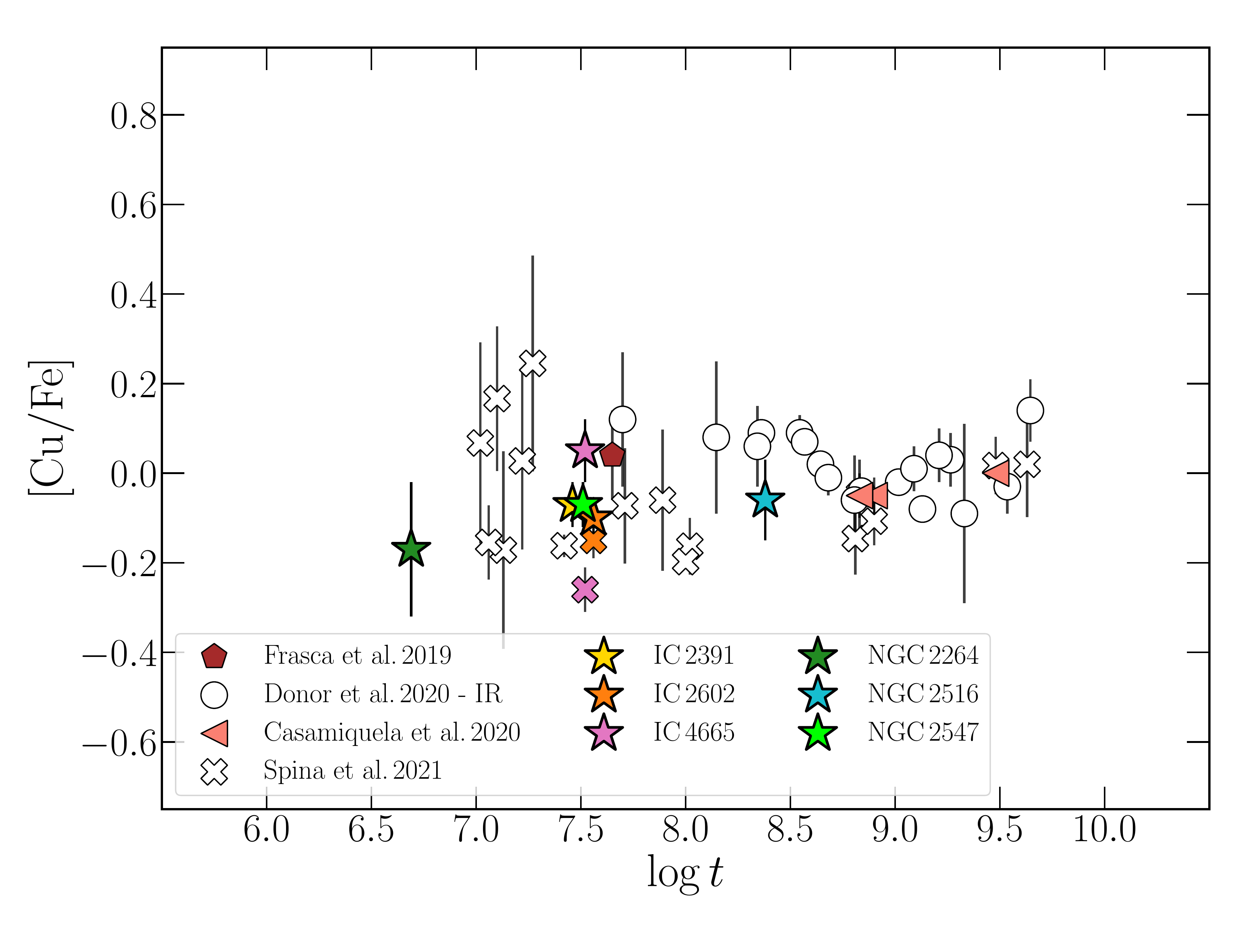}}
\caption{[Ba/Fe] (top left panel), [Y/Fe] (top right panel), [La/Fe] (bottom left panel), and [Cu/Fe] (bottom-right panel) as a function of the age of Galactic open clusters with 7.5$<$R$_{\rm{GC}<}$9\,kpc. The ages are from \cite{2020cantat-gaudin}: the typical uncertanties for $\log t$ young clusters is 0.15-0.25, while for old OCs it is 0.1-0.2. The stars represent the estimates derived in this work. The other symbols represent estimates from the literature:  red circles  from \cite{2009dorazi}; empty diamonds    from \cite{2011maiorca}; purple crosses   from \cite{2012dorazi}; inversed triangles are   \cite{2013desilva}; grey squares   from \cite{2015reddy}; the black diamond   from \cite{2017biazzo}; blue triangles   from \cite{2018magrini}; brown pentagon   from \cite{2019frasca}; light pink triangles are from \cite{2020casamiquela}; empty circles are from \cite{2020donor}, but only the measurements for those clusters with reliable membership determination (q$\_$flag=1,2); empty x-crosses   from \cite{2021spina}. } 
\label{Ba_Y_age_relation}
\end{figure*}

\subsection{The Galactic chemical evolution of $s$-process elements at young ages}\label{sec:evolution}

\begin{figure*}[!]
\centering
      \subfloat{\includegraphics[width=0.48\textwidth]{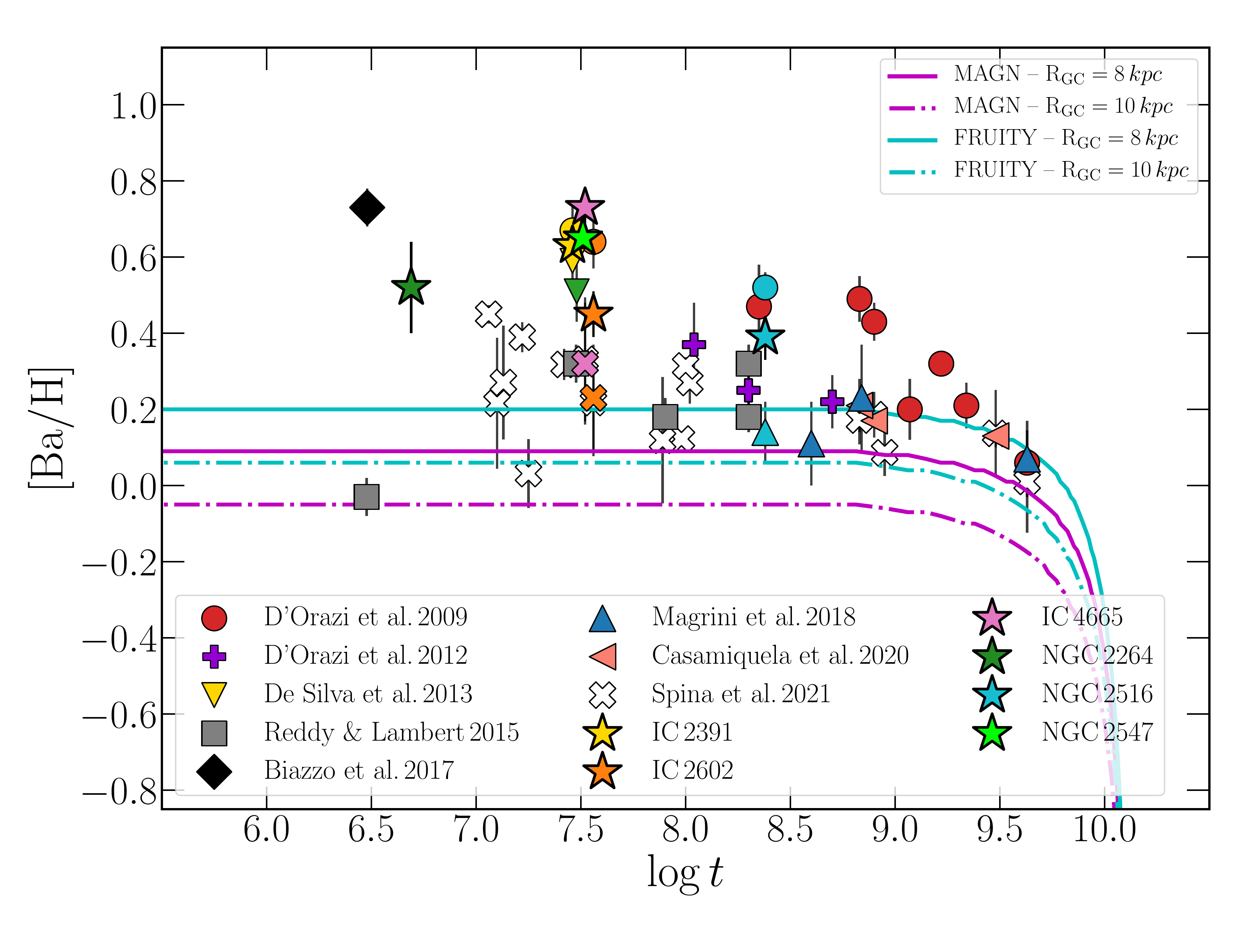}}
      \qquad
      \subfloat{\includegraphics[width=0.48\textwidth]{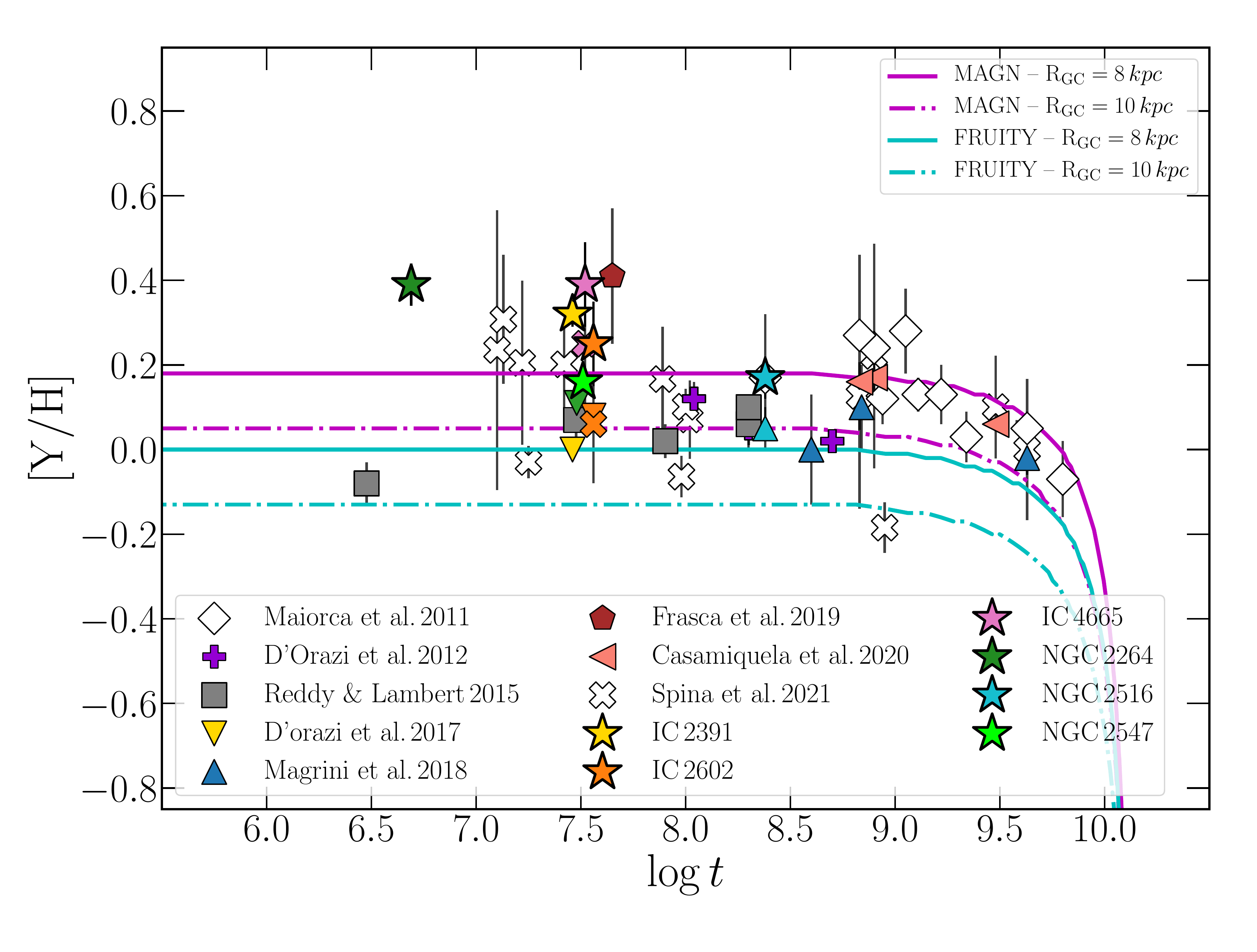}}
      \qquad
      \subfloat{\includegraphics[width=0.48\textwidth]{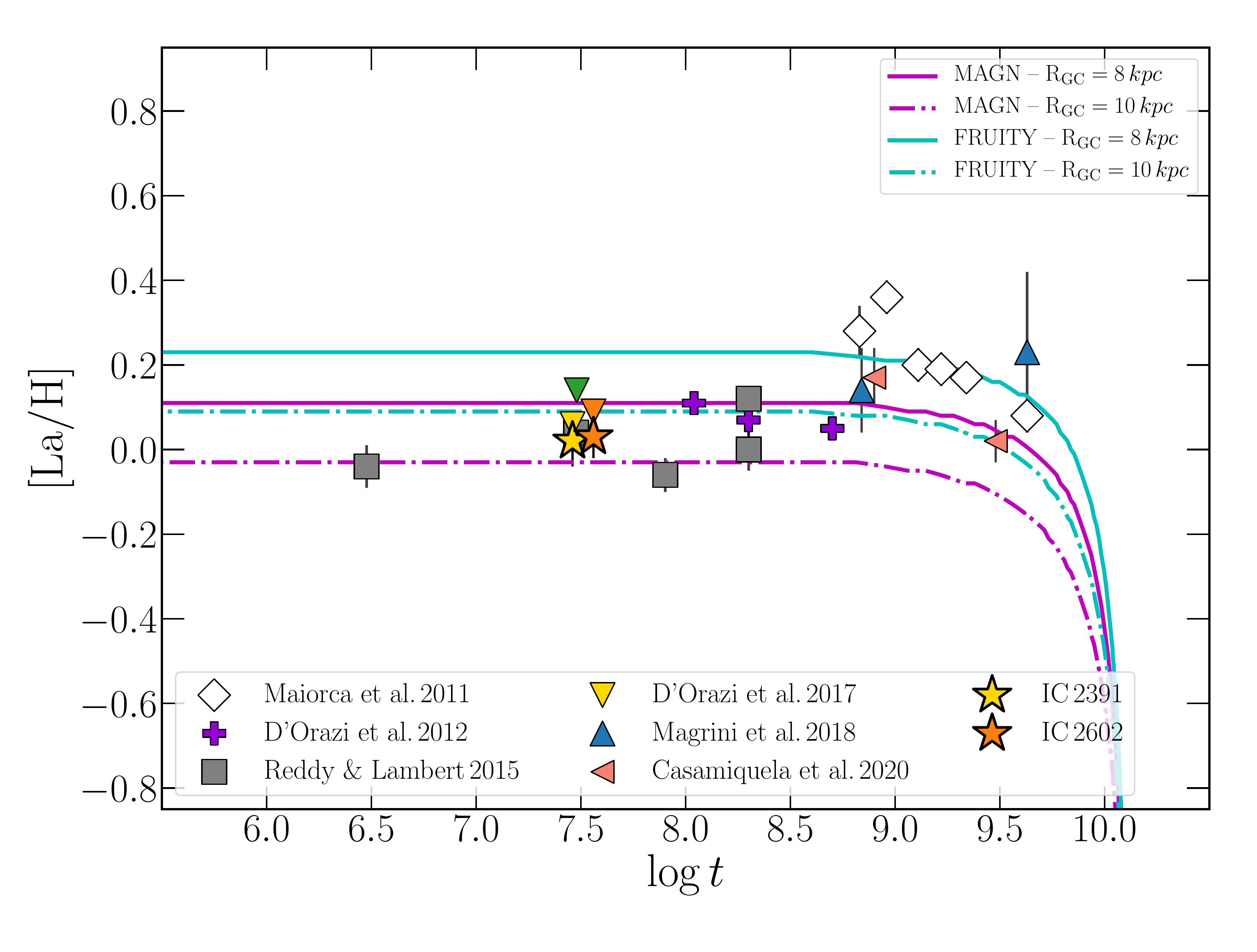}}
      \qquad
      \subfloat{\includegraphics[width=0.48\textwidth]{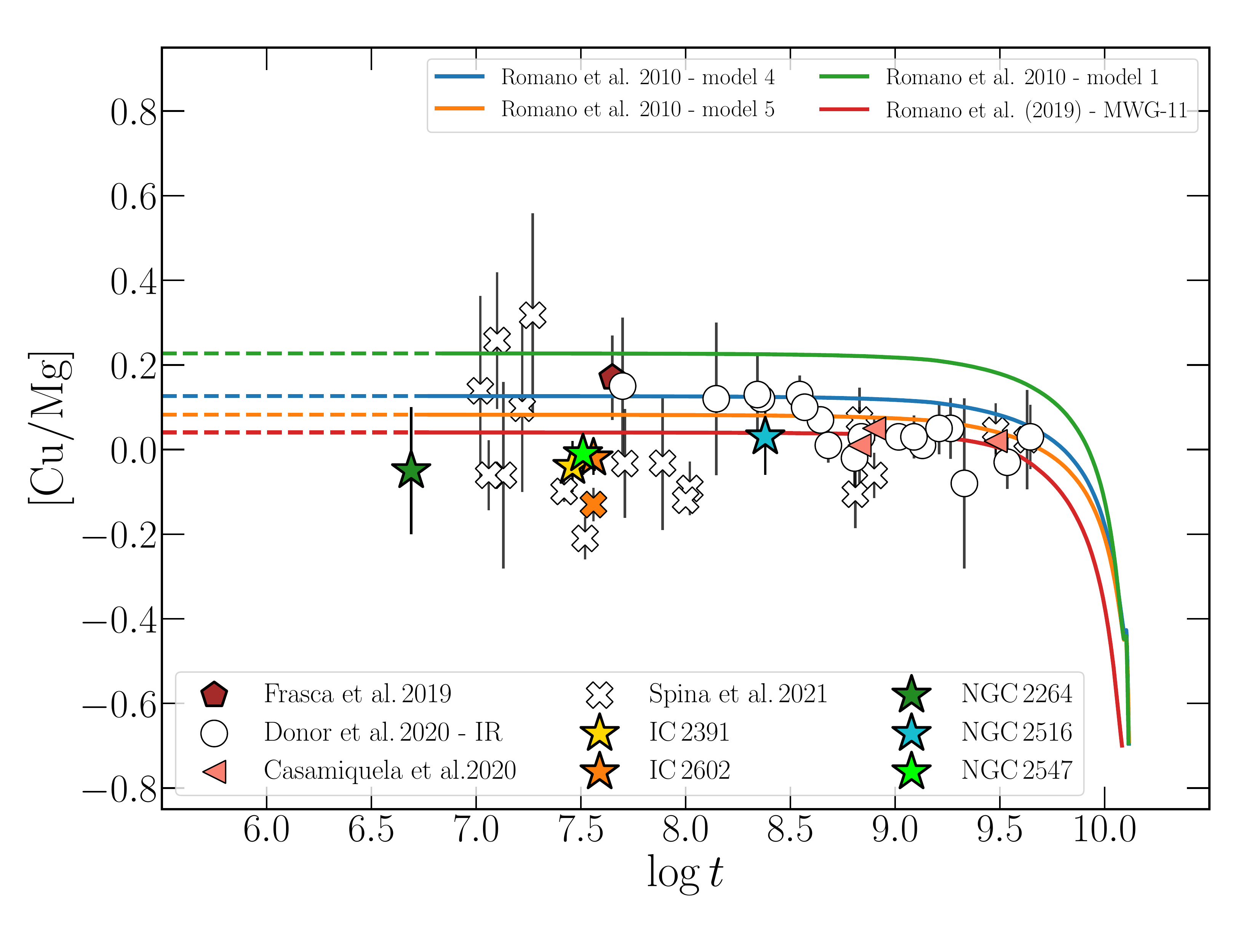}}
      \qquad
      \subfloat{\includegraphics[width=0.48\textwidth]{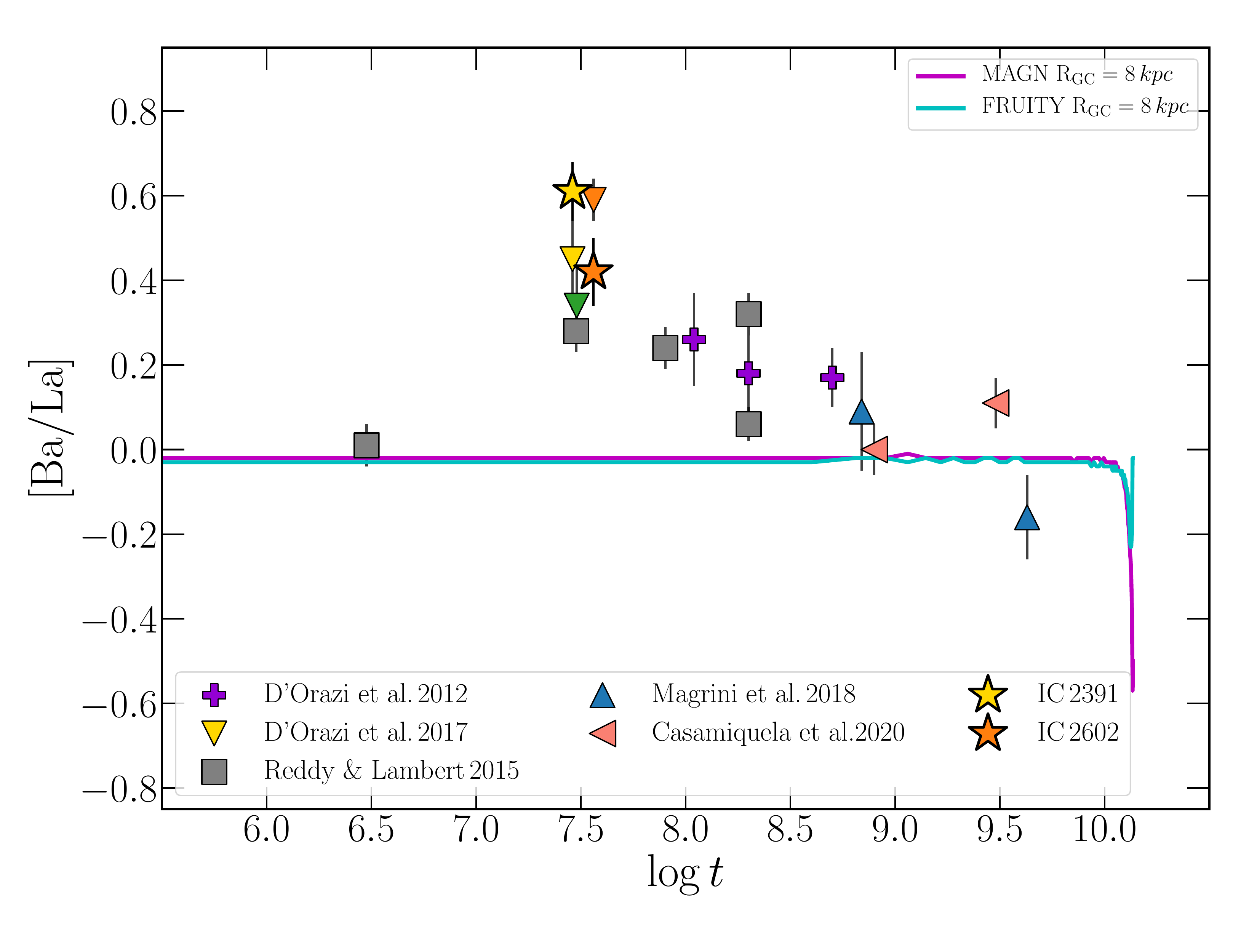}}
      \qquad
      \subfloat{\includegraphics[width=0.48\textwidth]{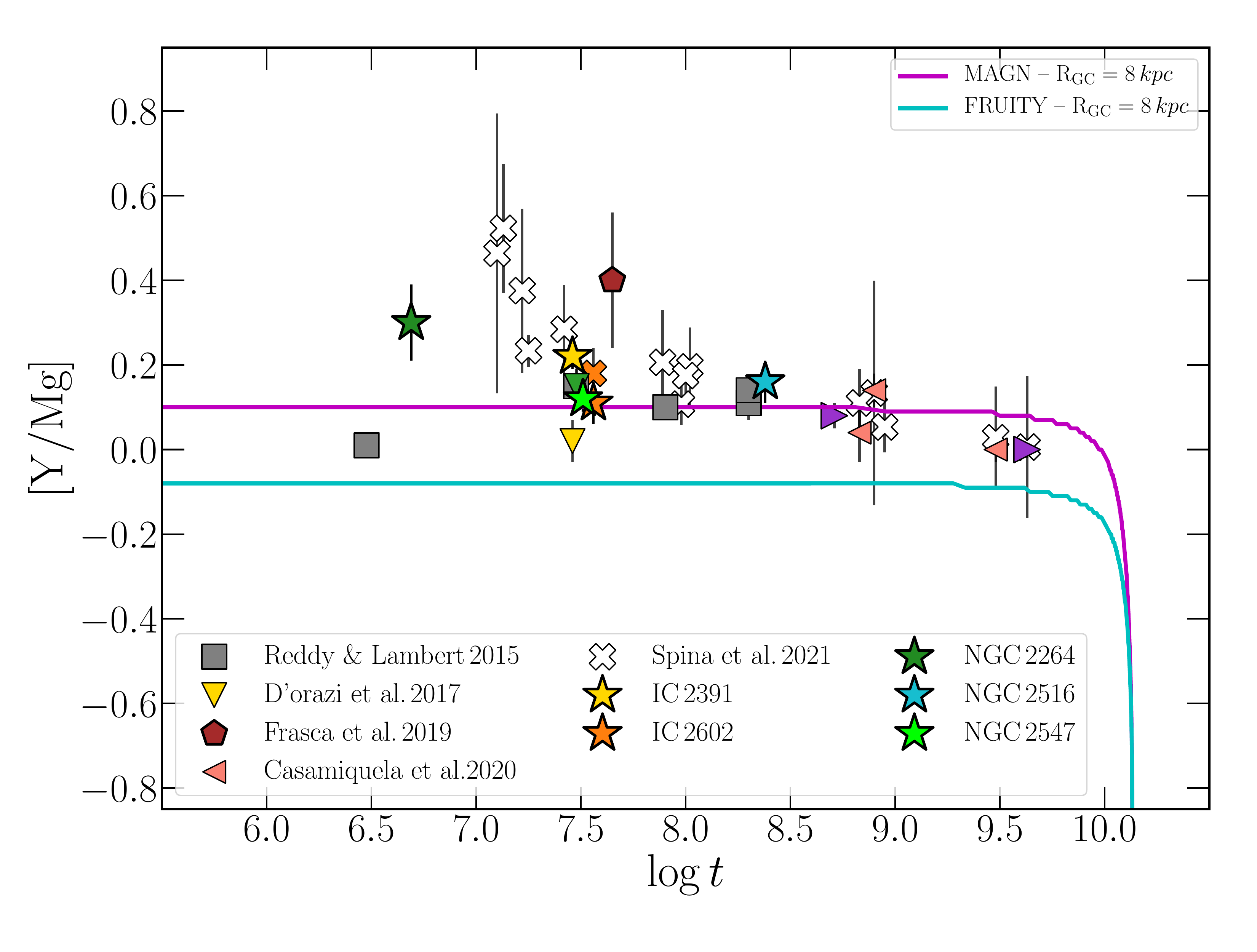}}
\caption{[Ba/H] (top left panel), [Y/H] (top right panel), [La/H] (centre left panel), [Cu/Mg] (centre right panel), [Ba/La] (bottom left panel), and [Y/Mg] (bottom right panel) as a function of the age of Galactic open clusters with 7.5$<$R$_{\rm{GC}<}$9\,kpc. The cluster symbols are the same as in Fig.\ref{Ba_Y_age_relation}. The cyan and magenta lines are the GCE models described in \cite{2021magrini} with the MAGN stellar yields (\citealt{2020vescovi}; continuous curves) and FRUITY (\citealt{2009cristallo}; dot-dashed curves). In the [Cu/Mg] vs $\log t$ the models from \cite{2010romano} with different stellar yields are considered: model 1 (green curve) with the \cite{1995woolsey} yields, model 5 with yields from \cite{2006kobayashi} with (orange line) or model 4 without (blue line) hypernovae contribution, and  \cite{2019romano} (red line) with yields from \cite{limongi2018}. See the text for further details. }
\label{elements_age_models}
\end{figure*}

In Fig.\,\ref{Ba_Y_age_relation}  we plot the Ba, Y, La, and Cu  abundance ratios as a function of the age of the Galactic OCs taken from different sources in the literature. For all the clusters in the different samples we considered the ages from \cite{2020cantat-gaudin}, who report that the uncertainties in $\log t$ for young clusters ranges from 0.15 to 0.25, while for old clusters from 0.1 to 0.2.  

The [Ba/Fe] time evolution, with increasing values at decreasing ages, is  confirmed by the observational data and we can now also detect a significant scatter at young ages. In particular, for the SFR \object{NGC\,2264} (age $\approx 5$\,Myr), the Lupus region (age $\sim$3\,Myr), and the Orion subgroup Ic (age $\sim$3\,Myr) the values are [Ba/Fe]$\approx$+0.4\,dex (this study), [Ba/Fe]$\approx$0.7\,dex \citep{2017biazzo}, and [Ba/Fe]$\approx$0.1 dex \citep{2015reddy}, respectively. This large scatter between the different [Ba/Fe] at $\log t \sim 6.5$ cannot be fully explained with the adopted microturbulence values, and it certainly reflects fundamental issues in the analysis of such young stars. The solution of magnetic intensification proposed by \cite{2020spina} can only partly explain the problem: our analysis suggests that we are witnessing an additional effect. A similar rising trend, although of much smaller extent, emerges when Y is considered. Like Ba, La belongs to the second peak of the $s$-process elements. As  can be seen in the bottom left panel of Fig.\,\ref{Ba_Y_age_relation}, our measurements confirm solar values even at ages where [Ba/Fe] is extremely enhanced, as indicated in previous works \citep[e.g.][]{2012dorazi,2015reddy,2015mishenina}.  
Finally, we find in the literature only a few studies where the Cu abundance was derived for OCs, namely \cite{2019frasca}, \cite{2020casamiquela}, \cite{2020donor} (near-infrared measurements), and \cite{2021spina}. From the bottom right panel of Fig.\,\ref{Ba_Y_age_relation}, it is evident that [Cu/Fe] is solar (within the uncertainties) at all ages; we note the large scatter of the measurements, especially at younger ages.

In Fig.\,\ref{elements_age_models} we plot the individual [X/H] of Ba, Y, and La and the [Cu/Mg], [Ba/La], and [Y/Mg] ratios as a function of age, and we compare them with the predictions from the GCE models of \cite{2021magrini}. The recent production of the first-peak, Y, and second peak, La and Ba, $s$-process elements is mainly driven by the evolution of AGB stars, with lower percentages coming from massive stars in the early Galactic epochs (see e.g. \citealt{2014cescutti} for a summary of the variety of possible scenarios). In low-mass AGB stars, the neutrons necessary for the production of $s$-process elements are mainly provided by the so-called $^{13}$C pocket, which forms at the bottom of the convective envelope after each third dredge-up episode \citep{2009cristallo}. The extension of the $^{13}$C pocket plays a major role in the final production of neutron capture elements, and it can be parametrised in different ways. The GCE models adopted here consider the $s$-process yields from the FRUITY models, calculated by applying a simple exponentially decreasing profile of the convective velocities at the inner border of the convective envelope \citep{2009cristallo}, and the MAGN models, a recent revision of the FRUITY yields which include the mixing triggered by magnetic fields \citep{2020vescovi}, which could explain the behaviour of [Y/Mg] in open clusters at different Galactocentric distances \citep{2021magrini}. 

As  can be seen, the GCE models considered here cannot reconcile the time evolution of Ba with that of La.  
In Fig.\,\ref{elements_age_models} we show the [Ba/La] time evolution; as expected, the production of  Ba and La in the models is the same, thus neither of them can predict the (apparent) massive Ba production and the observed [Ba/La] rise in the last 100\,Myr. As discussed in \cite{2015mishenina}, according to any $s$-process predictions high Ba yields should always be accompanied by high La and Ce yields, due to the presence of the magic number of neutrons (82) corresponding to  these elements (e.g. \citealt{2001busso}). This is at odds with what is observed in OCs. \cite{2015mishenina} proposed that the intermediate neutron-capture ($i$-) process, which proceeds along a different path of neutron captures than the $s$-process, is an additional source of Ba. According to their analysis, a combination of $s$-, $r$-, and $i$- processes may be able to reproduce the [Ba/La]$> +0.20$ dex observed in OCs for [Eu/La] ranging from $-0.4$ and $+0.4$\,dex (see Fig.5 and Fig.6 in \citealt{2015mishenina}). We cannot confirm this analysis in more detail because we cannot measure Eu in our stellar sample since the available Eu lines are too weak to be detected in stars with mild rotations as those in our sample. 
There are still large uncertainties concerning what stellar sources can host the $i$-process, and their efficiency in producing heavy elements. In particular, to explain the Ba excess in the YOCs a site that has become relevant only in the last 200\,Myr would be needed, and that was not yet effective in contributing to the Ba abundance in the solar system. The discovery that the Ba enhancement could be explained by some observational issues (i.e. alteration of the spectral line) would help  to solve the Ba puzzle without affecting our present understanding of the nucleosynthesis of Ba and La in the Galactic disc. From our discussion presented above we cannot make this conclusion either, and therefore the Ba puzzle remain unsolved.     

Since our analysis has provided no clear answer for the excess of Ba, in order to trace the $s$-process element abundances at young ages, in particular for ages less than 500\,Myr, we
suggest   looking for other elements and investigating further La and Ce.
Theoretical models for Galactic chemical enrichment of heavy elements and theoretical GCE models including only the $s$-process  should consider Ba and Y with extreme caution. The anomalous abundances of Y can also have an impact on the use of some chemical clocks,  [Y/Mg] for example, as  age indicators for young stars. As different studies have demonstrated (see e.g. \citealt{2015nissen,2018spina,2021casamiquela} for solar twin stars), the [Y/Mg] ratio properly traces the age up to 500-700\,Myr. Unfortunately, the two studies do not consider ages as young as the clusters we have analysed in this work. Our analysis suggests that the effects of alteration of the spectral lines could affect the relation [Y/Mg] versus age, with larger impact below 100\,Myr, but probably also between 100 and 500\,Myr. This is particularly evident in the bottom right panel in Fig.\,\ref{elements_age_models}, where it is clear that the adopted GCE models cannot reproduce the increased abundance at young ages. Thus, we suggest that caution should be applied when using Y as a tracer of the $s$-process and as an age indicator below 500\,Myr (e.g. [Y/Mg] and all the other ratios based on the Y abundances).

\subsection{The time evolution of Cu}

To follow the evolution of Cu, we adopt the GCE models presented by \cite{2007romano} and \cite{2010romano,2019romano}. We have seen that most   Cu production on Galactic scales is due to the weak $s$-process acting in massive stars. This mechanism depends on the initial metallicity of the stars. The neutrons originate mainly from the reaction $^{22}$Ne($\alpha$,n)$^{25}$Mg; the large abundance of $^{22}$Ne during He-burning cores derives from the original CNO nuclei transmuted into $^{14}$N in the H-burning ashes, followed by double $\alpha$-capture on $^{14}$N. Copper produced in this way is thus a secondary element. A small primary yield of Cu, 5 to 10$\%$ of its solar abundance derives from explosive nucleosynthesis in the inner regions of core-collapse supernovae (see e.g. \citealt{1995woolsey}, \citealt{rauscher:02}, \citealt{2010pignatari}). SN\,Ia models predict a negligible production of Cu during thermonuclear explosions (e.g. \citealt{1999iwamoto}; \citealt{2005travaglio}). Low- and intermediate-mass stars produce minor quantities of Cu as well. Thus, the models adopted in this paper assume that almost all Cu comes from massive stars, with a minor contribution from SNe\,Ia. After a short early phase in which the primary contribution from explosive nucleosynthesis in core-collapse SNe dominates, the evolution of Cu is regulated by the weak $s$-process. The models shown in Fig.\ref{elements_age_models} include massive star yields from (i) \cite{1995woolsey} (green curve), (ii) \cite{2006kobayashi} with (orange curve), or (iii) without (blue curve) the contribution from hypernovae (cf. models 4 and 5 of \citealt{2010romano}), and (iv) the yields from rotating (for [Fe/H]$<-1$) and non-rotating (for [Fe/H]$>-1$) core-collapse SN progenitor provided by \cite{2018limongi} (red line, corresponding to model MWG-11 of \citealt{2019romano} to which we refer  for more details).

We note that some data in the range $\log t \sim 7.5-8$ (from this work and from \citealt{2021spina}) fall below the predictions of the GCE models by about 10-60\% (corresponding to 0.05-0.2\,dex). Considering the uncertainties on the atomic physics affecting the measured abundances from Cu or Mg (or both), all the abundance ratios could be increased by 20-30\%. Both Mg and Cu are mostly made in CCSNe, but they are produced through different nucleosynthesis processes. For instance, the present uncertainties in the $^{22}$Ne($\alpha$,n)$^{25}$Mg reaction (responsible for the Cu production) could justify a reduction of the [Cu/Mg] curve (see Fig.18 in \citealt{2016talwar}). Additional variations of the order of the discrepancy observed could derive from the limitations of the GCE simulations adopted here, which predict average trends and cannot deal with local inhomogeinities that could play a role at this level of variation. Overall, taking into account the observational uncertainties and the limitations of the models, the red line (model MWG-11 adopting the yields by \citealt{2018limongi}) is in fair agreement with the full dataset.

\section{Concluding remarks}\label{sec:conclusions}

In this work we expanded our investigation of the chemical composition of young OCs, started in \cite{2020baratellaA}, to shed light on the behaviour of the $s$-process elements at very young ages ($t \lesssim 200$\,Myr). In particular, we derived abundances of Cu\,{\sc i}, Sr\,{\sc i} and {\sc ii}, Y\,{\sc ii}, Zr\,{\sc ii}, Ba\,{\sc ii}, La\,{\sc ii,} and Ce\,{\sc ii}. For all the clusters we reported the very first determinations of [Cu/Fe]. Regarding \object{IC\,2391} and \object{IC\,2602}, which are the most studied clusters in our sample, we presented the first determinations of [Sr/Fe]. On the other hand, we presented for the first time heavy-element abundances for \object{NGC\,2264} and \object{NGC\,2547}.

Our measurements confirm the super-solar (0.25-0.70\,dex) [Ba/Fe] abundances in the youngest population, a mild enhancement of [Y/Fe] (between 0 and 0.30\,dex), and a solar-scaled abundance pattern for all the other $s$-process elements. We investigated several aspects in order to envisage possible solutions to the anomalous behaviour of the $s$-process element Ba.

From the comparison of spectral lines in the Sun and a solar analogue of $\sim30$\,Myr, we note that the lines of some elements, for example Fe\,{\sc ii}, Ca\,{\sc i}, Ba\,{\sc ii}, and Y\,{\sc ii}, are stronger in the young star than in the Sun. On the other hand, La, Sr, and relatively weak lines of other elements, like Sc\,{\sc i} or Sc\,{\sc ii}, are almost identical. Sc lines can form at different depths to the other elements for which we observe a remarkable difference. Looking at the results obtained for the GBS, which are old and quiet stars, we do not reveal any anomalous trend. It is not clear what is altering the structure of the photosphere in very young stars and how this can modify the profiles of the spectral lines. From our analysis, the situation appears rather complex: the magnetic intensification is not sufficient to fully explain  the large abundances. Both ionised and neutral lines are altered in the same way, but this alteration may vary with the optical depth, in the sense that lines at smaller \logtau (i.e. in the upper layers) are more affected than lines forming deeper in the photosphere.  

The solution proposed by \cite{2020spina} of magnetic intensification and a pure photospheric, 1D-LTE treatment involving microturbulence does not seem to be   sufficient to account for the observed pattern since in our case we are dealing with much younger stars ($t \lesssim 200$ Myr). Recently, \cite{2021senavci} analysed the young ($\sim$ 30 Myr), active, and relatively fast rotator ($\sim 17$\kms), the  solar analogue EK Draconis. They derived precise atmospheric parameters and chemical abundances, and they studied the spot distribution on the stellar surface. They found a significant overabundance of Ba ([Ba/H]=$+0.63$\,dex) and values of [Cu/H], [Sr/H], [Y/H], and [Ce/H] that are solar within the uncertainties. Following \cite{2017reddy}, they concluded that the Ba overabundance is likely due to the assumption of depth-independent microturbulence velocity, but our analysis suggests that the explanation may be far more complex.   

Overall, the anomalous behaviour of the $s$-process elements, in particular of Ba and Y with respect to La, cannot be reconciled with only nucleosynthesis models and Galactic chemical evolution predictions. NLTE effects cannot be similarly invoked since the corrections are not sufficiently large. 
Thus, we suggest that Ba should not be used as a tracer of the $s$-process elements for young stars along with Y. We instead promote the use of La or Ce as the most reliable tracer for the investigation of time evolution of the $s$-process elements (especially at recent Galactic ages). 
Possible solutions of the Ba puzzle, both from the spectra and the nucleosynthesis prospective, are still under investigation. \cite{2020masseron} reported anomalous Ba enhancements (ten times higher than the other $s$-process elements) in stars that are also anomalous in the lighter elements, specifically those rich in P \citep{2020masseron_nat}. As discussed above, this pattern cannot be reproduced by an $s$-process model because of the nuclear properties of the isotopes involved. It remains to be seen whether the Ba anomaly in these P-rich stars has any connection with the Ba anomaly in YOCs; while the P-rich stars have [Fe/H] values of roughly $-1$, and are therefore  not young, their overall puzzling nucleosynthetic pattern may represent a clue to the site of the $i$-process.

Finally, we note that Zr lines might provide us with reliable diagnostics for the first-peak elements because its lines form deeper in the photosphere than the two Y lines we have used here. However, we should also increase the number of spectroscopic observations of very young objects in the 3000-5000\,\AA, where the best La, Zr, and Ce lines are. Future  multi-object, high-resolution spectrographs in the blue wavelength domain will give fundamental contributions to this framework.

\begin{acknowledgements}
We thank the anonymous referee for very helpful comments and suggestions.
MB would like to thank Denise Piatti for enlightening discussions about this work. The authors also thank Barb\'ala Cseh for discussion on the Ba enhancements in P-rich stars.
This work is based on data products from observations made with ESO Telescopes at the La Silla Paranal Observatory under programme ID 188.B-3002. These data products have been processed by the Cambridge Astronomy Survey Unit (CASU) at the Institute of Astronomy, University of Cambridge, and by the FLAMES/UVES reduction team at INAF/Osservatorio Astrofisico di Arcetri. These data have been obtained from the Gaia-ESO Survey Data Archive, prepared and hosted by the Wide Field Astronomy Unit, Institute for Astronomy, University of Edinburgh, which is funded by the UK Science and Technology Facilities Council. AB acknowledges funding from MIUR Premiale 2016 MITiC.
F.J.E. acknowledges financial support from the Spanish MINECO/FEDER through the grant AYA2017-84089 and MDM-2017-0737 at Centro de Astrobiologia (CSIC-INTA), Unidad de Excelencia María de Maeztu, and from the European Union’s Horizon 2020 research and innovation programme under Grant Agreement no. 824064 through the ESCAPE - The European Science Cluster of Astronomy $\&$ Particle Physics ESFRI Research Infrastructures project.
T.B. acknowledges financial support by grant No. 2018-04857 from the Swedish Research Council.
M.L. acknowledges the financial support of the Hungarian National Research, Development and Innovation Office (NKFI), grant KH$\_18 130405$
MP acknowledges support of NuGrid and JINA-CEE (NSF Grant PHY-1430152), STFC (through the University of Hull’s Consolidated Grant ST/R000840/1), and ongoing access to {\tt viper}, the University of Hull High Performance Computing Facility. 
MP acknowledges the support from the "Lendulet-2014" Programme of the Hungarian Academy of Sciences (Hungary). 
MP thank the ChETEC COST Action (CA16117), supported by the European Cooperation in Science and Technology, and the US IReNA Accelnet network.
LM acknowledge the funding from MIUR Premiale 2016: MITiC and the funding from the INAF PRIN-SKA 2017 program 1.05.01.88.04,  the COST Action CA18104: MW-Gaia. MVdS and  LM  thank the WEAVE-Italia consortium.
\end{acknowledgements}

\bibliographystyle{aa} 
\bibliography{baratella} 

\begin{thebibliography}{141}
\expandafter\ifx\csname natexlab\endcsname\relax\def\natexlab#1{#1}\fi

\bibitem[{{Aleo} {et~al.}(2017){Aleo}, {Sobotka}, \& {Ram{\'\i}rez}}]{2017aleo}
{Aleo}, P.~D., {Sobotka}, A.~C., \& {Ram{\'\i}rez}, I. 2017, \apj, 846, 24

\bibitem[{{Arcones} \& {Montes}(2011)}]{arcones:11}
{Arcones}, A. \& {Montes}, F. 2011, \apj, 731, 5

\bibitem[{{Asplund} {et~al.}(2009){Asplund}, {Grevesse}, {Sauval}, \&
  {Scott}}]{2009asplund}
{Asplund}, M., {Grevesse}, N., {Sauval}, A.~J., \& {Scott}, P. 2009, \araa, 47,
  481

\bibitem[{{Banerjee} {et~al.}(2018){Banerjee}, {Qian}, \&
  {Heger}}]{banerjee:18}
{Banerjee}, P., {Qian}, Y.-Z., \& {Heger}, A. 2018, \apj, 865, 120

\bibitem[{{Baratella} {et~al.}(2020{\natexlab{a}}){Baratella}, {D'Orazi},
  {Biazzo}, {Desidera}, {Gratton}, {Benatti}, {Bignamini}, {Carleo}, {Cecconi},
  {Claudi}, {Cosentino}, {Ghedina}, {Harutyunyan}, {Lanza}, {Malavolta},
  {Maldonado}, {Mallonn}, {Messina}, {Micela}, {Molinari}, {Poretti},
  {Scandariato}, \& {Sozzetti}}]{2020baratellaB}
{Baratella}, M., {D'Orazi}, V., {Biazzo}, K., {et~al.} 2020{\natexlab{a}},
  \aap, 640, A123

\bibitem[{{Baratella} {et~al.}(2020{\natexlab{b}}){Baratella}, {D'Orazi},
  {Carraro}, {Desidera}, {Randich}, {Magrini}, {Adibekyan}, {Smiljanic},
  {Spina}, {Tsantaki}, {Tautvai{\v{s}}ien{\.{e}}}, {Sousa}, {Jofr{\'e}},
  {Jim{\'e}nez-Esteban}, {Delgado-Mena}, {Martell}, {Van der Swaelmen},
  {Roccatagliata}, {Gilmore}, {Alfaro}, {Bayo}, {Bensby}, {Bragaglia},
  {Franciosini}, {Gonneau}, {Heiter}, {Hourihane}, {Jeffries}, {Koposov},
  {Morbidelli}, {Prisinzano}, {Sacco}, {Sbordone}, {Worley}, {Zaggia}, \&
  {Lewis}}]{2020baratellaA}
{Baratella}, M., {D'Orazi}, V., {Carraro}, G., {et~al.} 2020{\natexlab{b}},
  \aap, 634, A34

\bibitem[{{Bensby} {et~al.}(2014){Bensby}, {Feltzing}, \& {Oey}}]{2014bensby}
{Bensby}, T., {Feltzing}, S., \& {Oey}, M.~S. 2014, \aap, 562, A71

\bibitem[{{Bergemann} {et~al.}(2012){Bergemann}, {Hansen}, {Bautista}, \&
  {Ruchti}}]{2012bergemann}
{Bergemann}, M., {Hansen}, C.~J., {Bautista}, M., \& {Ruchti}, G. 2012, \aap,
  546, A90

\bibitem[{{Bertolli} {et~al.}(2013){Bertolli}, {Herwig}, {Pignatari}, \&
  {Kawano}}]{2013bertolli}
{Bertolli}, M.~G., {Herwig}, F., {Pignatari}, M., \& {Kawano}, T. 2013, arXiv
  e-prints, arXiv:1310.4578

\bibitem[{{Biazzo} {et~al.}(2017){Biazzo}, {Frasca}, {Alcal{\'a}}, {Zusi},
  {Covino}, {Randich}, {Esposito}, {Manara}, {Antoniucci}, {Nisini},
  {Rigliaco}, \& {Getman}}]{2017biazzo}
{Biazzo}, K., {Frasca}, A., {Alcal{\'a}}, J.~M., {et~al.} 2017, \aap, 605, A66

\bibitem[{{Biazzo} {et~al.}(2011){Biazzo}, {Randich}, \& {Palla}}]{2011biazzoA}
{Biazzo}, K., {Randich}, S., \& {Palla}, F. 2011, \aap, 525, A35

\bibitem[{{Bisterzo} {et~al.}(2005){Bisterzo}, {Pompeia}, {Gallino},
  {Pignatari}, {Cunha}, {Heger}, \& {Smith}}]{bisterzo:05}
{Bisterzo}, S., {Pompeia}, L., {Gallino}, R., {et~al.} 2005, \nphysa, 758, 284

\bibitem[{{Bisterzo} {et~al.}(2014){Bisterzo}, {Travaglio}, {Gallino},
  {Wiescher}, \& {K{\"a}ppeler}}]{bisterzo:14}
{Bisterzo}, S., {Travaglio}, C., {Gallino}, R., {Wiescher}, M., \&
  {K{\"a}ppeler}, F. 2014, \apj, 787, 10

\bibitem[{{Blanco-Cuaresma} {et~al.}(2014{\natexlab{a}}){Blanco-Cuaresma},
  {Soubiran}, {Heiter}, \& {Jofr{\'e}}}]{2014blanco-ispec}
{Blanco-Cuaresma}, S., {Soubiran}, C., {Heiter}, U., \& {Jofr{\'e}}, P.
  2014{\natexlab{a}}, \aap, 569, A111

\bibitem[{{Blanco-Cuaresma} {et~al.}(2014{\natexlab{b}}){Blanco-Cuaresma},
  {Soubiran}, {Jofr{\'e}}, \& {Heiter}}]{2014blancocuaresma}
{Blanco-Cuaresma}, S., {Soubiran}, C., {Jofr{\'e}}, P., \& {Heiter}, U.
  2014{\natexlab{b}}, \aap, 566, A98

\bibitem[{{Buder} {et~al.}(2021){Buder}, {Sharma}, {Kos}, {Amarsi},
  {Nordlander}, {Lind}, {Martell}, {Asplund}, {Bland-Hawthorn}, {Casey}, {de
  Silva}, {D'Orazi}, {Freeman}, {Hayden}, {Lewis}, {Lin}, {Schlesinger},
  {Simpson}, {Stello}, {Zucker}, {Zwitter}, {Beeson}, {Buck}, {Casagrande},
  {Clark}, {{\v{C}}otar}, {da Costa}, {de Grijs}, {Feuillet}, {Horner},
  {Kafle}, {Khanna}, {Kobayashi}, {Liu}, {Montet}, {Nandakumar}, {Nataf},
  {Ness}, {Spina}, {Tepper-Garc{\'\i}a}, {Ting}, {Traven},
  {Vogrin{\v{c}}i{\v{c}}}, {Wittenmyer}, {Wyse}, {{\v{Z}}erjal},
  {{\v{Z}}erjal}, \& {Galah Collaboration}}]{2021buder}
{Buder}, S., {Sharma}, S., {Kos}, J., {et~al.} 2021, \mnras, 506, 150

\bibitem[{{Busso} {et~al.}(2001){Busso}, {Gallino}, {Lambert}, {Travaglio}, \&
  {Smith}}]{2001busso}
{Busso}, M., {Gallino}, R., {Lambert}, D.~L., {Travaglio}, C., \& {Smith},
  V.~V. 2001, \apj, 557, 802

\bibitem[{{Cantat-Gaudin} {et~al.}(2020){Cantat-Gaudin}, {Anders},
  {Castro-Ginard}, {Jordi}, {Romero-G{\'o}mez}, {Soubiran}, {Casamiquela},
  {Tarricq}, {Moitinho}, {Vallenari}, {Bragaglia}, {Krone-Martins}, \&
  {Kounkel}}]{2020cantat-gaudin}
{Cantat-Gaudin}, T., {Anders}, F., {Castro-Ginard}, A., {et~al.} 2020, \aap,
  640, A1

\bibitem[{{Carrera} {et~al.}(2019){Carrera}, {Bragaglia}, {Cantat-Gaudin},
  {Vallenari}, {Balaguer-N{\'u}{\~n}ez}, {Bossini}, {Casamiquela}, {Jordi},
  {Sordo}, \& {Soubiran}}]{2019carrera}
{Carrera}, R., {Bragaglia}, A., {Cantat-Gaudin}, T., {et~al.} 2019, \aap, 623,
  A80

\bibitem[{{Casagrande} {et~al.}(2010){Casagrande}, {Ram{\'\i}rez},
  {Mel{\'e}ndez}, {Bessell}, \& {Asplund}}]{2010casagrande}
{Casagrande}, L., {Ram{\'\i}rez}, I., {Mel{\'e}ndez}, J., {Bessell}, M., \&
  {Asplund}, M. 2010, \aap, 512, A54

\bibitem[{{Casali} {et~al.}(2020){Casali}, {Spina}, {Magrini}, {Karakas},
  {Kobayashi}, {Casey}, {Feltzing}, {Van der Swaelmen}, {Tsantaki},
  {Jofr{\'e}}, {Bragaglia}, {Feuillet}, {Bensby}, {Biazzo}, {Gonneau},
  {Tautvai{\v{s}}ien{\.{e}}}, {Baratella}, {Roccatagliata}, {Pancino}, {Sousa},
  {Adibekyan}, {Martell}, {Bayo}, {Jackson}, {Jeffries}, {Gilmore}, {Randich},
  {Alfaro}, {Koposov}, {Korn}, {Recio-Blanco}, {Smiljanic}, {Franciosini},
  {Hourihane}, {Monaco}, {Morbidelli}, {Sacco}, {Worley}, \&
  {Zaggia}}]{2020casali}
{Casali}, G., {Spina}, L., {Magrini}, L., {et~al.} 2020, \aap, 639, A127

\bibitem[{{Casamiquela} {et~al.}(2019){Casamiquela}, {Blanco-Cuaresma},
  {Carrera}, {Balaguer-N{\'u}{\~n}ez}, {Jordi}, {Anders}, {Chiappini},
  {Carbajo-Hijarrubia}, {Aguado}, {del Pino}, {D{\'\i}az-P{\'e}rez}, {Gallart},
  \& {Pancino}}]{2019casamiquela}
{Casamiquela}, L., {Blanco-Cuaresma}, S., {Carrera}, R., {et~al.} 2019, \mnras,
  490, 1821

\bibitem[{{Casamiquela} {et~al.}(2021){Casamiquela}, {Soubiran}, {Jofr{\'e}},
  {Chiappini}, {Lagarde}, {Tarricq}, {Carrera}, {Jordi},
  {Balaguer-N{\'u}{\~n}ez}, {Carbajo-Hijarrubia}, \&
  {Blanco-Cuaresma}}]{2021casamiquela}
{Casamiquela}, L., {Soubiran}, C., {Jofr{\'e}}, P., {et~al.} 2021, arXiv
  e-prints, arXiv:2103.14692

\bibitem[{{Casamiquela} {et~al.}(2020){Casamiquela}, {Tarricq}, {Soubiran},
  {Blanco-Cuaresma}, {Jofr{\'e}}, {Heiter}, \& {Tucci Maia}}]{2020casamiquela}
{Casamiquela}, L., {Tarricq}, Y., {Soubiran}, C., {et~al.} 2020, \aap, 635, A8

\bibitem[{{Cescutti} \& {Chiappini}(2014)}]{2014cescutti}
{Cescutti}, G. \& {Chiappini}, C. 2014, \aap, 565, A51

\bibitem[{{Choplin} {et~al.}(2021){Choplin}, {Siess}, \&
  {Goriely}}]{choplin:21}
{Choplin}, A., {Siess}, L., \& {Goriely}, S. 2021, \aap, 648, A119

\bibitem[{{Claret}(2019)}]{2019claret}
{Claret}, A. 2019, Research Notes of the American Astronomical Society, 3, 17

\bibitem[{{Clarkson} {et~al.}(2018){Clarkson}, {Herwig}, \&
  {Pignatari}}]{clarkson:18}
{Clarkson}, O., {Herwig}, F., \& {Pignatari}, M. 2018, \mnras, 474, L37

\bibitem[{{C{\^o}t{\'e}} {et~al.}(2018){C{\^o}t{\'e}}, {Denissenkov}, {Herwig},
  {Ruiter}, {Ritter}, {Pignatari}, \& {Belczynski}}]{cote:18}
{C{\^o}t{\'e}}, B., {Denissenkov}, P., {Herwig}, F., {et~al.} 2018, \apj, 854,
  105

\bibitem[{{Cowan} \& {Rose}(1977)}]{1977cowan}
{Cowan}, J.~J. \& {Rose}, W.~K. 1977, \apj, 212, 149

\bibitem[{{Cristallo} {et~al.}(2016){Cristallo}, {Karinkuzhi}, {Goswami},
  {Piersanti}, \& {Gobrecht}}]{cristallo:16}
{Cristallo}, S., {Karinkuzhi}, D., {Goswami}, A., {Piersanti}, L., \&
  {Gobrecht}, D. 2016, \apj, 833, 181

\bibitem[{{Cristallo} {et~al.}(2009){Cristallo}, {Straniero}, {Gallino},
  {Piersanti}, {Dom{\'\i}nguez}, \& {Lederer}}]{2009cristallo}
{Cristallo}, S., {Straniero}, O., {Gallino}, R., {et~al.} 2009, \apj, 696, 797

\bibitem[{{{\c{S}}enavc{\i}} {et~al.}(2021){{\c{S}}enavc{\i}},
  {K{\i}l{\i}{\c{c}}o{\u{g}}lu}, {I{\c{s}}{\i}k}, {Hussain}, {Montes}, {Bahar},
  \& {Solanki}}]{2021senavci}
{{\c{S}}enavc{\i}}, H.~V., {K{\i}l{\i}{\c{c}}o{\u{g}}lu}, T., {I{\c{s}}{\i}k},
  E., {et~al.} 2021, \mnras, 502, 3343

\bibitem[{{Cunha} {et~al.}(2016){Cunha}, {Frinchaboy}, {Souto}, {Thompson},
  {Zasowski}, {Allende Prieto}, {Carrera}, {Chiappini}, {Donor},
  {Garc{\'\i}a-Hern{\'a}ndez}, {Garc{\'\i}a P{\'e}rez}, {Hayden}, {Holtzman},
  {Jackson}, {Johnson}, {Majewski}, {M{\'e}sz{\'a}ros}, {Meyer}, {Nidever},
  {O'Connell}, {Schiavon}, {Schultheis}, {Shetrone}, {Simmons}, {Smith}, \& {et
  al.}}]{2016cunha}
{Cunha}, K., {Frinchaboy}, P.~M., {Souto}, D., {et~al.} 2016, Astronomische
  Nachrichten, 337, 922

\bibitem[{{Cutri} {et~al.}(2003){Cutri}, {Skrutskie}, {van Dyk}, {Beichman},
  {Carpenter}, {Chester}, {Cambresy}, {Evans}, {Fowler}, {Gizis}, {Howard},
  {Huchra}, {Jarrett}, {Kopan}, {Kirkpatrick}, {Light}, {Marsh}, {McCallon},
  {Schneider}, {Stiening}, {Sykes}, {Weinberg}, {Wheaton}, {Wheelock}, \&
  {Zacarias}}]{2003cutri}
{Cutri}, R.~M., {Skrutskie}, M.~F., {van Dyk}, S., {et~al.} 2003, {2MASS All
  Sky Catalog of point sources.}

\bibitem[{{De Silva} {et~al.}(2013){De Silva}, {D'Orazi}, {Melo}, {Torres},
  {Gieles}, {Quast}, \& {Sterzik}}]{2013desilva}
{De Silva}, G.~M., {D'Orazi}, V., {Melo}, C., {et~al.} 2013, \mnras, 431, 1005

\bibitem[{{Denissenkov} {et~al.}(2017){Denissenkov}, {Herwig}, {Battino},
  {Ritter}, {Pignatari}, {Jones}, \& {Paxton}}]{denissenkov:17}
{Denissenkov}, P.~A., {Herwig}, F., {Battino}, U., {et~al.} 2017, \apjl, 834,
  L10

\bibitem[{{Denissenkov} {et~al.}(2021){Denissenkov}, {Herwig}, {Perdikakis}, \&
  {Schatz}}]{denissenkov:21}
{Denissenkov}, P.~A., {Herwig}, F., {Perdikakis}, G., \& {Schatz}, H. 2021,
  \mnras [\eprint[arXiv]{2010.15798}]

\bibitem[{{Denissenkov} {et~al.}(2019){Denissenkov}, {Herwig}, {Woodward},
  {Andrassy}, {Pignatari}, \& {Jones}}]{denissenkov:19}
{Denissenkov}, P.~A., {Herwig}, F., {Woodward}, P., {et~al.} 2019, \mnras, 488,
  4258

\bibitem[{{Donor} {et~al.}(2020){Donor}, {Frinchaboy}, {Cunha}, {O'Connell},
  {Allende Prieto}, {Almeida}, {Anders}, {Beaton}, {Bizyaev}, {Brownstein},
  {Carrera}, {Chiappini}, {Cohen}, {Garc{\'\i}a-Hern{\'a}ndez}, {Geisler},
  {Hasselquist}, {J{\"o}nsson}, {Lane}, {Majewski}, {Minniti}, {Bidin}, {Pan},
  {Roman-Lopes}, {Sobeck}, \& {Zasowski}}]{2020donor}
{Donor}, J., {Frinchaboy}, P.~M., {Cunha}, K., {et~al.} 2020, \aj, 159, 199

\bibitem[{{Donor} {et~al.}(2018){Donor}, {Frinchaboy}, {Cunha}, {Thompson},
  {O'Connell}, {Zasowski}, {Jackson}, {Meyer McGrath}, {Almeida}, {Bizyaev},
  {Carrera}, {Garc{\'\i}a-Hern{\'a}ndez}, {Nitschelm}, {Pan}, \&
  {Zamora}}]{2018donor}
{Donor}, J., {Frinchaboy}, P.~M., {Cunha}, K., {et~al.} 2018, \aj, 156, 142

\bibitem[{{D'Orazi} {et~al.}(2012){D'Orazi}, {Biazzo}, {Desidera}, {Covino},
  {Andrievsky}, \& {Gratton}}]{2012dorazi}
{D'Orazi}, V., {Biazzo}, K., {Desidera}, S., {et~al.} 2012, \mnras, 423, 2789

\bibitem[{{D'Orazi} {et~al.}(2017){D'Orazi}, {De Silva}, \&
  {Melo}}]{2017dorazi}
{D'Orazi}, V., {De Silva}, G.~M., \& {Melo}, C.~F.~H. 2017, \aap, 598, A86

\bibitem[{{D'Orazi} {et~al.}(2009){D'Orazi}, {Magrini}, {Randich}, {Galli},
  {Busso}, \& {Sestito}}]{2009dorazi}
{D'Orazi}, V., {Magrini}, L., {Randich}, S., {et~al.} 2009, \apjl, 693, L31

\bibitem[{{Farouqi} {et~al.}(2009){Farouqi}, {Kratz}, {Mashonkina}, {Pfeiffer},
  {Cowan}, {Thielemann}, \& {Truran}}]{farouqi:09}
{Farouqi}, K., {Kratz}, K.~L., {Mashonkina}, L.~I., {et~al.} 2009, \apjl, 694,
  L49

\bibitem[{{Fenner} {et~al.}(2003){Fenner}, {Gibson}, {Lee}, {Karakas},
  {Lattanzio}, {Chieffi}, {Limongi}, \& {Yong}}]{2003fenner}
{Fenner}, Y., {Gibson}, B.~K., {Lee}, H.~c., {et~al.} 2003, \pasa, 20, 340

\bibitem[{{Frasca} {et~al.}(2019){Frasca}, {Alonso-Santiago}, {Catanzaro},
  {Bragaglia}, {Carretta}, {Casali}, {D'Orazi}, {Magrini}, {Andreuzzi},
  {Oliva}, {Origlia}, {Sordo}, \& {Vallenari}}]{2019frasca}
{Frasca}, A., {Alonso-Santiago}, J., {Catanzaro}, G., {et~al.} 2019, \aap, 632,
  A16

\bibitem[{{Frasca} {et~al.}(2015){Frasca}, {Biazzo}, {Lanzafame}, {Alcal{\'a}},
  {Brugaletta}, {Klutsch}, {Stelzer}, {Sacco}, {Spina}, {Jeffries}, {Montes},
  {Alfaro}, {Barentsen}, {Bonito}, {Gameiro}, {L{\'o}pez-Santiago}, {Pace},
  {Pasquini}, {Prisinzano}, {Sousa}, {Gilmore}, {Randich}, {Micela},
  {Bragaglia}, {Flaccomio}, {Bayo}, {Costado}, {Franciosini}, {Hill},
  {Hourihane}, {Jofr{\'e}}, {Lardo}, {Maiorca}, {Masseron}, {Morbidelli}, \&
  {Worley}}]{2015frasca}
{Frasca}, A., {Biazzo}, K., {Lanzafame}, A.~C., {et~al.} 2015, \aap, 575, A4

\bibitem[{{Gadun} \& {Sheminova}(1988)}]{1988gadun}
{Gadun}, A.~S. \& {Sheminova}, V.~A. 1988, Preprint of the Institute for
  Theoretical Physics of Academy of Sciences of USSR, 87, 3

\bibitem[{{Gallagher} {et~al.}(2020){Gallagher}, {Bergemann}, {Collet}, {Plez},
  {Leenaarts}, {Carlsson}, {Yakovleva}, \& {Belyaev}}]{2020gallagher}
{Gallagher}, A.~J., {Bergemann}, M., {Collet}, R., {et~al.} 2020, \aap, 634,
  A55

\bibitem[{{Gallino} {et~al.}(1998){Gallino}, {Arlandini}, {Busso}, {Lugaro},
  {Travaglio}, {Straniero}, {Chieffi}, \& {Limongi}}]{1998gallino}
{Gallino}, R., {Arlandini}, C., {Busso}, M., {et~al.} 1998, \apj, 497, 388

\bibitem[{{Gilmore} {et~al.}(2012){Gilmore}, {Randich}, {Asplund}, {Binney},
  {Bonifacio}, {Drew}, {Feltzing}, {Ferguson}, {Jeffries}, {Micela},
  {Negueruela}, {Prusti}, {Rix}, {Vallenari}, {Alfaro}, {Allende-Prieto},
  {Babusiaux}, {Bensby}, {Blomme}, {Bragaglia}, {Flaccomio}, {Fran{\c{c}}ois},
  {Irwin}, {Koposov}, {Korn}, {Lanzafame}, {Pancino}, {Paunzen},
  {Recio-Blanco}, {Sacco}, {Smiljanic}, {Van Eck}, {Walton}, {Aden}, {Aerts},
  {Affer}, {Alcala}, {Altavilla}, {Alves}, {Antoja}, {Arenou}, {Argiroffi},
  {Asensio Ramos}, {Bailer-Jones}, {Balaguer-Nunez}, {Bayo}, {Barbuy},
  {Barisevicius}, {Barrado y Navascues}, {Battistini}, {Bellas Velidis},
  {Bellazzini}, {Belokurov}, {Bergemann}, {Bertelli}, {Biazzo}, {Bienayme},
  {Bland-Hawthorn}, {Boeche}, {Bonito}, {Boudreault}, {Bouvier}, {Brandao},
  {Brown}, {de Bruijne}, {Burleigh}, {Caballero}, {Caffau}, {Calura},
  {Capuzzo-Dolcetta}, {Caramazza}, {Carraro}, {Casagrande}, {Casewell},
  {Chapman}, {Chiappini}, {Chorniy}, {Christlieb}, {Cignoni}, {Cocozza},
  {Colless}, {Collet}, {Collins}, {Correnti}, {Covino}, {Crnojevic}, {Cropper},
  {Cunha}, {Damiani}, {David}, {Delgado}, {Duffau}, {Edvardsson}, {Eldridge},
  {Enke}, {Eriksson}, {Evans}, {Eyer}, {Famaey}, {Fellhauer}, {Ferreras},
  {Figueras}, {Fiorentino}, {Flynn}, {Folha}, {Franciosini}, {Frasca},
  {Freeman}, {Fremat}, {Friel}, {Gaensicke}, {Gameiro}, {Garzon}, {Geier},
  {Geisler}, {Gerhard}, {Gibson}, {Gomboc}, {Gomez}, {Gonzalez-Fernandez},
  {Gonzalez Hernandez}, {Gosset}, {Grebel}, {Greimel}, {Groenewegen},
  {Grundahl}, {Guarcello}, {Gustafsson}, {Hadrava}, {Hatzidimitriou}, {Hambly},
  {Hammersley}, {Hansen}, {Haywood}, {Heber}, {Heiter}, {Held}, {Helmi},
  {Hensler}, {Herrero}, {Hill}, {Hodgkin}, {Huelamo}, {Huxor}, {Ibata},
  {Jackson}, {de Jong}, {Jonker}, {Jordan}, {Jordi}, {Jorissen}, {Katz},
  {Kawata}, {Keller}, {Kharchenko}, {Klement}, {Klutsch}, {Knude}, {Koch},
  {Kochukhov}, {Kontizas}, {Koubsky}, {Lallement}, {de Laverny}, {van Leeuwen},
  {Lemasle}, {Lewis}, {Lind}, {Lindstrom}, {Lobel}, {Lopez Santiago}, {Lucas},
  {Ludwig}, {Lueftinger}, {Magrini}, {Maiz Apellaniz}, {Maldonado}, {Marconi},
  {Marino}, {Martayan}, {Martinez-Valpuesta}, {Matijevic}, {McMahon},
  {Messina}, {Meyer}, {Miglio}, {Mikolaitis}, {Minchev}, {Minniti}, {Moitinho},
  {Momany}, {Monaco}, {Montalto}, {Monteiro}, {Monier}, {Montes}, {Mora},
  {Moraux}, {Morel}, {Mowlavi}, {Mucciarelli}, {Munari}, {Napiwotzki},
  {Nardetto}, {Naylor}, {Naze}, {Nelemans}, {Okamoto}, {Ortolani}, {Pace},
  {Palla}, {Palous}, {Parker}, {Penarrubia}, {Pillitteri}, {Piotto}, {Posbic},
  {Prisinzano}, {Puzeras}, {Quirrenbach}, {Ragaini}, {Read}, {Read}, {Reyle},
  {De Ridder}, {Robichon}, {Robin}, {Roeser}, {Romano}, {Royer}, {Ruchti},
  {Ruzicka}, {Ryan}, {Ryde}, {Santos}, {Sanz Forcada}, {Sarro Baro},
  {Sbordone}, {Schilbach}, {Schmeja}, {Schnurr}, {Schoenrich}, {Scholz},
  {Seabroke}, {Sharma}, {De Silva}, {Smith}, {Solano}, {Sordo}, {Soubiran},
  {Sousa}, {Spagna}, {Steffen}, {Steinmetz}, {Stelzer}, {Stempels},
  {Tabernero}, {Tautvaisiene}, {Thevenin}, {Torra}, {Tosi}, {Tolstoy}, {Turon},
  {Walker}, {Wambsganss}, {Worley}, {Venn}, {Vink}, {Wyse}, {Zaggia},
  {Zeilinger}, {Zoccali}, {Zorec}, {Zucker}, {Zwitter}, \& {Gaia-ESO Survey
  Team}}]{ges}
{Gilmore}, G., {Randich}, S., {Asplund}, M., {et~al.} 2012, The Messenger, 147,
  25

\bibitem[{{Gray}(1992)}]{1992gray}
{Gray}, D.~F. 1992, {The observation and analysis of stellar photospheres.},
  Vol.~20

\bibitem[{{Grevesse} {et~al.}(2015){Grevesse}, {Scott}, {Asplund}, \&
  {Sauval}}]{2015grevesse}
{Grevesse}, N., {Scott}, P., {Asplund}, M., \& {Sauval}, A.~J. 2015, \aap, 573,
  A27

\bibitem[{{Gurtovenko} {et~al.}(1974){Gurtovenko}, {Ratnikova}, \& {de
  Jager}}]{1974gurtovenko}
{Gurtovenko}, E., {Ratnikova}, V., \& {de Jager}, C. 1974, \solphys, 37, 43

\bibitem[{{Gurtovenko} \& {Sheminova}(2015)}]{2015gurtovenko}
{Gurtovenko}, E.~A. \& {Sheminova}, V.~A. 2015, arXiv e-prints,
  arXiv:1505.00975

\bibitem[{{Gustafsson} {et~al.}(2008){Gustafsson}, {Edvardsson}, {Eriksson},
  {J{\o}rgensen}, {Nordlund}, \& {Plez}}]{marcs}
{Gustafsson}, B., {Edvardsson}, B., {Eriksson}, K., {et~al.} 2008, \aap, 486,
  951

\bibitem[{{Hannaford} {et~al.}(1982){Hannaford}, {Lowe}, {Grevesse}, {Biemont},
  \& {Whaling}}]{1982hannaford}
{Hannaford}, P., {Lowe}, R.~M., {Grevesse}, N., {Biemont}, E., \& {Whaling}, W.
  1982, \apj, 261, 736

\bibitem[{{Heil} {et~al.}(2008){Heil}, {K{\"a}ppeler}, {Uberseder}, {Gallino},
  \& {Pignatari}}]{heil:08}
{Heil}, M., {K{\"a}ppeler}, F., {Uberseder}, E., {Gallino}, R., \& {Pignatari},
  M. 2008, \prc, 77, 015808

\bibitem[{{Heiter} {et~al.}(2020){Heiter}, {Lind}, {Bergemann}, {Asplund},
  {Mikolaitis}, {Barklem}, {Masseron}, {de Laverny}, {Magrini}, {Edvardsson},
  {J{\"o}nsson}, {Pickering}, {Ryde}, {Bayo Ar{\'a}n}, {Bensby}, {Casey},
  {Feltzing}, {Jofr{\'e}}, {Korn}, {Pancino}, {Damiani}, {Lanzafame}, {Lardo},
  {Monaco}, {Morbidelli}, {Smiljanic}, {Worley}, {Zaggia}, {Randich}, \&
  {Gilmore}}]{geslinelist}
{Heiter}, U., {Lind}, K., {Bergemann}, M., {et~al.} 2020, arXiv e-prints,
  arXiv:2011.02049

\bibitem[{{Herwig} {et~al.}(2011){Herwig}, {Pignatari}, {Woodward}, {Porter},
  {Rockefeller}, {Fryer}, {Bennett}, \& {Hirschi}}]{herwig:11}
{Herwig}, F., {Pignatari}, M., {Woodward}, P.~R., {et~al.} 2011, \apj, 727, 89

\bibitem[{{Iwamoto} {et~al.}(1999){Iwamoto}, {Brachwitz}, {Nomoto},
  {Kishimoto}, {Umeda}, {Hix}, \& {Thielemann}}]{1999iwamoto}
{Iwamoto}, K., {Brachwitz}, F., {Nomoto}, K., {et~al.} 1999, \apjs, 125, 439

\bibitem[{{Jacobson} \& {Friel}(2013)}]{2013jacobson}
{Jacobson}, H.~R. \& {Friel}, E.~D. 2013, \aj, 145, 107

\bibitem[{{James} {et~al.}(2006){James}, {Melo}, {Santos}, \&
  {Bouvier}}]{2006james}
{James}, D.~J., {Melo}, C., {Santos}, N.~C., \& {Bouvier}, J. 2006, \aap, 446,
  971

\bibitem[{{Jofr{\'e}} {et~al.}(2015{\natexlab{a}}){Jofr{\'e}}, {Petrucci},
  {Saffe}, {Saker}, {Artur de la Villarmois}, {Chavero}, {G{\'o}mez}, \&
  {Mauas}}]{2015jofre_emiliano}
{Jofr{\'e}}, E., {Petrucci}, R., {Saffe}, C., {et~al.} 2015{\natexlab{a}},
  \aap, 574, A50

\bibitem[{{Jofr{\'e}} {et~al.}(2015{\natexlab{b}}){Jofr{\'e}}, {Heiter},
  {Soubiran}, {Blanco-Cuaresma}, {Masseron}, {Nordlander}, {Chemin}, {Worley},
  {Van Eck}, {Hourihane}, {Gilmore}, {Adibekyan}, {Bergemann}, {Cantat-Gaudin},
  {Delgado-Mena}, {Gonz{\'a}lez Hern{\'a}ndez}, {Guiglion}, {Lardo}, {de
  Laverny}, {Lind}, {Magrini}, {Mikolaitis}, {Montes}, {Pancino},
  {Recio-Blanco}, {Sordo}, {Sousa}, {Tabernero}, \& {Vallenari}}]{2015jofre}
{Jofr{\'e}}, P., {Heiter}, U., {Soubiran}, C., {et~al.} 2015{\natexlab{b}},
  \aap, 582, A81

\bibitem[{{Jofr{\'e}} {et~al.}(2018){Jofr{\'e}}, {Heiter}, {Tucci Maia},
  {Soubiran}, {Worley}, {Hawkins}, {Blanco-Cuaresma}, \& {Rodrigo}}]{2018jofre}
{Jofr{\'e}}, P., {Heiter}, U., {Tucci Maia}, M., {et~al.} 2018, Research Notes
  of the American Astronomical Society, 2, 152

\bibitem[{{Jofr{\'e}} {et~al.}(2017){Jofr{\'e}}, {Heiter}, {Worley},
  {Blanco-Cuaresma}, {Soubiran}, {Masseron}, {Hawkins}, {Adibekyan}, {Buder},
  {Casamiquela}, {Gilmore}, {Hourihane}, \& {Tabernero}}]{2017jofre_paula}
{Jofr{\'e}}, P., {Heiter}, U., {Worley}, C.~C., {et~al.} 2017, \aap, 601, A38

\bibitem[{{Jones} {et~al.}(2019){Jones}, {C{\^o}t{\'e}}, {R{\"o}pke}, \&
  {Wanajo}}]{jones:19}
{Jones}, S., {C{\^o}t{\'e}}, B., {R{\"o}pke}, F.~K., \& {Wanajo}, S. 2019,
  \apj, 882, 170

\bibitem[{{Jones} {et~al.}(2016){Jones}, {Ritter}, {Herwig}, {Fryer},
  {Pignatari}, {Bertolli}, \& {Paxton}}]{jones:16}
{Jones}, S., {Ritter}, C., {Herwig}, F., {et~al.} 2016, \mnras, 455, 3848

\bibitem[{{K{\"a}ppeler} {et~al.}(2011){K{\"a}ppeler}, {Gallino}, {Bisterzo},
  \& {Aoki}}]{2011kappeler}
{K{\"a}ppeler}, F., {Gallino}, R., {Bisterzo}, S., \& {Aoki}, W. 2011, Reviews
  of Modern Physics, 83, 157

\bibitem[{{Karakas} \& {Lattanzio}(2014)}]{karakas2014}
{Karakas}, A.~I. \& {Lattanzio}, J.~C. 2014, \pasa, 31, e030

\bibitem[{{Kaufer} {et~al.}(1999){Kaufer}, {Stahl}, {Tubbesing},
  {N{\o}rregaard}, {Avila}, {Francois}, {Pasquini}, \& {Pizzella}}]{1999kaufer}
{Kaufer}, A., {Stahl}, O., {Tubbesing}, S., {et~al.} 1999, The Messenger, 95, 8

\bibitem[{{Kobayashi} {et~al.}(2020){Kobayashi}, {Karakas}, \&
  {Lugaro}}]{2020kobayashi}
{Kobayashi}, C., {Karakas}, A.~I., \& {Lugaro}, M. 2020, \apj, 900, 179

\bibitem[{{Kobayashi} {et~al.}(2006){Kobayashi}, {Umeda}, {Nomoto}, {Tominaga},
  \& {Ohkubo}}]{2006kobayashi}
{Kobayashi}, C., {Umeda}, H., {Nomoto}, K., {Tominaga}, N., \& {Ohkubo}, T.
  2006, \apj, 653, 1145

\bibitem[{{Korotin} {et~al.}(2015){Korotin}, {Andrievsky}, {Hansen}, {Caffau},
  {Bonifacio}, {Spite}, {Spite}, \& {Fran{\c{c}}ois}}]{2015korotin}
{Korotin}, S.~A., {Andrievsky}, S.~M., {Hansen}, C.~J., {et~al.} 2015, \aap,
  581, A70

\bibitem[{{Kurucz}(2011)}]{2011kurucz}
{Kurucz}, R.~L. 2011, Canadian Journal of Physics, 89, 417

\bibitem[{{Laming}(2015)}]{2015laming}
{Laming}, J.~M. 2015, Living Reviews in Solar Physics, 12, 2

\bibitem[{{Landi Degl'Innocenti}(1982)}]{1982deglinnocenti}
{Landi Degl'Innocenti}, E. 1982, \solphys, 79, 291

\bibitem[{{Lawler} {et~al.}(2001){Lawler}, {Bonvallet}, \& {Sneden}}]{LBS}
{Lawler}, J.~E., {Bonvallet}, G., \& {Sneden}, C. 2001, Astrophys. J., 556,
  452, (LBS)

\bibitem[{{Lawler} {et~al.}(2009){Lawler}, {Sneden}, {Cowan}, {Ivans}, \& {Den
  Hartog}}]{LSCI}
{Lawler}, J.~E., {Sneden}, C., {Cowan}, J.~J., {Ivans}, I.~I., \& {Den Hartog},
  E.~A. 2009, Astrophys. J. Suppl. Ser., 182, 51, (LSCI)

\bibitem[{{Limongi} \& {Chieffi}(2018{\natexlab{a}})}]{limongi2018}
{Limongi}, M. \& {Chieffi}, A. 2018{\natexlab{a}}, \apjs, 237, 13

\bibitem[{{Limongi} \& {Chieffi}(2018{\natexlab{b}})}]{2018limongi}
{Limongi}, M. \& {Chieffi}, A. 2018{\natexlab{b}}, \apjs, 237, 13

\bibitem[{{Ljung} {et~al.}(2006){Ljung}, {Nilsson}, {Asplund}, \&
  {Johansson}}]{LNAJ}
{Ljung}, G., {Nilsson}, H., {Asplund}, M., \& {Johansson}, S. 2006, \aap, 456,
  1181

\bibitem[{{Lodders}(2019)}]{2019lodders}
{Lodders}, K. 2019, arXiv e-prints, arXiv:1912.00844

\bibitem[{{Luck}(2018)}]{2018luck}
{Luck}, R.~E. 2018, \aj, 155, 111

\bibitem[{{Lugaro} {et~al.}(2015){Lugaro}, {Campbell}, {Van Winckel}, {De
  Smedt}, {Karakas}, \& {K{\"a}ppeler}}]{lugaro:15}
{Lugaro}, M., {Campbell}, S.~W., {Van Winckel}, H., {et~al.} 2015, \aap, 583,
  A77

\bibitem[{{Lugaro} {et~al.}(2012){Lugaro}, {Karakas}, {Stancliffe}, \&
  {Rijs}}]{2012lugaro}
{Lugaro}, M., {Karakas}, A.~I., {Stancliffe}, R.~J., \& {Rijs}, C. 2012, \apj,
  747, 2

\bibitem[{{Magrini} {et~al.}(2018){Magrini}, {Spina}, {Randich}, {Friel},
  {Kordopatis}, {Worley}, {Pancino}, {Bragaglia}, {Donati},
  {Tautvai{\v{s}}ien{\.{e}}}, {Bagdonas}, {Delgado-Mena}, {Adibekyan}, {Sousa},
  {Jim{\'e}nez-Esteban}, {Sanna}, {Roccatagliata}, {Bonito}, {Sbordone},
  {Duffau}, {Gilmore}, {Feltzing}, {Jeffries}, {Vallenari}, {Alfaro}, {Bensby},
  {Francois}, {Koposov}, {Korn}, {Recio-Blanco}, {Smiljanic}, {Bayo},
  {Carraro}, {Casey}, {Costado}, {Damiani}, {Franciosini}, {Frasca},
  {Hourihane}, {Jofr{\'e}}, {de Laverny}, {Lewis}, {Masseron}, {Monaco},
  {Morbidelli}, {Prisinzano}, {Sacco}, \& {Zaggia}}]{2018magrini}
{Magrini}, L., {Spina}, L., {Randich}, S., {et~al.} 2018, \aap, 617, A106

\bibitem[{{Magrini} {et~al.}(2021){Magrini}, {Vescovi}, {Casali}, {Cristallo},
  {Viscasillas V{\'a}zquez}, {Cescutti}, {Spina}, {Van Der Swaelmen}, \&
  {Randich}}]{2021magrini}
{Magrini}, L., {Vescovi}, D., {Casali}, G., {et~al.} 2021, \aap, 646, L2

\bibitem[{{Maiorca} {et~al.}(2011){Maiorca}, {Randich}, {Busso}, {Magrini}, \&
  {Palmerini}}]{2011maiorca}
{Maiorca}, E., {Randich}, S., {Busso}, M., {Magrini}, L., \& {Palmerini}, S.
  2011, \apj, 736, 120

\bibitem[{{Mamajek} \& {Hillenbrand}(2008)}]{2008mamajek}
{Mamajek}, E.~E. \& {Hillenbrand}, L.~A. 2008, \apj, 687, 1264

\bibitem[{{Masseron} {et~al.}(2020{\natexlab{a}}){Masseron},
  {Garc{\'\i}a-Hern{\'a}ndez}, {Santove{\~n}a}, {Manchado}, {Zamora},
  {Manteiga}, \& {Dafonte}}]{2020masseron_nat}
{Masseron}, T., {Garc{\'\i}a-Hern{\'a}ndez}, D.~A., {Santove{\~n}a}, R.,
  {et~al.} 2020{\natexlab{a}}, Nature Communications, 11, 3759

\bibitem[{{Masseron} {et~al.}(2020{\natexlab{b}}){Masseron},
  {Garc{\'\i}a-Hern{\'a}ndez}, {Zamora}, \& {Manchado}}]{2020masseron}
{Masseron}, T., {Garc{\'\i}a-Hern{\'a}ndez}, D.~A., {Zamora}, O., \&
  {Manchado}, A. 2020{\natexlab{b}}, \apjl, 904, L1

\bibitem[{{Matteucci} {et~al.}(1993){Matteucci}, {Raiteri}, {Busson},
  {Gallino}, \& {Gratton}}]{matteucci:93}
{Matteucci}, F., {Raiteri}, C.~M., {Busson}, M., {Gallino}, R., \& {Gratton},
  R. 1993, \aap, 272, 421

\bibitem[{{Mayor} {et~al.}(2003){Mayor}, {Pepe}, {Queloz}, {Bouchy},
  {Rupprecht}, {Lo Curto}, {Avila}, {Benz}, {Bertaux}, {Bonfils}, {Dall},
  {Dekker}, {Delabre}, {Eckert}, {Fleury}, {Gilliotte}, {Gojak}, {Guzman},
  {Kohler}, {Lizon}, {Longinotti}, {Lovis}, {Megevand}, {Pasquini}, {Reyes},
  {Sivan}, {Sosnowska}, {Soto}, {Udry}, {van Kesteren}, {Weber}, \&
  {Weilenmann}}]{2003mayor}
{Mayor}, M., {Pepe}, F., {Queloz}, D., {et~al.} 2003, The Messenger, 114, 20

\bibitem[{{McWilliam}(1998)}]{1998mcwilliam_Ba}
{McWilliam}, A. 1998, \aj, 115, 1640

\bibitem[{{Minchev} {et~al.}(2013){Minchev}, {Chiappini}, \&
  {Martig}}]{2013minchev}
{Minchev}, I., {Chiappini}, C., \& {Martig}, M. 2013, \aap, 558, A9

\bibitem[{{Mishenina} {et~al.}(2015){Mishenina}, {Pignatari}, {Carraro},
  {Kovtyukh}, {Monaco}, {Korotin}, {Shereta}, {Yegorova}, \&
  {Herwig}}]{2015mishenina}
{Mishenina}, T., {Pignatari}, M., {Carraro}, G., {et~al.} 2015, \mnras, 446,
  3651

\bibitem[{{Nissen}(2015)}]{2015nissen}
{Nissen}, P.~E. 2015, \aap, 579, A52

\bibitem[{{Pasquini} {et~al.}(2002){Pasquini}, {Avila}, {Blecha}, {Cacciari},
  {Cayatte}, {Colless}, {Damiani}, {de Propris}, {Dekker}, {di Marcantonio},
  {Farrell}, {Gillingham}, {Guinouard}, {Hammer}, {Kaufer}, {Hill}, {Marteaud},
  {Modigliani}, {Mulas}, {North}, {Popovic}, {Rossetti}, {Royer}, {Santin},
  {Schmutzer}, {Simond}, {Vola}, {Waller}, \& {Zoccali}}]{uves}
{Pasquini}, L., {Avila}, G., {Blecha}, A., {et~al.} 2002, The Messenger, 110, 1

\bibitem[{{Pignatari} {et~al.}(2010){Pignatari}, {Gallino}, {Heil}, {Wiescher},
  {K{\"a}ppeler}, {Herwig}, \& {Bisterzo}}]{2010pignatari}
{Pignatari}, M., {Gallino}, R., {Heil}, M., {et~al.} 2010, \apj, 710, 1557

\bibitem[{{Randich} {et~al.}(2013){Randich}, {Gilmore}, \& {Gaia-ESO
  Consortium}}]{2013randich}
{Randich}, S., {Gilmore}, G., \& {Gaia-ESO Consortium}. 2013, The Messenger,
  154, 47

\bibitem[{{Rauscher} {et~al.}(2002){Rauscher}, {Heger}, {Hoffman}, \&
  {Woosley}}]{rauscher:02}
{Rauscher}, T., {Heger}, A., {Hoffman}, R.~D., \& {Woosley}, S.~E. 2002, \apj,
  576, 323

\bibitem[{{Reddy} \& {Lambert}(2015)}]{2015reddy}
{Reddy}, A. B.~S. \& {Lambert}, D.~L. 2015, \mnras, 454, 1976

\bibitem[{{Reddy} \& {Lambert}(2017)}]{2017reddy}
{Reddy}, A. B.~S. \& {Lambert}, D.~L. 2017, \apj, 845, 151

\bibitem[{{Roberts} {et~al.}(2010){Roberts}, {Woosley}, \&
  {Hoffman}}]{roberts:10}
{Roberts}, L.~F., {Woosley}, S.~E., \& {Hoffman}, R.~D. 2010, \apj, 722, 954

\bibitem[{{Roederer} {et~al.}(2016){Roederer}, {Karakas}, {Pignatari}, \&
  {Herwig}}]{roederer:16}
{Roederer}, I.~U., {Karakas}, A.~I., {Pignatari}, M., \& {Herwig}, F. 2016,
  \apj, 821, 37

\bibitem[{{Romano} {et~al.}(2010){Romano}, {Karakas}, {Tosi}, \&
  {Matteucci}}]{2010romano}
{Romano}, D., {Karakas}, A.~I., {Tosi}, M., \& {Matteucci}, F. 2010, \aap, 522,
  A32

\bibitem[{{Romano} \& {Matteucci}(2007)}]{2007romano}
{Romano}, D. \& {Matteucci}, F. 2007, \mnras, 378, L59

\bibitem[{{Romano} {et~al.}(2019){Romano}, {Matteucci}, {Zhang}, {Ivison}, \&
  {Ventura}}]{2019romano}
{Romano}, D., {Matteucci}, F., {Zhang}, Z.-Y., {Ivison}, R.~J., \& {Ventura},
  P. 2019, \mnras, 490, 2838

\bibitem[{{Sacco} {et~al.}(2014){Sacco}, {Morbidelli}, {Franciosini},
  {Maiorca}, {Randich}, {Modigliani}, {Gilmore}, {Asplund}, {Binney},
  {Bonifacio}, {Drew}, {Feltzing}, {Ferguson}, {Jeffries}, {Micela},
  {Negueruela}, {Prusti}, {Rix}, {Vallenari}, {Alfaro}, {Allende Prieto},
  {Babusiaux}, {Bensby}, {Blomme}, {Bragaglia}, {Flaccomio}, {Francois},
  {Hambly}, {Irwin}, {Koposov}, {Korn}, {Lanzafame}, {Pancino}, {Recio-Blanco},
  {Smiljanic}, {Van Eck}, {Walton}, {Bergemann}, {Costado}, {de Laverny},
  {Heiter}, {Hill}, {Hourihane}, {Jackson}, {Jofre}, {Lewis}, {Lind}, {Lardo},
  {Magrini}, {Masseron}, {Prisinzano}, \& {Worley}}]{2014sacco}
{Sacco}, G.~G., {Morbidelli}, L., {Franciosini}, E., {et~al.} 2014, \aap, 565,
  A113

\bibitem[{{Santos} {et~al.}(2008){Santos}, {Melo}, {James}, {Gameiro},
  {Bouvier}, \& {Gomes}}]{2008santos}
{Santos}, N.~C., {Melo}, C., {James}, D.~J., {et~al.} 2008, \aap, 480, 889

\bibitem[{{Schuler} {et~al.}(2010){Schuler}, {Plunkett}, {King}, \&
  {Pinsonneault}}]{2010schuler}
{Schuler}, S.~C., {Plunkett}, A.~L., {King}, J.~R., \& {Pinsonneault}, M.~H.
  2010, \pasp, 122, 766

\bibitem[{{Scott} {et~al.}(2015){Scott}, {Asplund}, {Grevesse}, {Bergemann}, \&
  {Sauval}}]{2015scott}
{Scott}, P., {Asplund}, M., {Grevesse}, N., {Bergemann}, M., \& {Sauval}, A.~J.
  2015, \aap, 573, A26

\bibitem[{{Sheminova}(2019)}]{2019sheminova}
{Sheminova}, V.~A. 2019, Kinematics and Physics of Celestial Bodies, 35, 129

\bibitem[{{Sheminova} \& {Solanki}(1999)}]{1999Sheminova}
{Sheminova}, V.~A. \& {Solanki}, S.~K. 1999, \aap, 351, 701

\bibitem[{{Shi} {et~al.}(2014){Shi}, {Gehren}, {Zeng}, {Mashonkina}, \&
  {Zhao}}]{2014shi}
{Shi}, J.~R., {Gehren}, T., {Zeng}, J.~L., {Mashonkina}, L., \& {Zhao}, G.
  2014, \apj, 782, 80

\bibitem[{{Sneden} {et~al.}(2008){Sneden}, {Cowan}, \& {Gallino}}]{sneden:08}
{Sneden}, C., {Cowan}, J.~J., \& {Gallino}, R. 2008, \araa, 46, 241

\bibitem[{{Sneden}(1973)}]{1973sneden}
{Sneden}, C.~A. 1973, PhD thesis, THE UNIVERSITY OF TEXAS AT AUSTIN.

\bibitem[{{Sobeck} {et~al.}(2011){Sobeck}, {Kraft}, {Sneden}, {Preston},
  {Cowan}, {Smith}, {Thompson}, {Shectman}, \& {Burley}}]{2011sobeck}
{Sobeck}, J.~S., {Kraft}, R.~P., {Sneden}, C., {et~al.} 2011, \aj, 141, 175

\bibitem[{{Sousa} {et~al.}(2015){Sousa}, {Santos}, {Adibekyan}, {Delgado-Mena},
  \& {Israelian}}]{2015sousa}
{Sousa}, S.~G., {Santos}, N.~C., {Adibekyan}, V., {Delgado-Mena}, E., \&
  {Israelian}, G. 2015, \aap, 577, A67

\bibitem[{{Spina} {et~al.}(2018){Spina}, {Mel{\'e}ndez}, {Karakas}, {dos
  Santos}, {Bedell}, {Asplund}, {Ram{\'\i}rez}, {Yong}, {Alves-Brito}, {Bean},
  \& {Dreizler}}]{2018spina}
{Spina}, L., {Mel{\'e}ndez}, J., {Karakas}, A.~I., {et~al.} 2018, \mnras, 474,
  2580

\bibitem[{{Spina} {et~al.}(2020){Spina}, {Nordlander}, {Casey}, {Bedell},
  {D'Orazi}, {Mel{\'e}ndez}, {Karakas}, {Desidera}, {Baratella}, {Yana
  Galarza}, \& {Casali}}]{2020spina}
{Spina}, L., {Nordlander}, T., {Casey}, A.~R., {et~al.} 2020, \apj, 895, 52

\bibitem[{{Spina} {et~al.}(2017){Spina}, {Randich}, {Magrini}, {Jeffries},
  {Friel}, {Sacco}, {Pancino}, {Bonito}, {Bravi}, {Franciosini}, {Klutsch},
  {Montes}, {Gilmore}, {Vallenari}, {Bensby}, {Bragaglia}, {Flaccomio},
  {Koposov}, {Korn}, {Lanzafame}, {Smiljanic}, {Bayo}, {Carraro}, {Casey},
  {Costado}, {Damiani}, {Donati}, {Frasca}, {Hourihane}, {Jofr{\'e}}, {Lewis},
  {Lind}, {Monaco}, {Morbidelli}, {Prisinzano}, {Sousa}, {Worley}, \&
  {Zaggia}}]{2017spina}
{Spina}, L., {Randich}, S., {Magrini}, L., {et~al.} 2017, \aap, 601, A70

\bibitem[{{Spina} {et~al.}(2014{\natexlab{a}}){Spina}, {Randich}, {Palla},
  {Biazzo}, {Sacco}, {Alfaro}, {Franciosini}, {Magrini}, {Morbidelli},
  {Frasca}, {Adibekyan}, {Delgado-Mena}, {Sousa}, {Gonz{\'a}lez Hern{\'a}ndez},
  {Montes}, {Tabernero}, {Tautvai{\v{s}}ien{\.{e}}}, {Bonito}, {Lanzafame},
  {Gilmore}, {Jeffries}, {Vallenari}, {Bensby}, {Bragaglia}, {Flaccomio},
  {Korn}, {Pancino}, {Recio-Blanco}, {Smiljanic}, {Bergemann}, {Costado},
  {Damiani}, {Hill}, {Hourihane}, {Jofr{\'e}}, {de Laverny}, {Lardo},
  {Masseron}, {Prisinzano}, \& {Worley}}]{2014spinaB}
{Spina}, L., {Randich}, S., {Palla}, F., {et~al.} 2014{\natexlab{a}}, \aap,
  568, A2

\bibitem[{{Spina} {et~al.}(2014{\natexlab{b}}){Spina}, {Randich}, {Palla},
  {Sacco}, {Magrini}, {Franciosini}, {Morbidelli}, {Prisinzano}, {Alfaro},
  {Biazzo}, {Frasca}, {Gonz{\'a}lez Hern{\'a}ndez}, {Sousa}, {Adibekyan},
  {Delgado-Mena}, {Montes}, {Tabernero}, {Klutsch}, {Gilmore}, {Feltzing},
  {Jeffries}, {Micela}, {Vallenari}, {Bensby}, {Bragaglia}, {Flaccomio},
  {Koposov}, {Lanzafame}, {Pancino}, {Recio-Blanco}, {Smiljanic}, {Costado},
  {Damiani}, {Hill}, {Hourihane}, {Jofr{\'e}}, {de Laverny}, {Masseron}, \&
  {Worley}}]{2014spinaA}
{Spina}, L., {Randich}, S., {Palla}, F., {et~al.} 2014{\natexlab{b}}, \aap,
  567, A55

\bibitem[{{Spina} {et~al.}(2021){Spina}, {Ting}, {De Silva}, {Frankel},
  {Sharma}, {Cantat-Gaudin}, {Joyce}, {Stello}, {Karakas}, {Asplund},
  {Nordlander}, {Casagrande}, {D'Orazi}, {Casey}, {Cottrell},
  {Tepper-Garc{\'\i}a}, {Baratella}, {Kos}, {{\v{C}}otar}, {Bland-Hawthorn},
  {Buder}, {Freeman}, {Hayden}, {Lewis}, {Lin}, {Lind}, {Martell},
  {Schlesinger}, {Simpson}, {Zucker}, \& {Zwitter}}]{2021spina}
{Spina}, L., {Ting}, Y.-S., {De Silva}, G.~M., {et~al.} 2021, \mnras
  [\eprint[arXiv]{2011.02533}]

\bibitem[{{Sukhbold} {et~al.}(2016){Sukhbold}, {Ertl}, {Woosley}, {Brown}, \&
  {Janka}}]{sukhbold:16}
{Sukhbold}, T., {Ertl}, T., {Woosley}, S.~E., {Brown}, J.~M., \& {Janka}, H.~T.
  2016, \apj, 821, 38

\bibitem[{{Talwar} {et~al.}(2016){Talwar}, {Adachi}, {Berg}, {Bin}, {Bisterzo},
  {Couder}, {deBoer}, {Fang}, {Fujita}, {Fujita}, {G{\"o}rres}, {Hatanaka},
  {Itoh}, {Kadoya}, {Long}, {Miki}, {Patel}, {Pignatari}, {Shimbara}, {Tamii},
  {Wiescher}, {Yamamoto}, \& {Yosoi}}]{2016talwar}
{Talwar}, R., {Adachi}, T., {Berg}, G.~P.~A., {et~al.} 2016, \prc, 93, 055803

\bibitem[{{The} {et~al.}(2007){The}, {El Eid}, \& {Meyer}}]{the:07}
{The}, L.-S., {El Eid}, M.~F., \& {Meyer}, B.~S. 2007, \apj, 655, 1058

\bibitem[{{Travaglio} {et~al.}(1999){Travaglio}, {Galli}, {Gallino}, {Busso},
  {Ferrini}, \& {Straniero}}]{1999travaglio}
{Travaglio}, C., {Galli}, D., {Gallino}, R., {et~al.} 1999, \apj, 521, 691

\bibitem[{{Travaglio} {et~al.}(2005){Travaglio}, {Hillebrandt}, \&
  {Reinecke}}]{2005travaglio}
{Travaglio}, C., {Hillebrandt}, W., \& {Reinecke}, M. 2005, \aap, 443, 1007

\bibitem[{{Tsantaki} {et~al.}(2019){Tsantaki}, {Santos}, {Sousa},
  {Delgado-Mena}, {Adibekyan}, \& {Andreasen}}]{2019tsantaki}
{Tsantaki}, M., {Santos}, N.~C., {Sousa}, S.~G., {et~al.} 2019, \mnras, 485,
  2772

\bibitem[{{Uns{\"o}ld}(1932)}]{1932unsold}
{Uns{\"o}ld}, A. 1932, \zap, 4, 339

\bibitem[{{Velichko} {et~al.}(2010){Velichko}, {Mashonkina}, \&
  {Nilsson}}]{2010velichko}
{Velichko}, A.~B., {Mashonkina}, L.~I., \& {Nilsson}, H. 2010, Astronomy
  Letters, 36, 664

\bibitem[{{Vescovi} {et~al.}(2020){Vescovi}, {Cristallo}, {Busso}, \&
  {Liu}}]{2020vescovi}
{Vescovi}, D., {Cristallo}, S., {Busso}, M., \& {Liu}, N. 2020, \apjl, 897, L25

\bibitem[{{Wanajo} {et~al.}(2011){Wanajo}, {Janka}, \&
  {M{\"u}ller}}]{wanajo:11}
{Wanajo}, S., {Janka}, H.-T., \& {M{\"u}ller}, B. 2011, \apjl, 726, L15

\bibitem[{{Woosley} \& {Weaver}(1995)}]{1995woolsey}
{Woosley}, S.~E. \& {Weaver}, T.~A. 1995, \apjs, 101, 181

\bibitem[{{Yana Galarza} {et~al.}(2019){Yana Galarza}, {Mel{\'e}ndez},
  {Lorenzo-Oliveira}, {Valio}, {Reggiani}, {Carlos}, {Ponte}, {Spina},
  {Haywood}, \& {Gandolfi}}]{2019galarza}
{Yana Galarza}, J., {Mel{\'e}ndez}, J., {Lorenzo-Oliveira}, D., {et~al.} 2019,
  \mnras, 490, L86

\bibitem[{{Yong} {et~al.}(2012){Yong}, {Carney}, \& {Friel}}]{2012yong}
{Yong}, D., {Carney}, B.~W., \& {Friel}, E.~D. 2012, \aj, 144, 95

\end{thebibliography}

\begin{appendix}
\section{Additional table and figures}

\begin{table}[!h]
\begin{minipage}[b]{1.0\textwidth}\centering
\caption{Atmospheric parameters of the whole stellar sample, as derived in \cite{2020baratellaA}.}
\setlength\tabcolsep{7pt}
\begin{tabular}{lccccccr}
\toprule
CNAME & \teff\,(K) & \logg & $\xi$\,(\kms) & [Fe/H]$_{\rm{I}}$ & [Fe/H]$_{\rm{II}}$\\
\midrule
&&&\textit{GBS}\\
Sun & 5790 $\pm$ 50 & 4.47 $\pm$ 0.05 & 1.00$\pm0.10$ & \\
\object{$\alpha$\,Cen\,A} & 5830 $\pm$ 75 & 4.45 $\pm$ 0.10 & 1.09 $\pm$ 0.20 & +0.23$\pm0.02$ & +0.21 $\pm$ 0.05 \\
\object{$\tau$\,Cet} & 5401 $\pm$ 75 & 4.38 $\pm$ 0.10 & 0.89 $\pm$ 0.15 & $-0.44\pm0.02$ & $-0.44\pm$0.05 \\
\object{$\beta$\,Hyi} & 5870 $\pm$ 100 & 3.95 $\pm$ 0.10 & 1.35 $\pm$ 0.10 & $-0.09\pm0.02$ & $-0.09\pm0.05$ \\
\object{18\,Sco} & 5875 $\pm$ 100 & 4.55 $\pm$ 0.10 & 1.15 $\pm$ 0.15 & +0.06 $\pm$ 0.02 & +0.05 $\pm$ 0.04 \\
\\
&& &\textit{\object{IC\,2391}}\\
08365498-5308342 & 5215 $\pm$ 100 & 4.35 $\pm$ 0.10 & 0.85 $\pm$ 0.10 & 0.00 $\pm$ 0.01 $\pm$ 0.06 & 0.09 $\pm$ 0.03 $\pm$ 0.08  \\
08440521-5253171 & 5471 $\pm$ 100 & 4.20 $\pm$ 0.10 & 0.88 $\pm$ 0.04$^{\ast}$ & 0.00 $\pm$ 0.02 $\pm$ 0.07 & 0.05 $\pm$ 0.03 $\pm$ 0.06  \\
\\
&&&\textit{\object{IC\,2602}}\\
10440681-6359351  & 5525 $\pm$ 75 & 4.38 $\pm$ 0.15 & 1.00 $\pm$ 0.20 & 0.00 $\pm$ 0.01 $\pm$ 0.06 & 0.08 $\pm$ 0.03 $\pm$ 0.06 \\
10442256-6415301  & 5775 $\pm$ 75 & 4.49 $\pm$ 0.10 & 1.15 $\pm$ 0.10 & 0.04 $\pm$ 0.01 $\pm$ 0.07 & 0.05 $\pm$ 0.02 $\pm$ 0.05 \\
10481856-6409537  & 5680 $\pm$ 100 & 4.10 $\pm$ 0.10 & 1.09 $\pm$ 0.05$^{\ast}$ & 0.03 $\pm$ 0.02 $\pm$ 0.07 & 0.06 $\pm$ 0.03 $\pm$ 0.06 \\
\\
&&&\textit{\object{IC\,4665}}\\
17442711+0547196  & 5380 $\pm$ 75 & 4.48 $\pm$ 0.10 & 0.80 $\pm$ 0.04$^{\ast}$ & 0.14 $\pm$ 0.02 $\pm$ 0.05 & 0.17 $\pm$ 0.02 $\pm$ 0.06\\
17445810+0551329  & 5575 $\pm$ 75 & 4.47 $\pm$ 0.10 & 0.96 $\pm$ 0.10 & 0.12 $\pm$ 0.01 $\pm$ 0.07 & 0.13 $\pm$ 0.04 $\pm$ 0.04 \\
17452508+0551388  & 5300 $\pm$ 100 & 4.27 $\pm$ 0.15 & 1.03 $\pm$ 0.10 & 0.05 $\pm$ 0.03 $\pm$ 0.06 & 0.11 $\pm$ 0.03 $\pm$ 0.08 \\
\\
&&&\textit{\object{NGC\,2264}}\\
06405694+0948407  & 6150 $\pm$ 75 & 4.05 $\pm$ 0.10 & 1.29 $\pm$ 0.08$^{\ast}$ & 0.10 $\pm$ 0.02 $\pm$ 0.07 & 0.11 $\pm$ 0.02 $\pm$ 0.05\\
\\
&&&\textit{\object{NGC\,2516}}\\
07544342-6024437  & 5430 $\pm$ 100 & 4.51 $\pm$ 0.10 & 1.05 $\pm$ 0.10 & 0.03 $\pm$ 0.01 $\pm$ 0.07 & 0.04 $\pm$ 0.03 $\pm$ 0.08 \\
07550592-6104294  & 5550 $\pm$ 75 & 4.20 $\pm$ 0.10 & 0.95 $\pm$ 0.10 & 0.05 $\pm$ 0.03 $\pm$ 0.07 & 0.08 $\pm$ 0.02 $\pm$ 0.06\\
07551977-6104200  & 6050 $\pm$ 100 & 4.62 $\pm$ 0.10 & 1.15 $\pm$ 0.10$^{\ast}$ & 0.12 $\pm$ 0.02 $\pm$ 0.06 & 0.12 $\pm$ 0.03 $\pm$ 0.05 \\
07553236-6023094  & 5700 $\pm$ 75 & 4.52 $\pm$ 0.10 & 1.15 $\pm$ 0.15 & 0.09 $\pm$ 0.01 $\pm$ 0.07 & 0.13 $\pm$ 0.01 $\pm$ 0.07 \\
07564410-6034523  & 5650 $\pm$ 75 & 4.45 $\pm$ 0.10 & 1.02 $\pm$ 0.10 & 0.13 $\pm$ 0.02 $\pm$ 0.06 & 0.14 $\pm$ 0.04 $\pm$ 0.05 \\
07573608-6048128  & 5625 $\pm$ 100 & 4.55 $\pm$ 0.10 & 1.03 $\pm$ 0.10 & 0.11 $\pm$ 0.01 $\pm$ 0.07 & 0.10 $\pm$ 0.02 $\pm$ 0.08 \\
07574792-6056131  & 5580 $\pm$ 75 & 4.57 $\pm$ 0.10 & 1.08 $\pm$ 0.10 & 0.08 $\pm$ 0.01 $\pm$ 0.07 & 0.10 $\pm$ 0.03 $\pm$ 0.07 \\
07575215-6100318  & 5275 $\pm$ 100 & 4.54 $\pm$ 0.10 & 0.98 $\pm$ 0.10 & 0.07 $\pm$ 0.02 $\pm$ 0.05 & 0.10 $\pm$ 0.05 $\pm$ 0.08 \\
07583485-6103121  & 5758 $\pm$ 100 & 4.43 $\pm$ 0.10 & 1.05 $\pm$ 0.15 & 0.02 $\pm$ 0.02 $\pm$ 0.06 & 0.04 $\pm$ 0.03 $\pm$ 0.03 \\
07584257-6040199 & 5550 $\pm$ 75 & 4.48 $\pm$ 0.10 & 0.98 $\pm$ 0.10 & 0.09 $\pm$ 0.01 $\pm$ 0.06 & 0.08 $\pm$ 0.04 $\pm$ 0.06 \\
08000944-6033355  & 5675 $\pm$ 100 & 4.38 $\pm$ 0.10 & 0.90 $\pm$ 0.10 & 0.05 $\pm$ 0.02 $\pm$ 0.05 & 0.06 $\pm$ 0.02 $\pm$ 0.05 \\
08013658-6059021  & 5585 $\pm$ 100 & 4.32 $\pm$ 0.10 & 0.92 $\pm$ 0.05$^{\ast}$ & 0.06 $\pm$ 0.02 $\pm$ 0.06 & 0.07 $\pm$ 0.02 $\pm$ 0.05 \\
\\
&&&\textit{\object{NGC\,2547}}\\
08102854-4856518  & 5800 $\pm$ 100 & 4.20 $\pm$ 0.10 & 1.04 $\pm$ 0.07$^{\ast}$ & 0.11 $\pm$ 0.02 $\pm$ 0.07 & 0.14 $\pm$ 0.03 $\pm$ 0.05 \\
08110139-4900089  & 5375 $\pm$ 100 & 4.50 $\pm$ 0.10 & 1.05 $\pm$ 0.10 & 0.05 $\pm$ 0.01 $\pm$ 0.06 & 0.10 $\pm$ 0.04 $\pm$ 0.08 \\
\\
\bottomrule
\end{tabular}
\tablefoot{ The values of the $\xi$ parameter flagged with an asterisk are the input values computed from the photometric estimates of \teff and the trigonometric \logg.}
\label{atmospheric_param}
\end{minipage}
\end{table}

\begin{figure*}[!htb]
\centering
\includegraphics[scale=0.27]{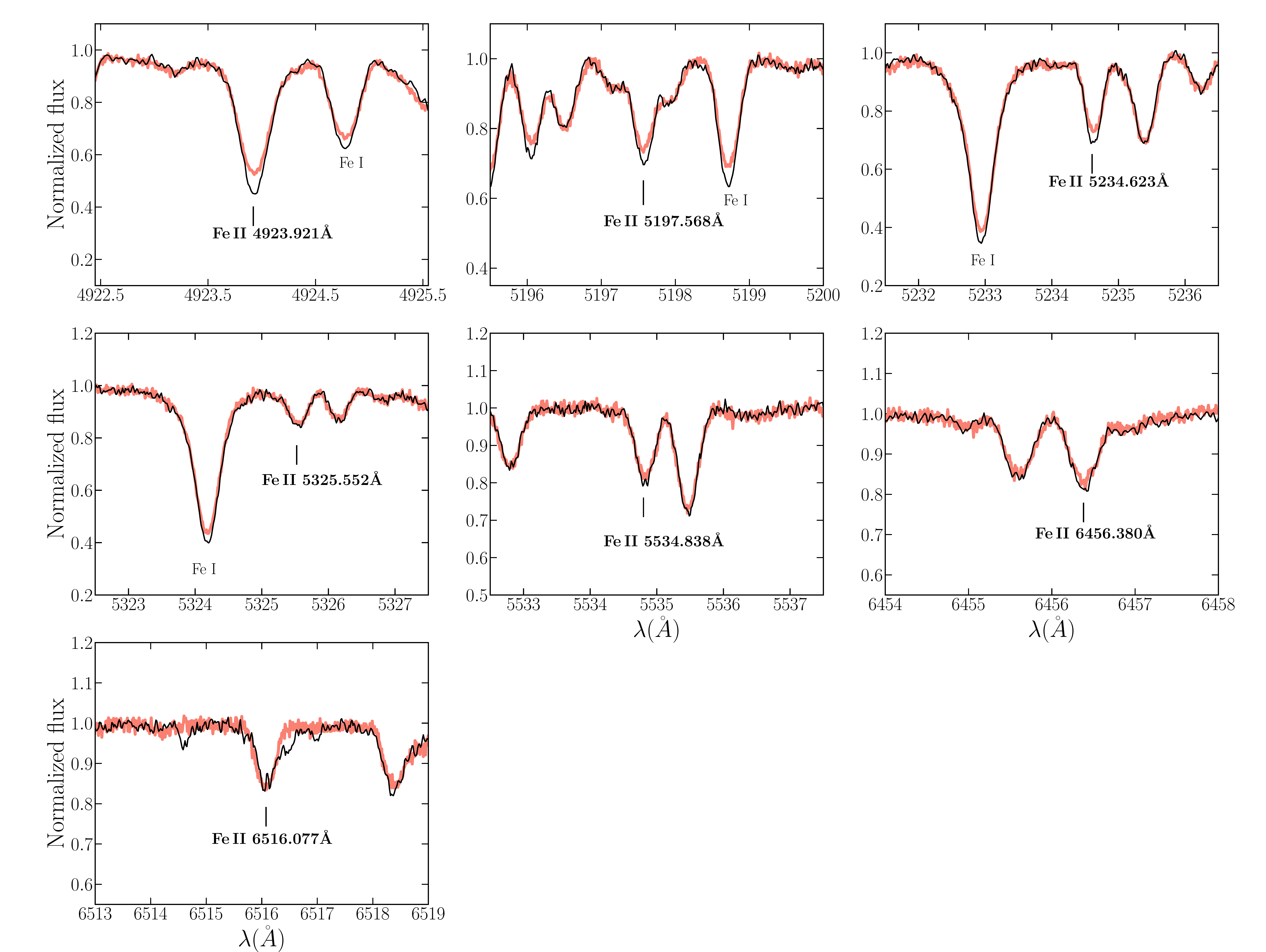}
\caption{Comparison of Fe\,{\sc ii} line profiles in the Sun (light pink), with a rotational profile of 11\,\vsini\,\kms, and star 10442256$-$6415301 (black). }
\label{comp_FeII_lines_v11}
\end{figure*}

\begin{figure*}[!htb]
     \centering
      \subfloat{\includegraphics[scale=0.28]{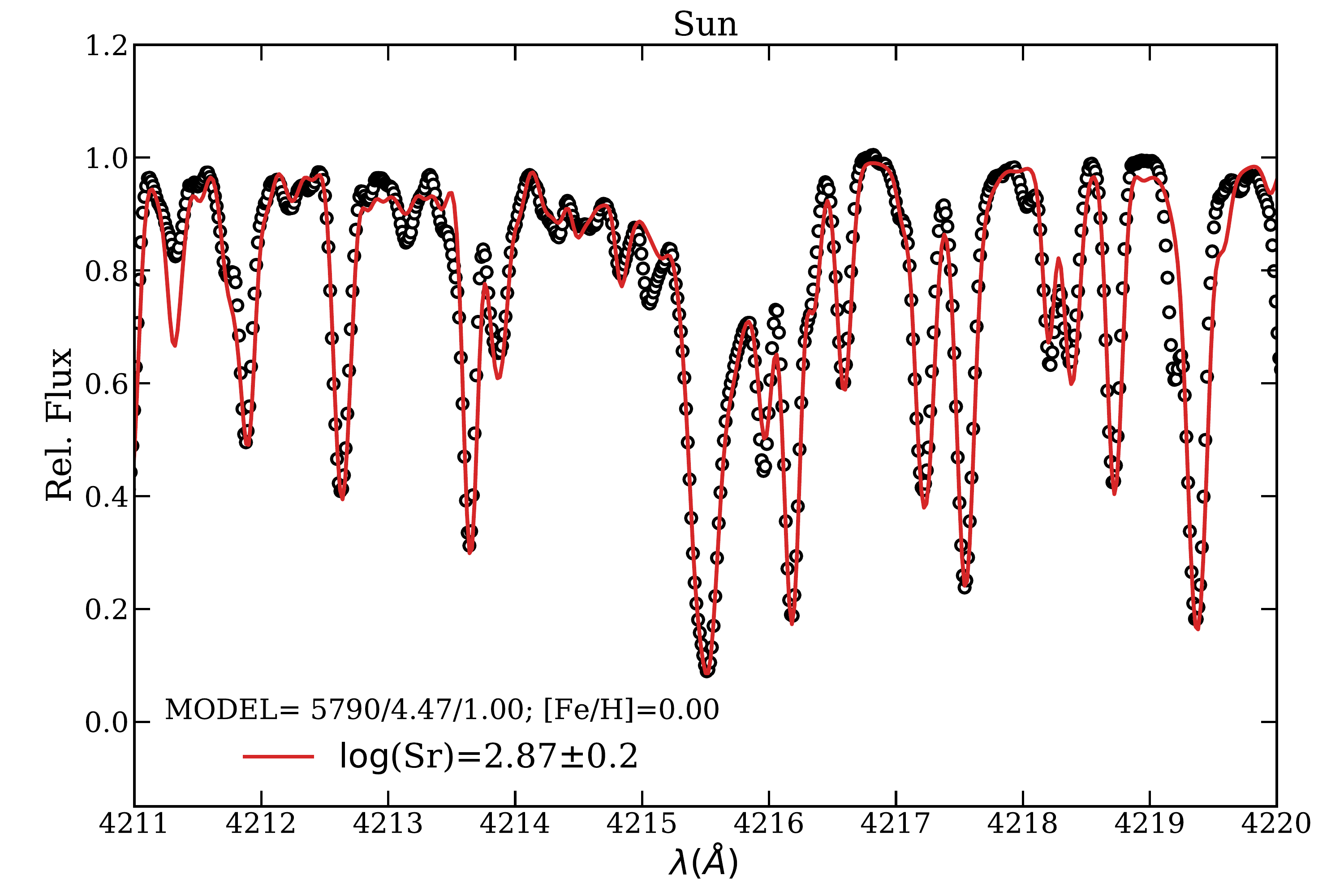}}
      \qquad
      \subfloat{\includegraphics[scale=0.28]{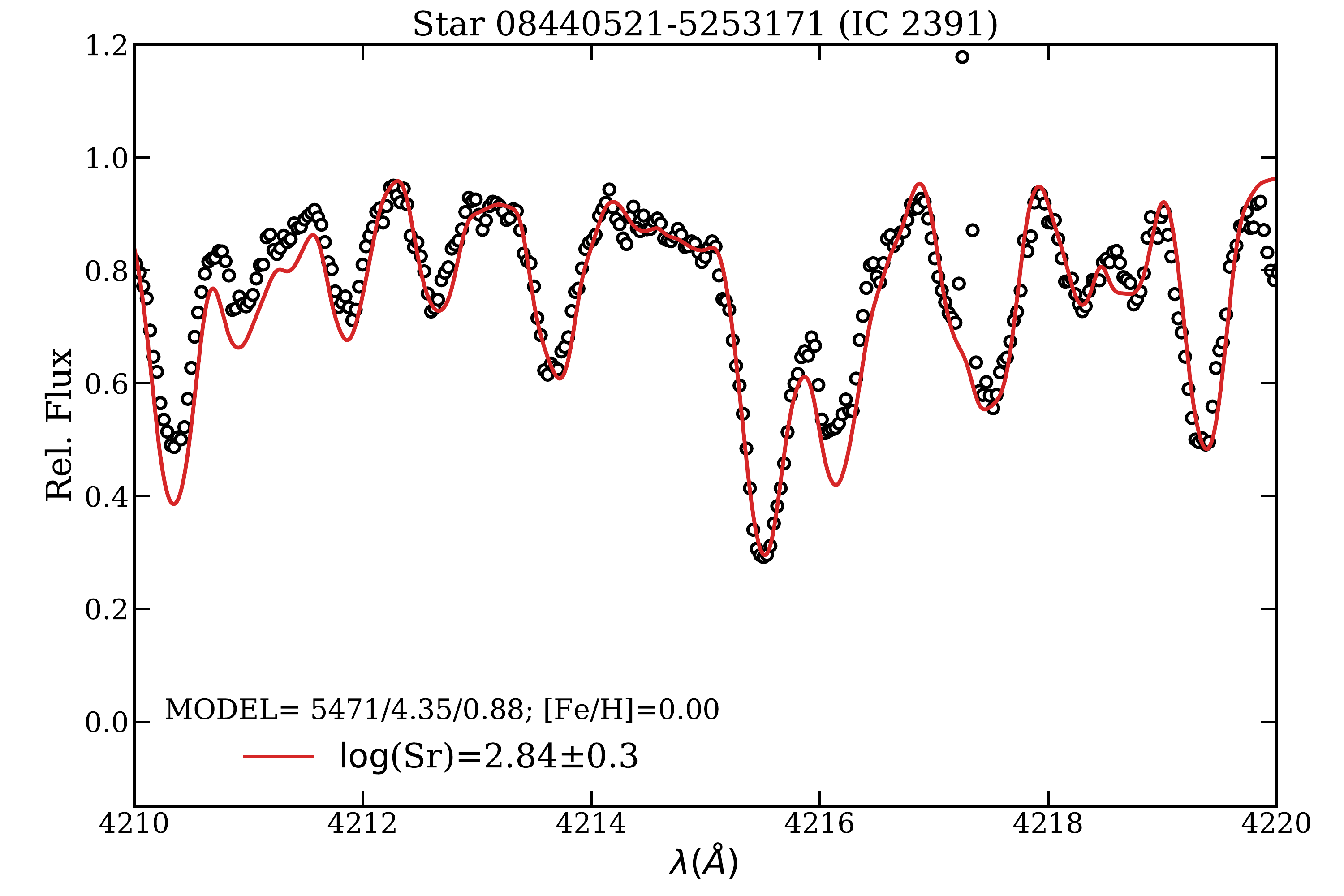}}
      \qquad
      \subfloat{\includegraphics[scale=0.28]{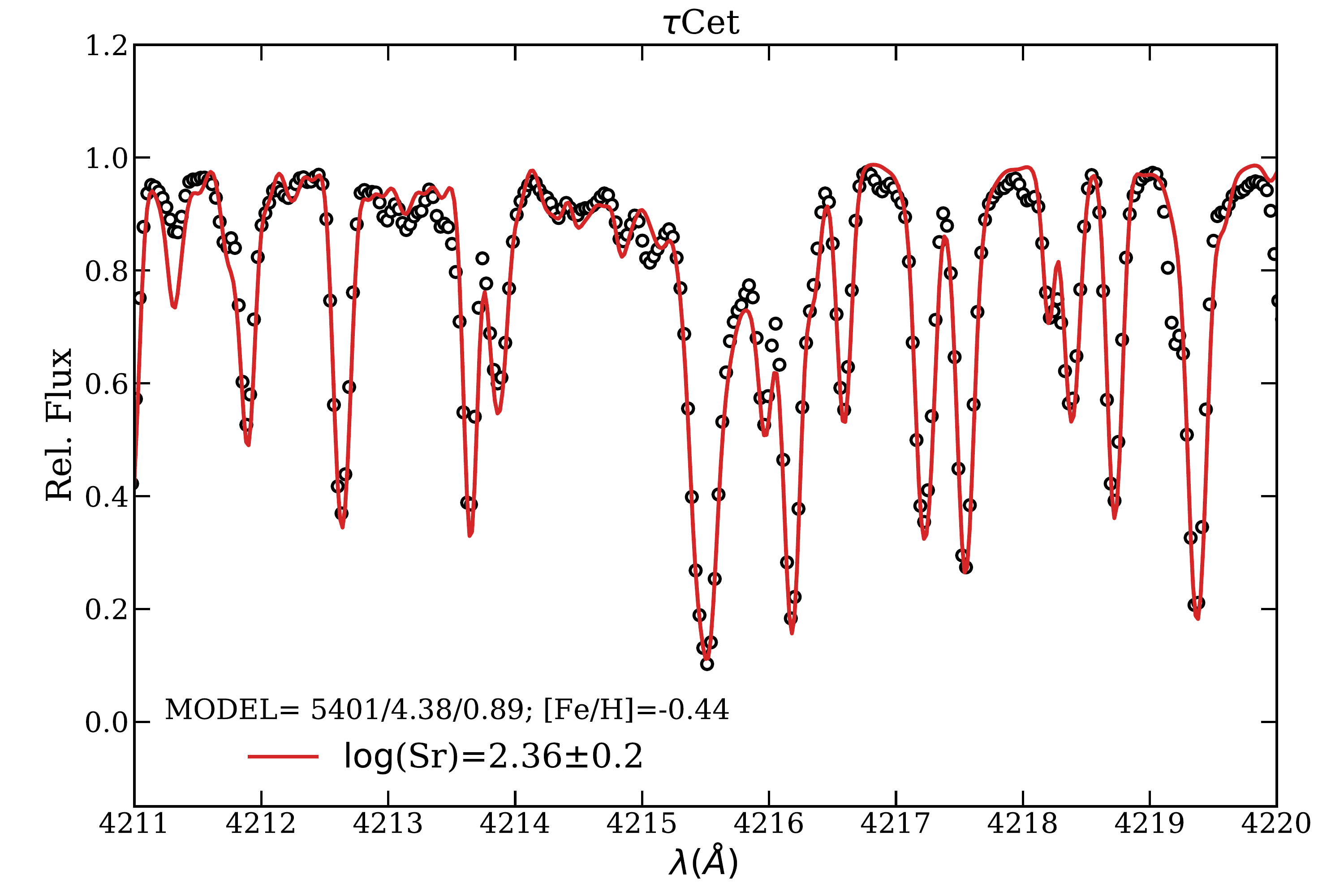}}
      \caption{ Best fit models (solid red lines) of Sr\,{\sc ii} lines in the Sun (UVES spectrum,  top left panel), star 08440521-5253171 of \object{IC\,2391} (FEROS spectrum,  top right panel), and \object{$\tau$\,Cet} (FEROS spectrum,  bottom panel). The open circles represent the observed spectra. }
        \label{best_fit_sr4215}
\end{figure*}

\begin{figure*}[!hb]
     \centering
      \subfloat{\includegraphics[scale=0.30]{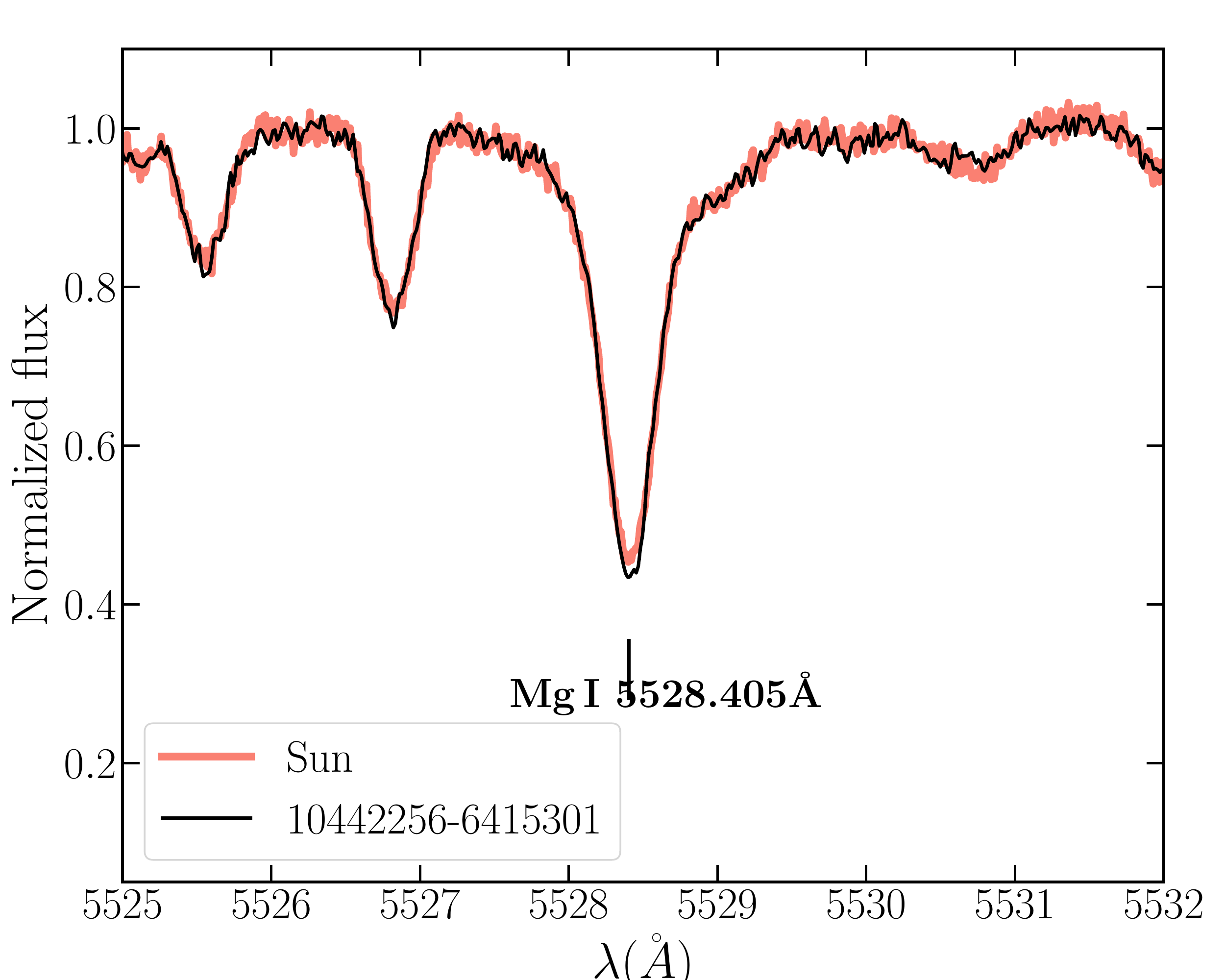}}
      \qquad
      \subfloat{\includegraphics[scale=0.30]{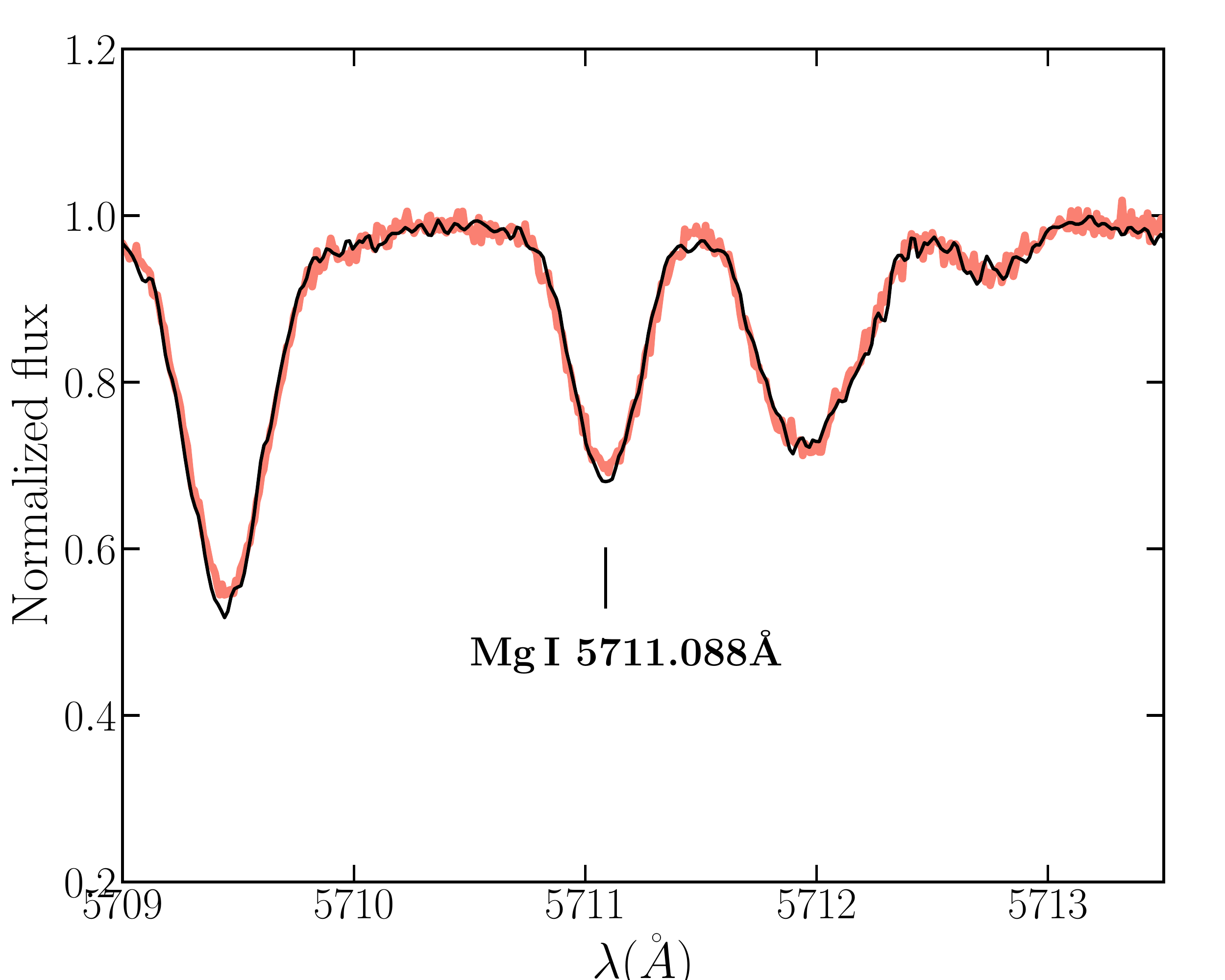}}
      \caption{ Comparison of Mg lines in the Sun (light pink) and star 10442256-6415301 (black). }
        \label{comp_Mg}
\end{figure*}

\begin{figure*}[!htb]
     \centering
      \subfloat{\includegraphics[scale=0.27]{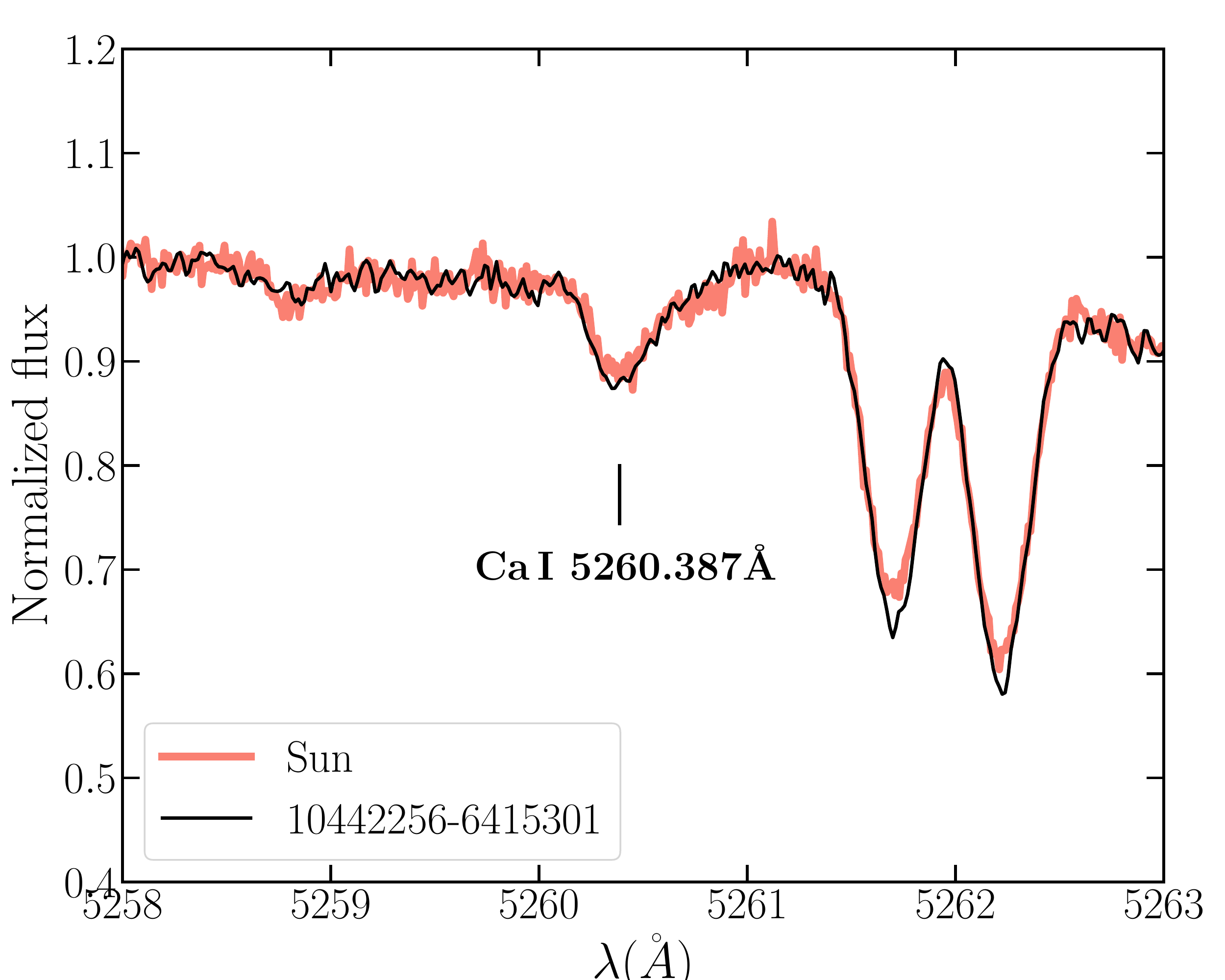}}
      \qquad
      \subfloat{\includegraphics[scale=0.27]{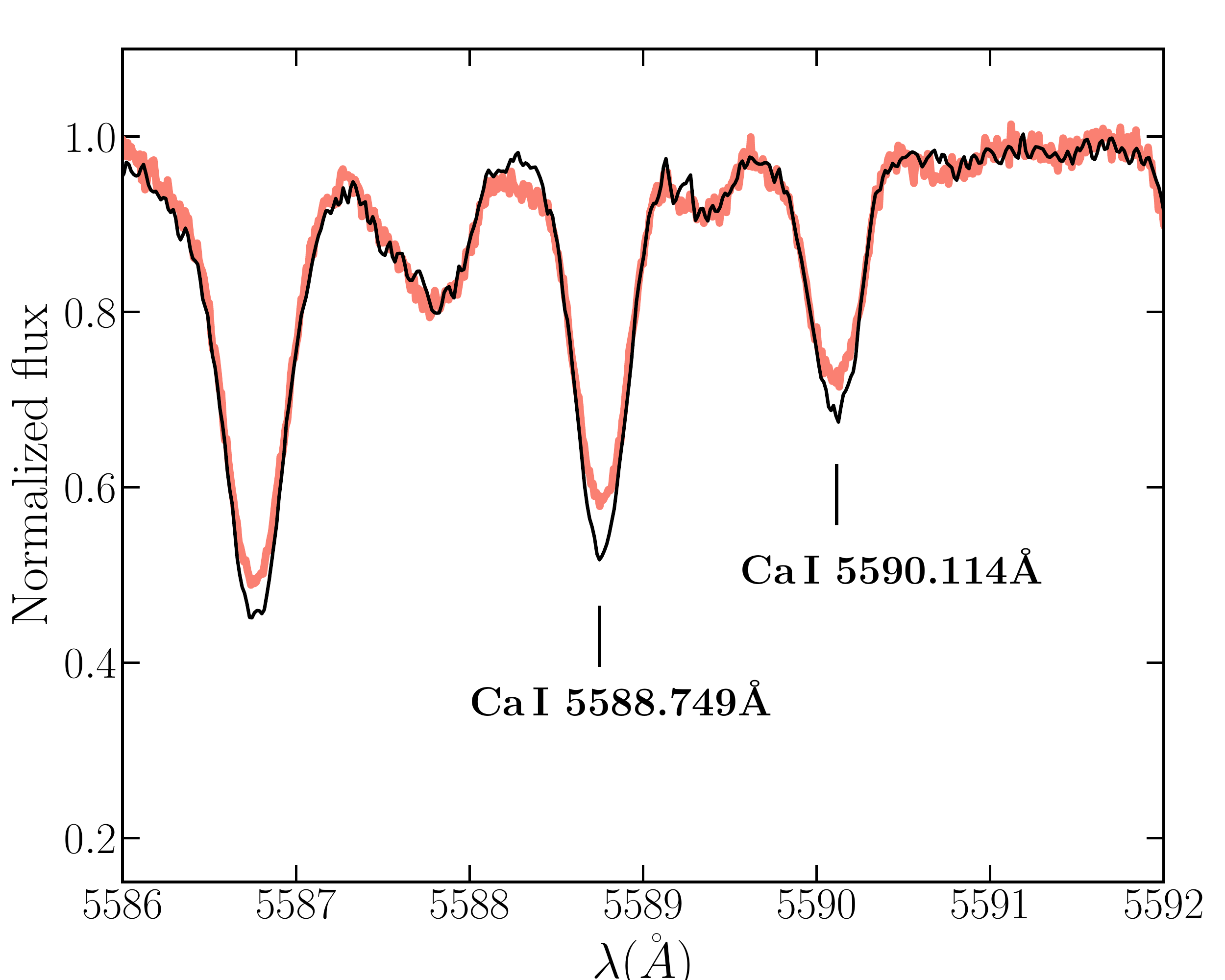}}
      \qquad
      \subfloat{\includegraphics[scale=0.27]{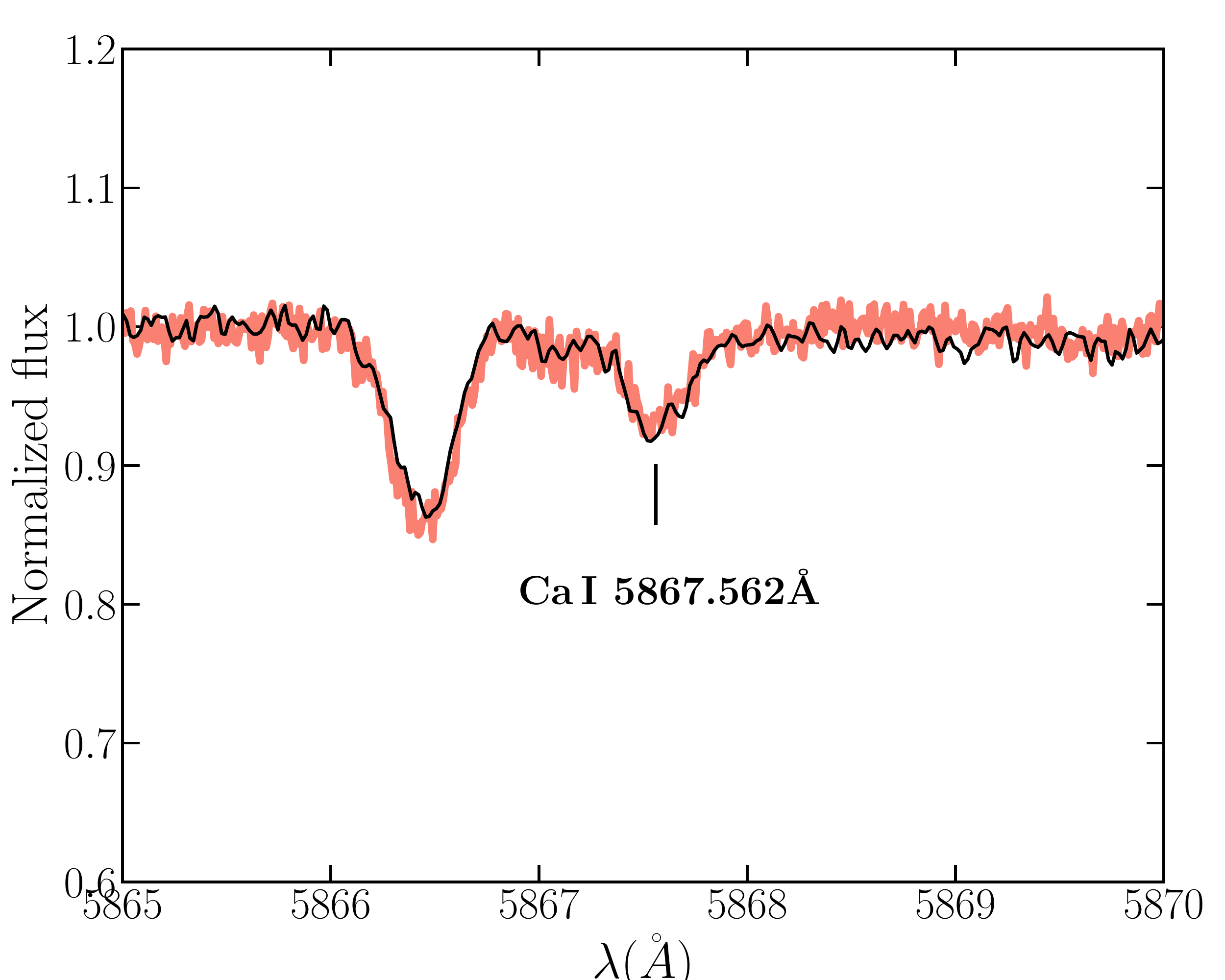}}
      \qquad
      \subfloat{\includegraphics[scale=0.27]{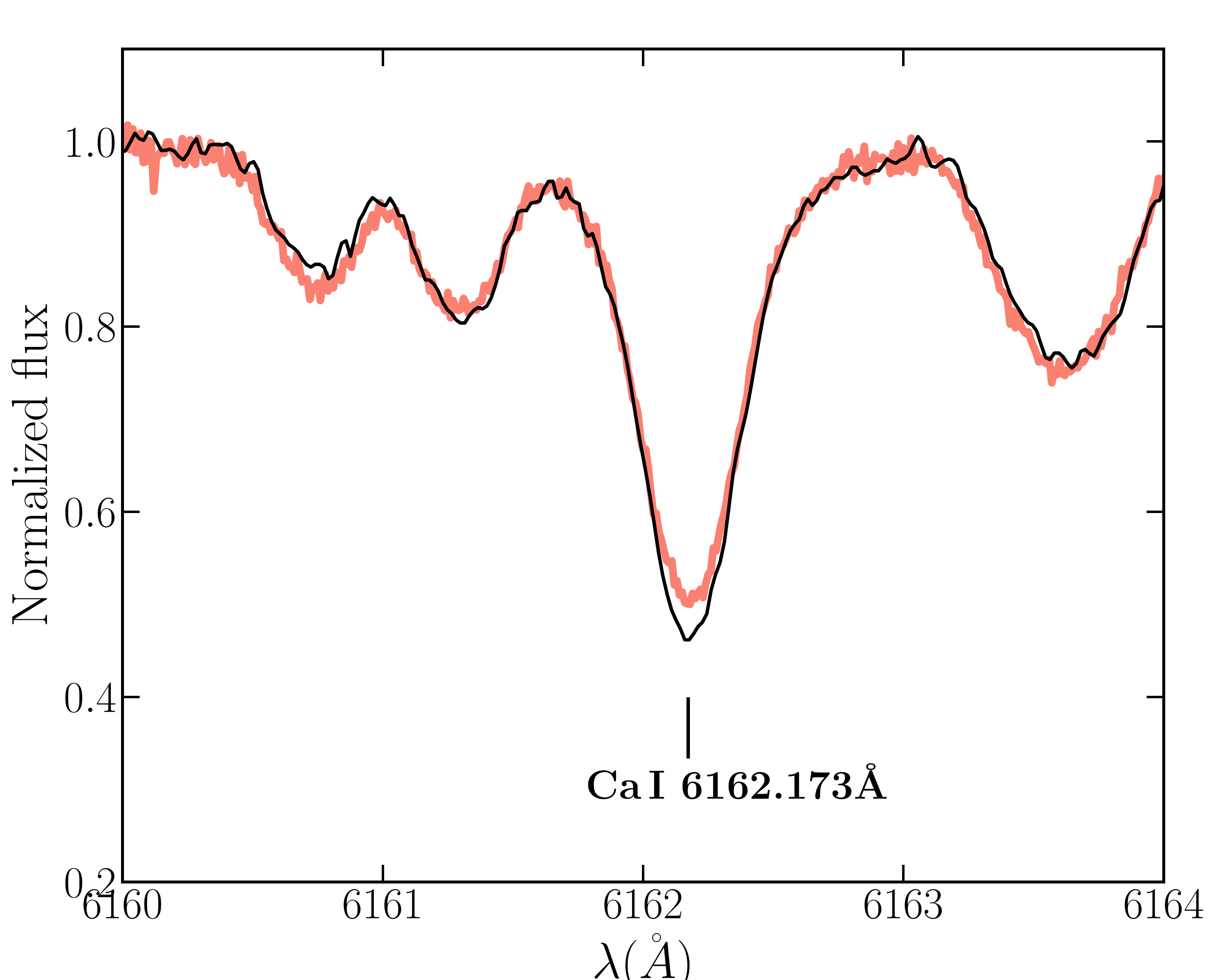}}
      \qquad
      \subfloat{\includegraphics[scale=0.27]{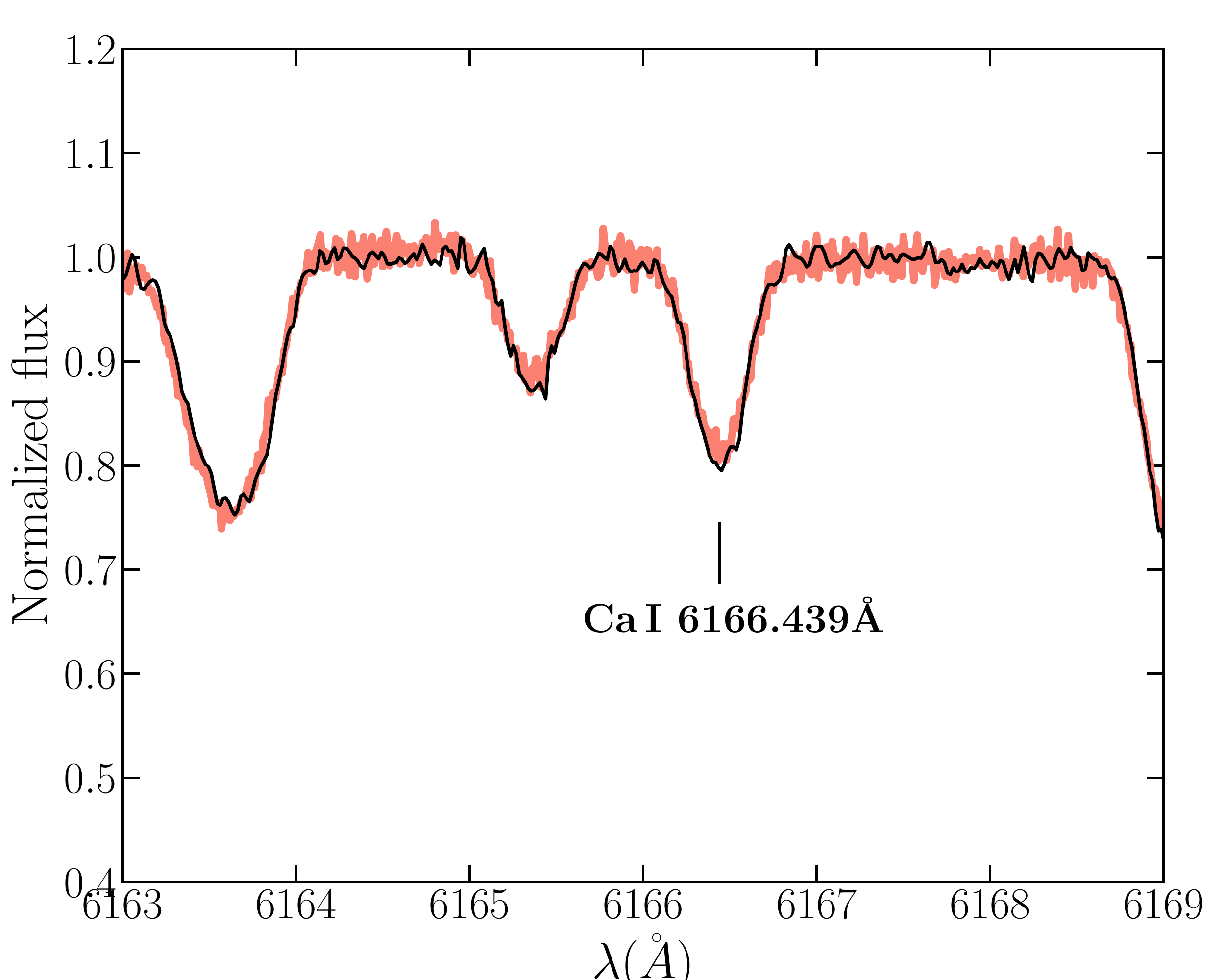}}
      \qquad
      \subfloat{\includegraphics[scale=0.27]{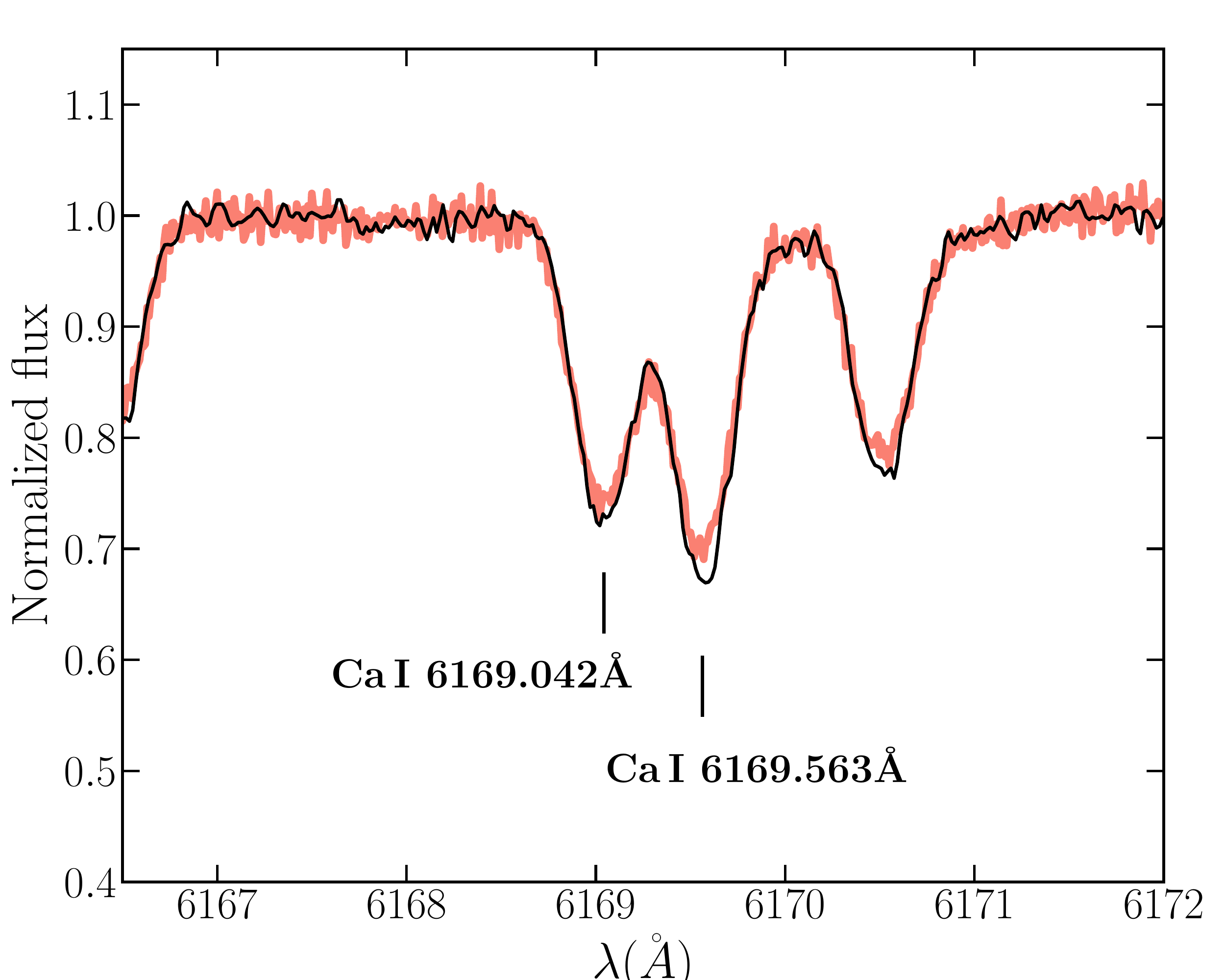}}
      \qquad
      \subfloat{\includegraphics[scale=0.27]{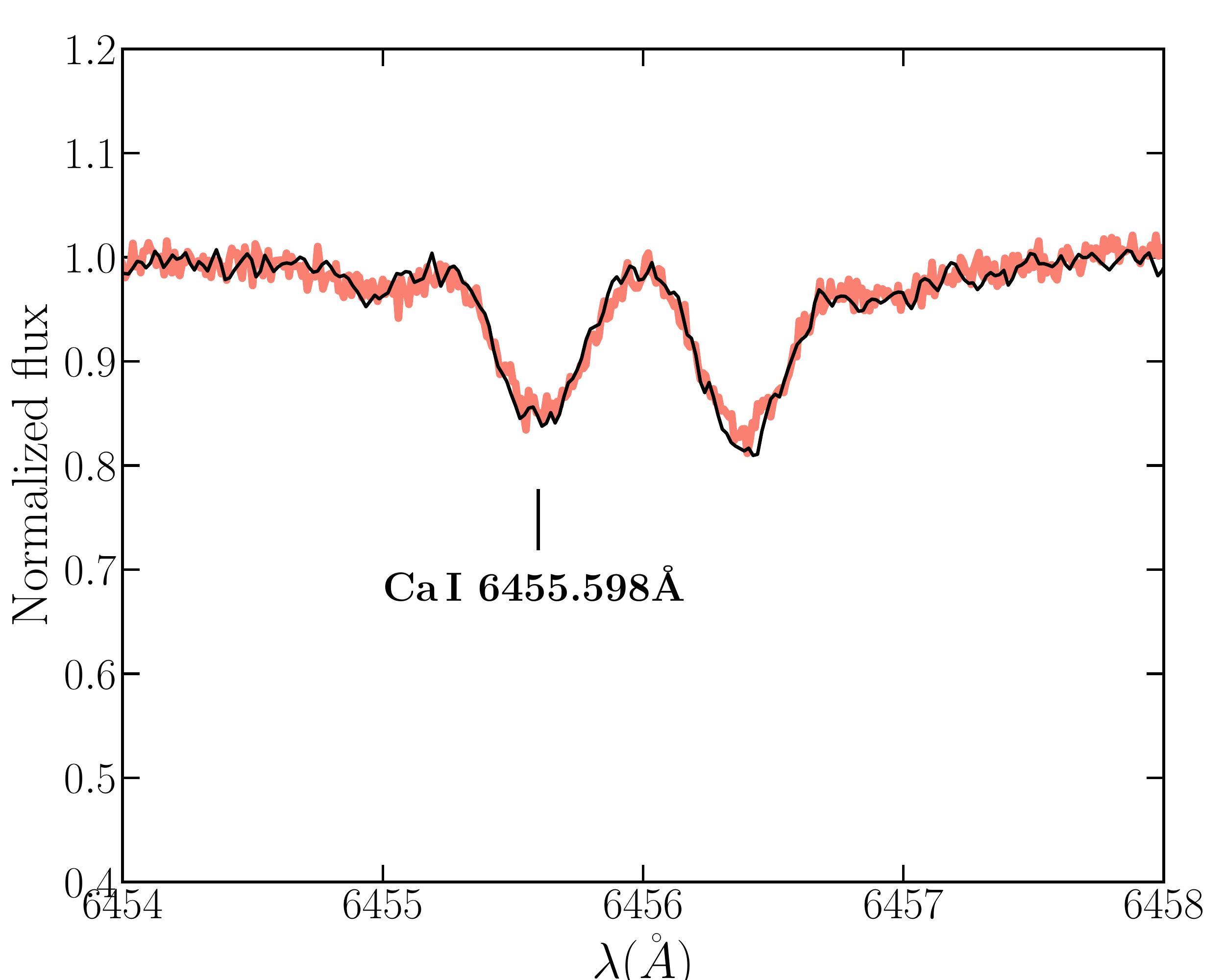}}
      \qquad
      \subfloat{\includegraphics[scale=0.27]{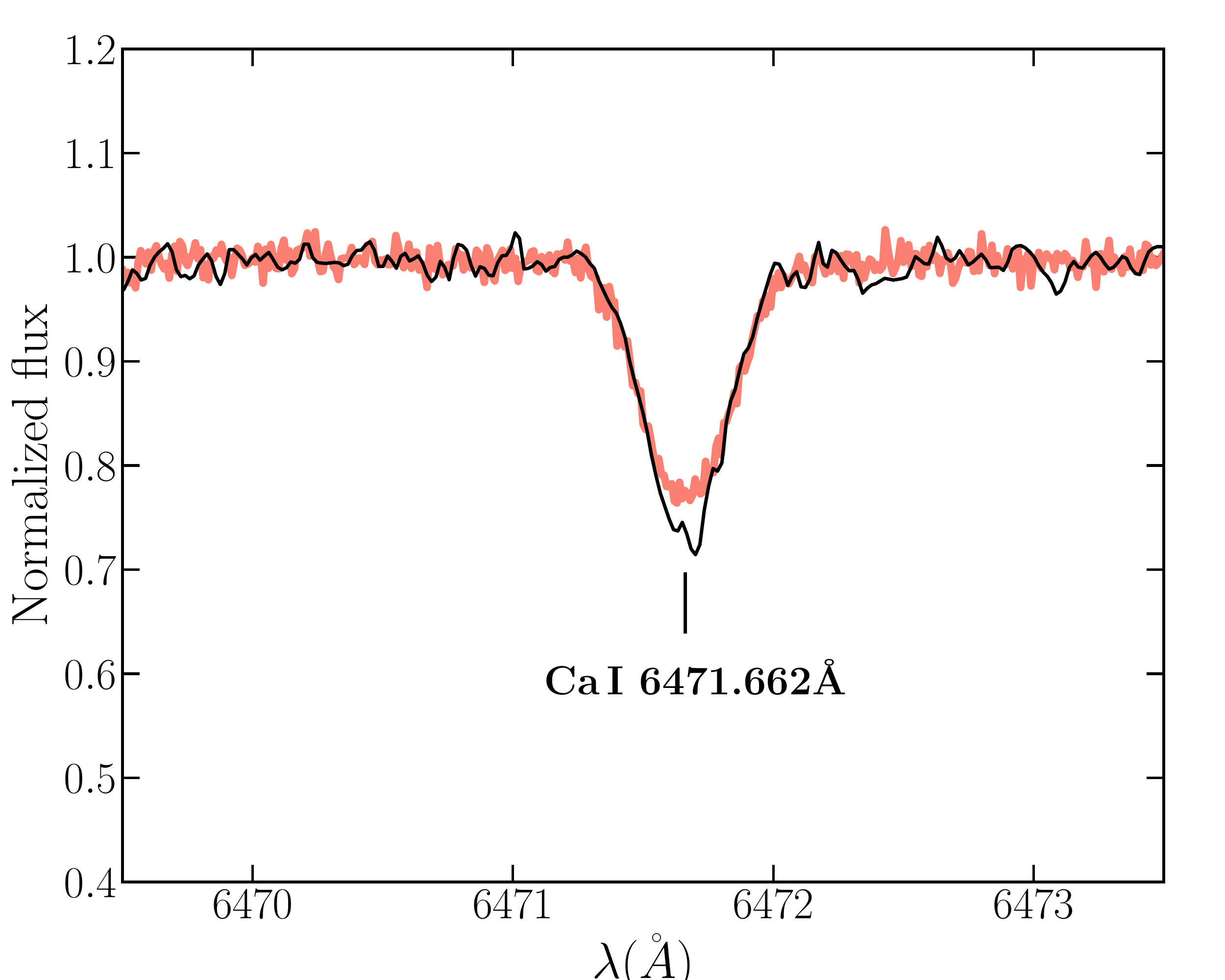}}
      \caption{ Comparison of Ca lines in the Sun (light pink) and star 10442256-6415301 (black).  }
        \label{comp_Ca}
\end{figure*}

\begin{figure*}[!htb]
     \centering
      \subfloat{\includegraphics[scale=0.22]{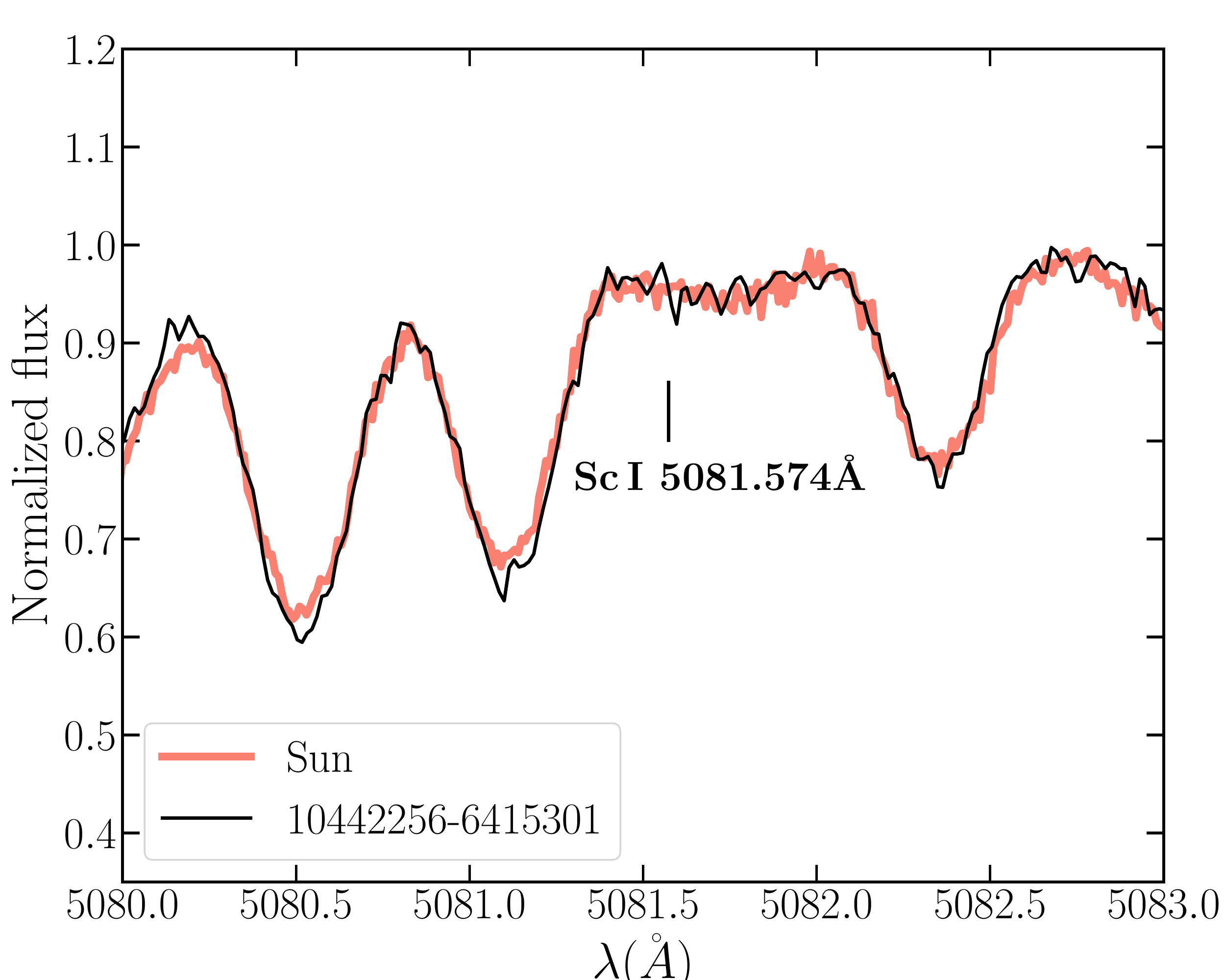}}
      \qquad
      \subfloat{\includegraphics[scale=0.22]{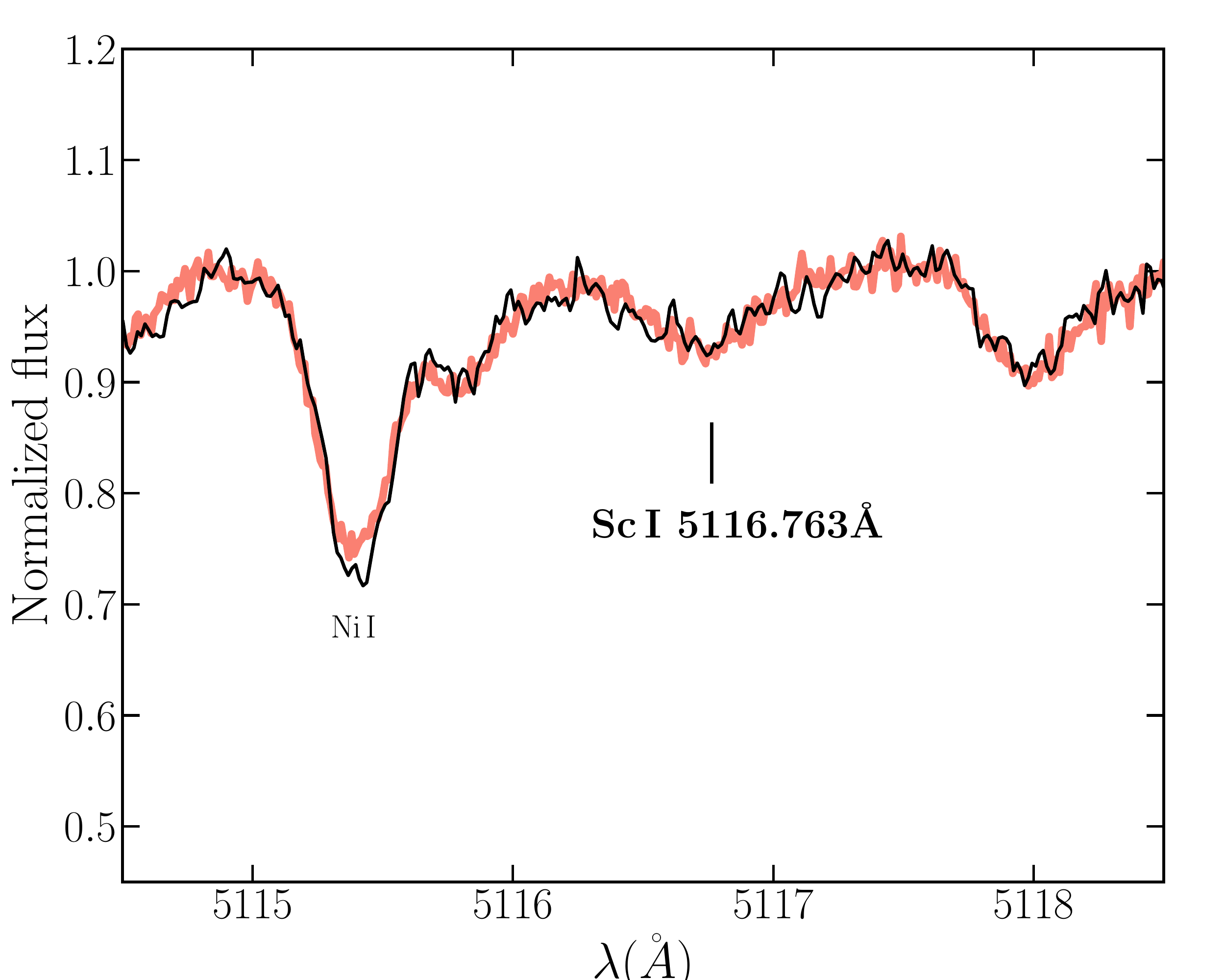}}
      \qquad
      \subfloat{\includegraphics[scale=0.22]{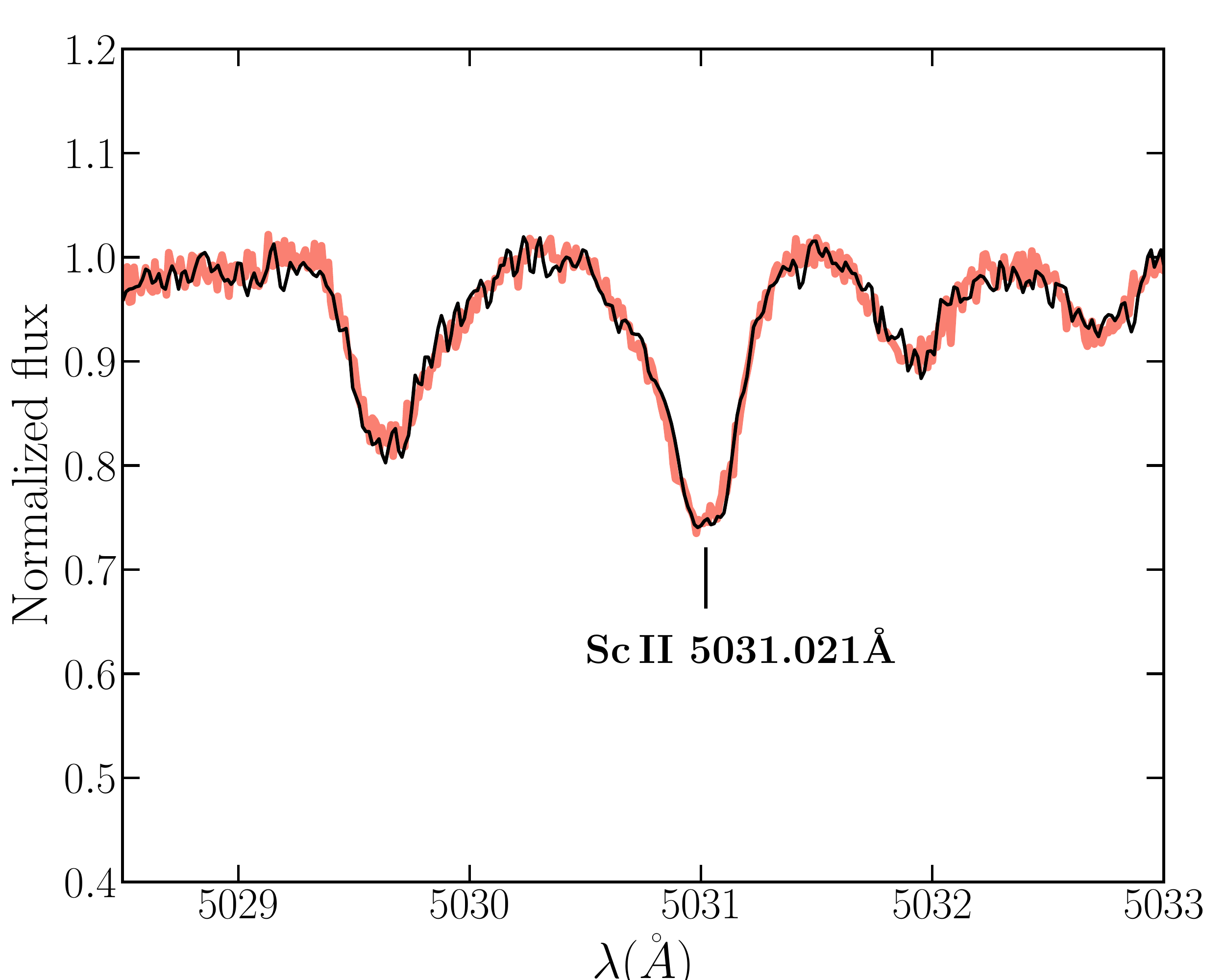}}
      \qquad
      \subfloat{\includegraphics[scale=0.22]{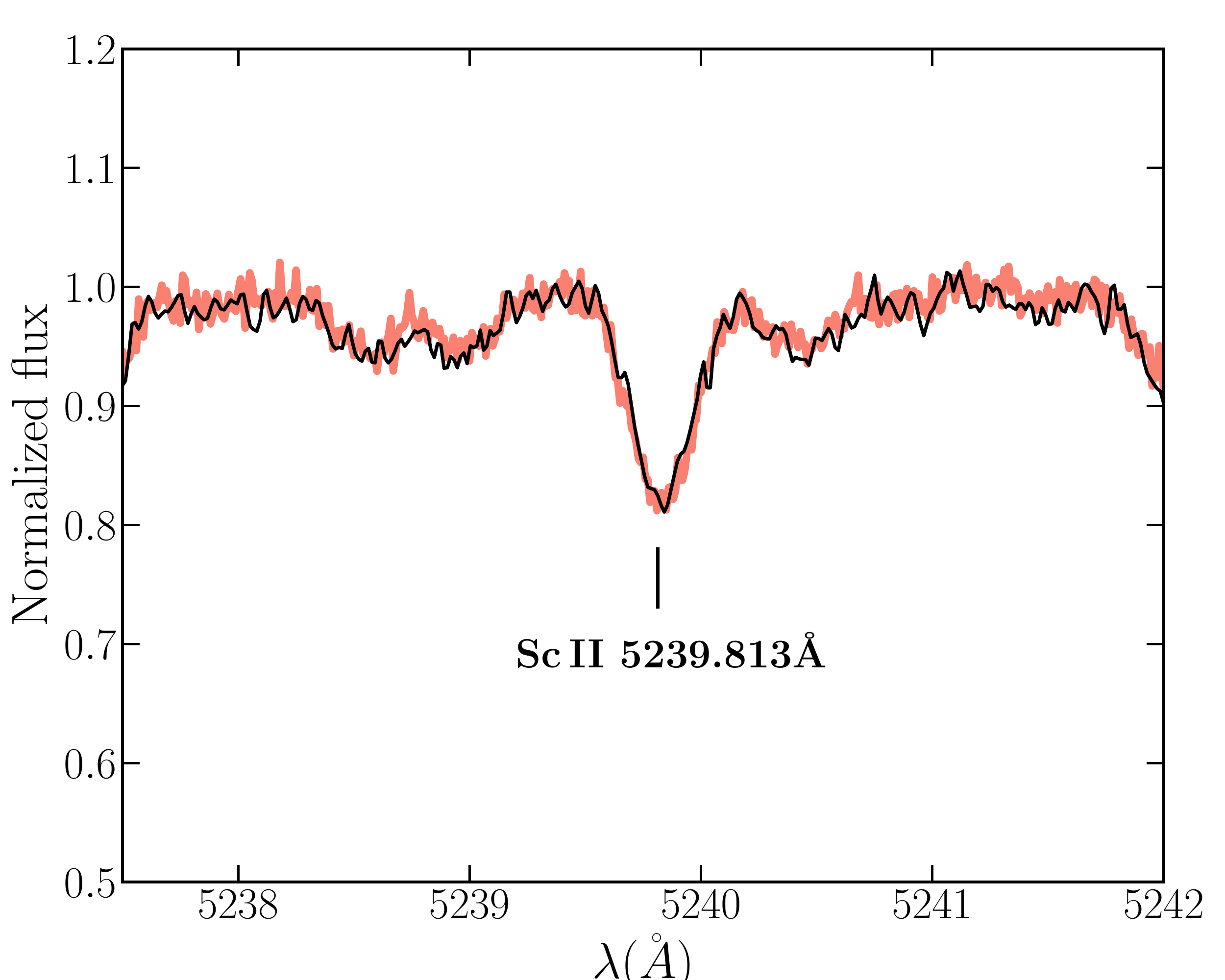}}
      \qquad
      \subfloat{\includegraphics[scale=0.22]{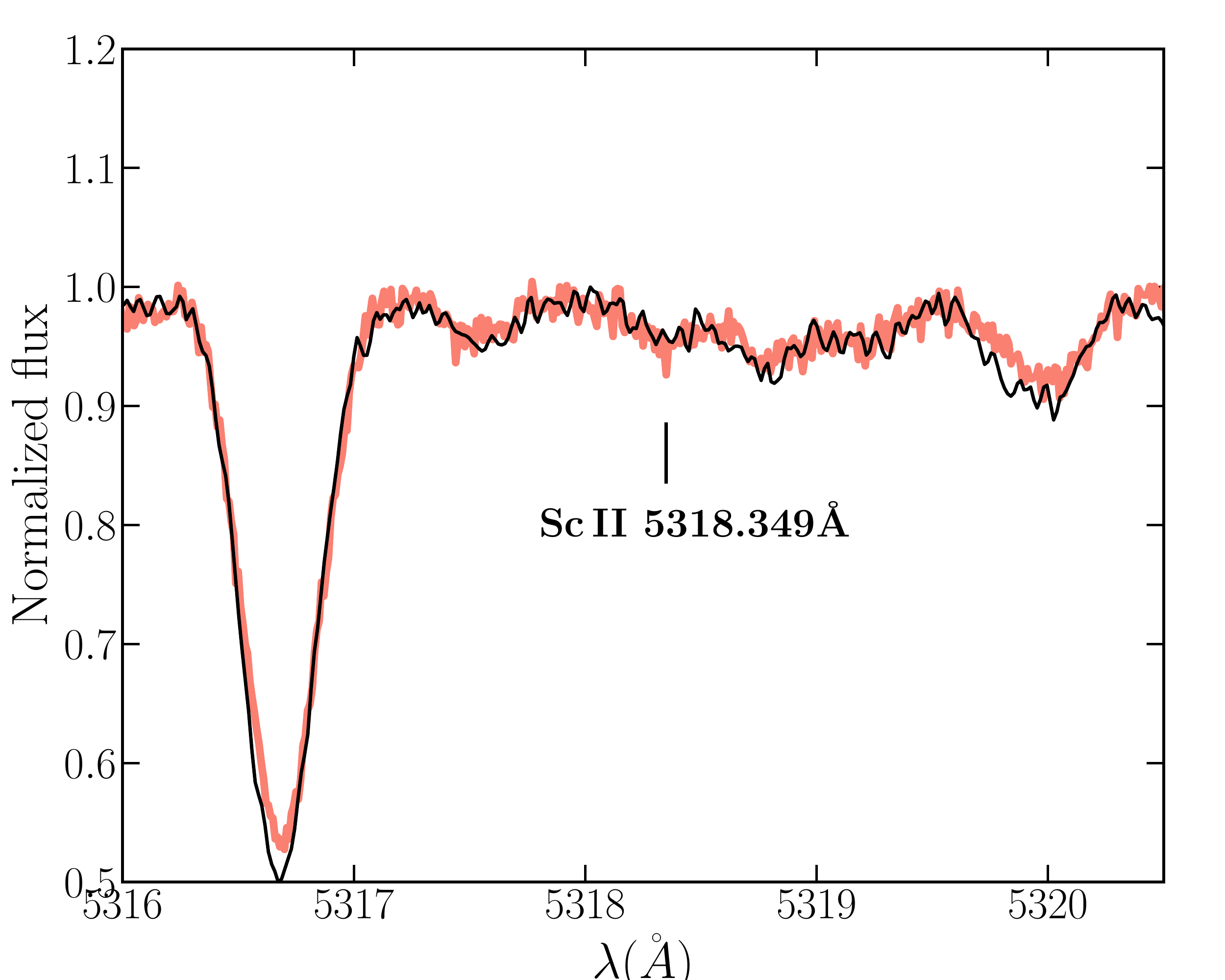}}
      \qquad
      \subfloat{\includegraphics[scale=0.22]{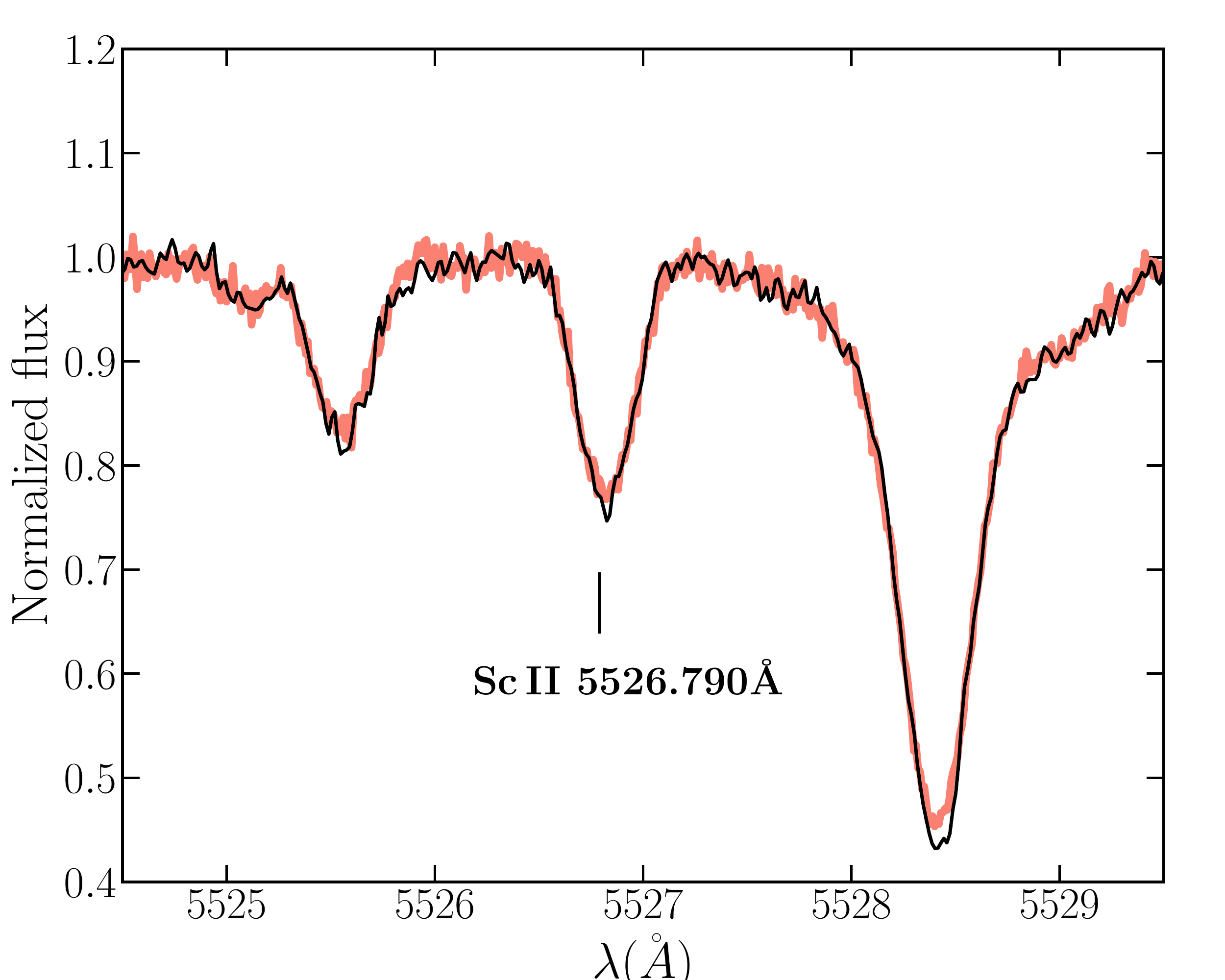}}
      \qquad
      \subfloat{\includegraphics[scale=0.22]{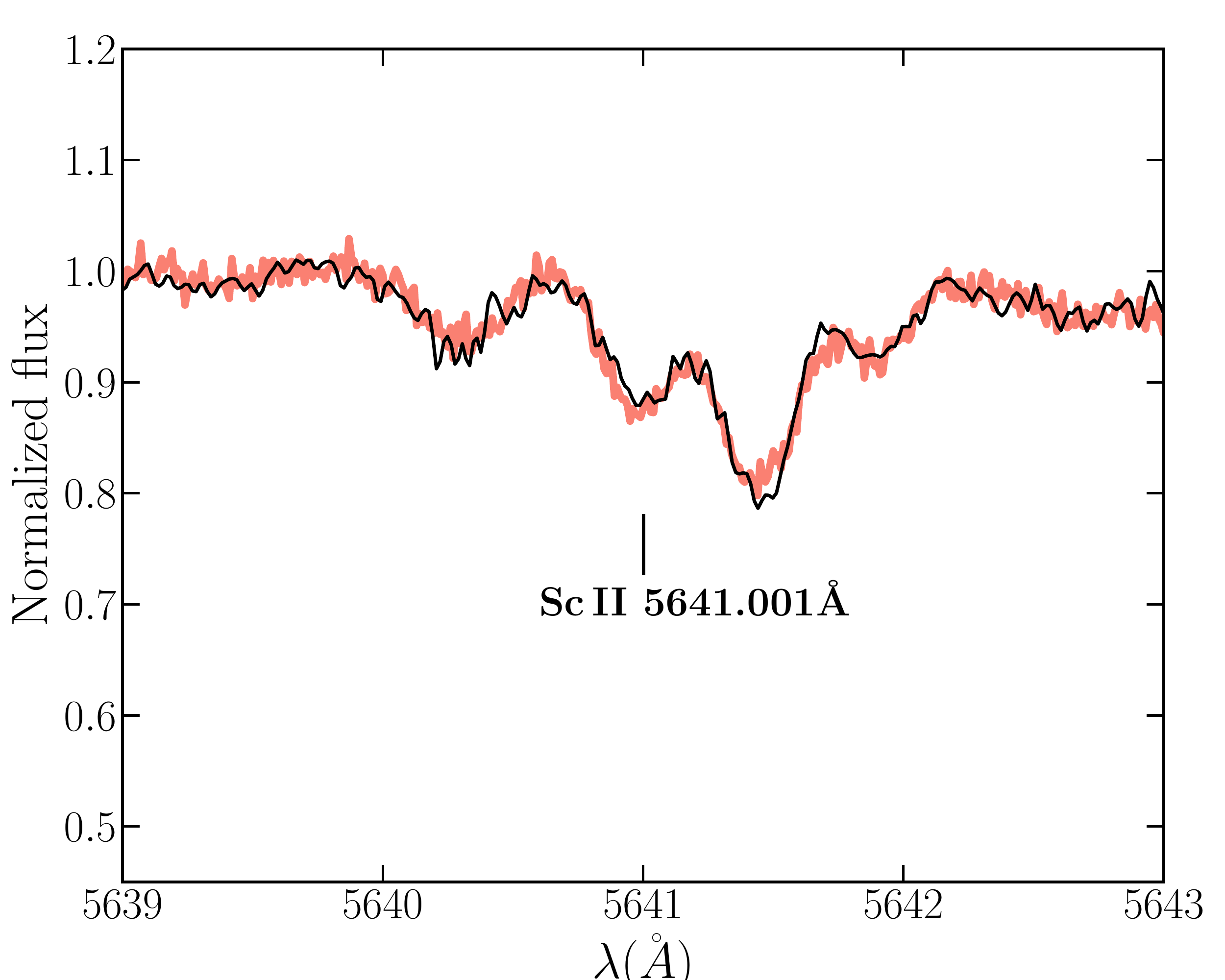}}
      \qquad
      \subfloat{\includegraphics[scale=0.22]{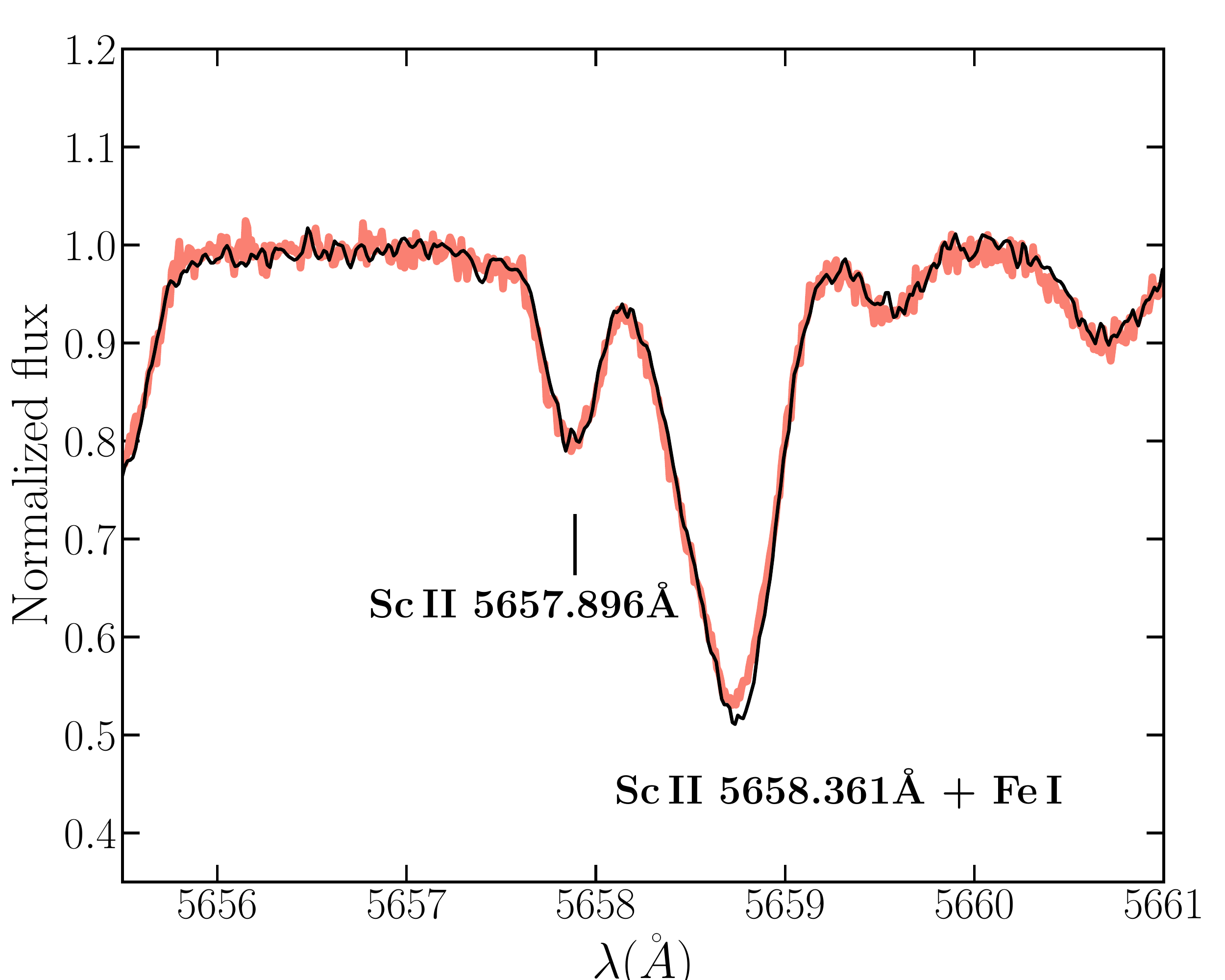}}
      \qquad
      \subfloat{\includegraphics[scale=0.22]{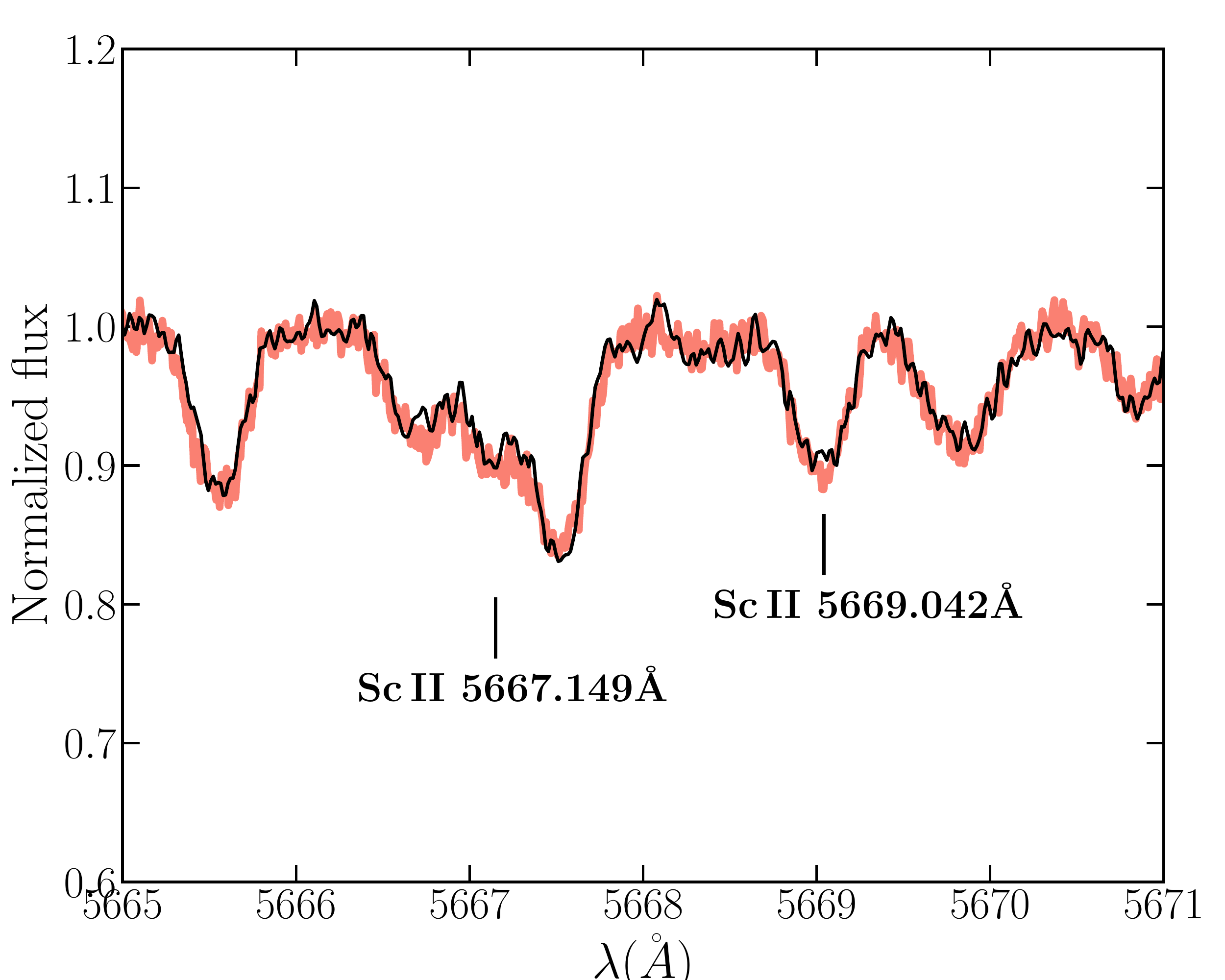}}
      \qquad
      \subfloat{\includegraphics[scale=0.22]{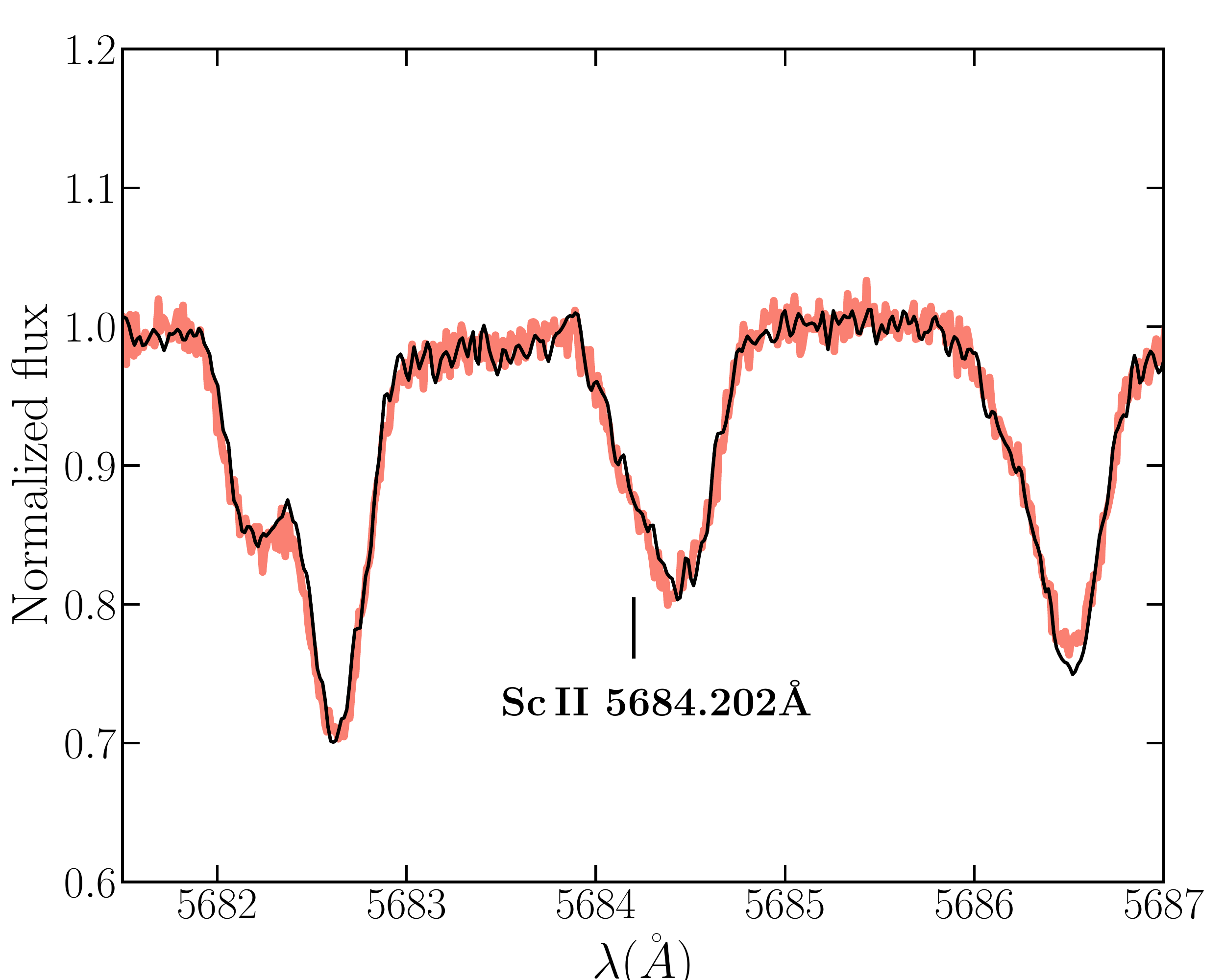}}
      \qquad
      \subfloat{\includegraphics[scale=0.22]{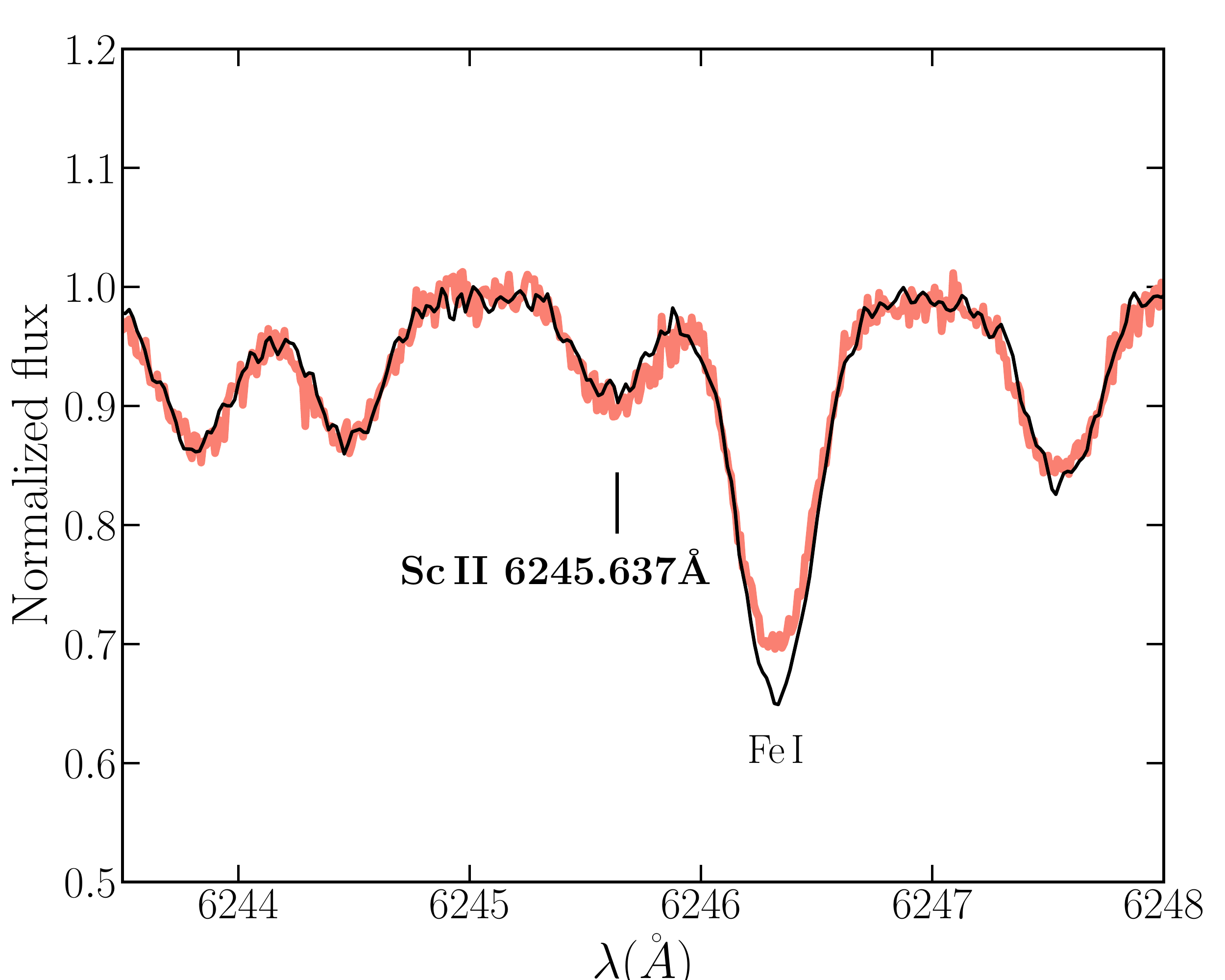}}
      \qquad
      \subfloat{\includegraphics[scale=0.22]{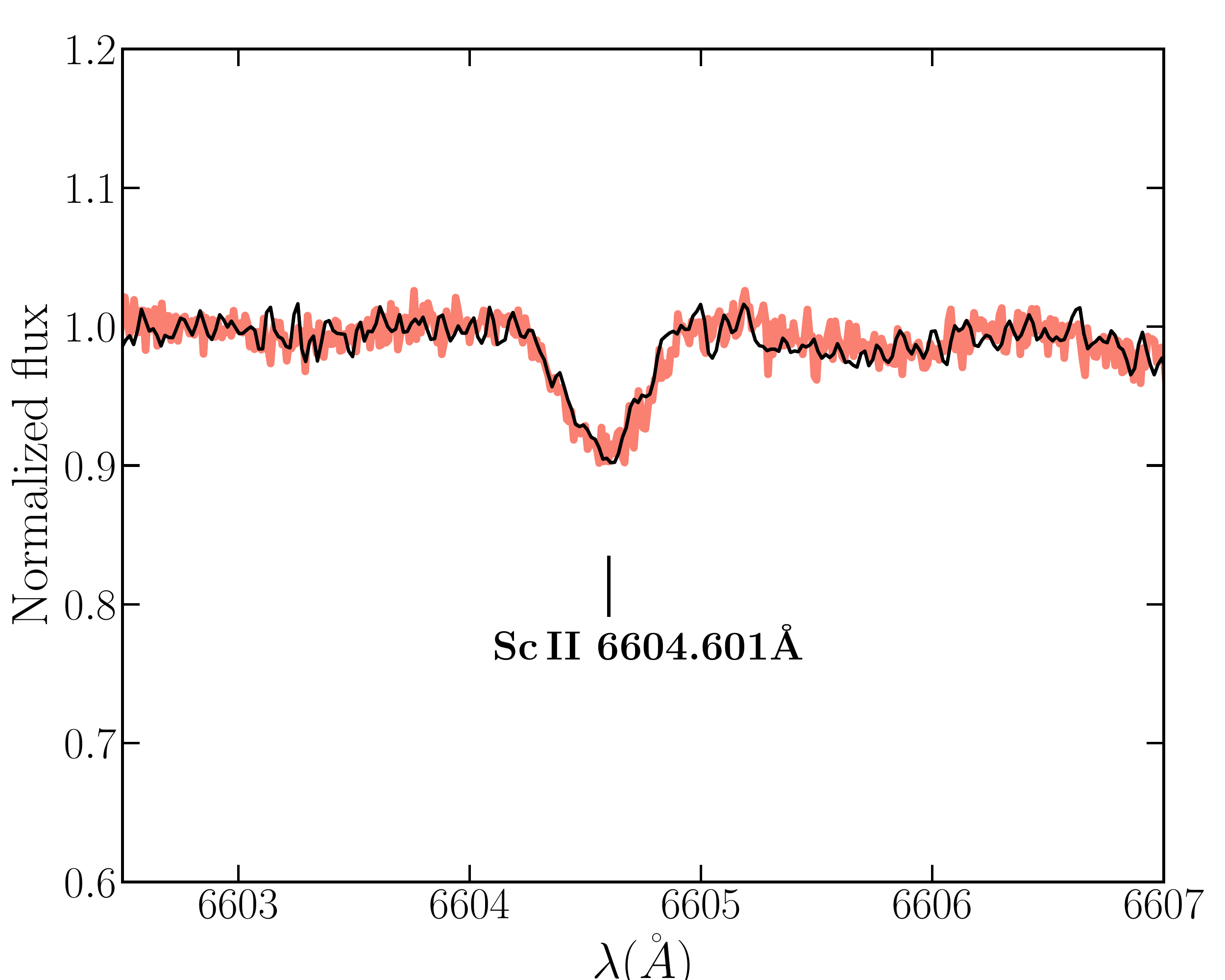}}
      \caption{ Comparison of Sc line profiles in the Sun (light pink line), with a rotational broadened profile of 11 \kms, and star 10442256-6415301 (black line).  }
        \label{comp_Sc_lines_v11}
\end{figure*}

\end{appendix}

\end{document}